\documentclass[11pt,a4paper,english]{article}
\pdfoutput=1
\usepackage{jheppub}

\usepackage[T1]{fontenc}

\def\equationautorefname~#1\null{eq.~(#1)\null}

\usepackage{chngpage}
\usepackage{subfig}

\usepackage{babel}
\usepackage{array}
\usepackage{longtable}
\usepackage{float}
\usepackage{url}
\usepackage{amsmath}
\usepackage{amssymb}

\usepackage{microtype}
\hypersetup{pdftitle={Cutting massless four-loop propagators},
 pdfauthor={Vitaly Magerya + Andrey Pikelner}}

\newcommand{\noun}[1]{\textsc{#1}}
\providecommand*{\strong}[1]{\textbf{#1}}
\providecommand*{\code}[1]{\texttt{#1}}

\newcommand\vvvv[1]{\ensuremath{\mathrm{VVVV}_{#1}}}
\newcommand\vvvr[1]{\ensuremath{\mathrm{VVVR}_{#1}}}
\newcommand\vvrv[1]{\ensuremath{\mathrm{VVRV}_{#1}}}
\newcommand\vvrr[1]{\ensuremath{\mathrm{VVRR}_{#1}}}
\newcommand\vrrv[1]{\ensuremath{\mathrm{VRRV}_{#1}}}
\newcommand\vrrr[1]{\ensuremath{\mathrm{VRRR}_{#1}}}
\newcommand\rrrr[1]{\ensuremath{\mathrm{RRRR}_{#1}}}

\newcommand{\fn}[2]{#1\!\left(#2\right)}

\newcommand\ep{\ensuremath{\varepsilon}}
\newcommand\homH{\ensuremath{\mathcal{H}}}
\newcommand\partR{\ensuremath{\mathcal{R}}}

\usepackage{tikz}

\definecolor{SapGreen}{HTML}{3B7B3B}
\definecolor{PhtaloGreen}{HTML}{2b7a76}
\definecolor{EmeraldGreen}{HTML}{1ea78d}
\definecolor{EnglishRed}{HTML}{b02427}
\definecolor{IronOxideRed}{HTML}{a13634}
\definecolor{CadmiumRed}{HTML}{df2b3f}
\definecolor{Vermillion}{HTML}{e14235}
\hypersetup{
    urlcolor=EnglishRed,
    linkcolor=EnglishRed,
    citecolor=EmeraldGreen
}

\usepackage{dsfont}

\allowdisplaybreaks[1]

\ifcsname pdfmdfivesum\endcsname
\else
    \ifcsname mdfivesum\endcsname
        
    \else
        \errmessage{Neither pdfmdfivesum nor mdfivesum commands are present. Fail}
    \fi
\fi

\usepackage{physics}
\usetikzlibrary{calc}
\usetikzlibrary{external}
\tikzset{baseline=-0.6ex}
\tikzset{inner sep=0}

\pgfdeclarelayer{nodelayer}
\pgfdeclarelayer{edgelayer}
\pgfsetlayers{edgelayer,nodelayer}
\tikzstyle{none}=[]

\tikzstyle{dot}=[fill=black, shape=circle, draw=black, inner sep=1pt]
\tikzstyle{blob}=[fill=white, draw={rgb,255: red,176; green,36; blue,39}, dashed, shape=circle, line width=1pt, inner sep=5pt]

\tikzstyle{edge}=[-, draw={rgb,255: red,176; green,36; blue,39}, line width=1pt, preaction={{draw=white,line width=2pt}}, line cap=rect]
\tikzstyle{massive edge}=[-, draw={rgb,255: red,133; green,119; blue,181}, line width=2.5pt, preaction={{draw=white,line width=3pt}}, line cap=rect]
\tikzstyle{cut edge}=[-, draw={rgb,255: red,147; green,151; blue,152}, line width=0.5pt, densely dashed, line cap=rect]
\tikzstyle{incoming edge}=[line width=1pt, line cap=rect, draw={rgb,255: red,102; green,102; blue,102}, {|-}]
\tikzstyle{outgoing edge}=[line width=1pt, line cap=rect, draw={rgb,255: red,102; green,102; blue,102}, ->]

\newcommand*{\smallfig}[1]{\raisebox{0.15ex}{\scalebox{0.75}{%
    \scalebox{0.75}{\input{fig/#1.tikz}}%
}}}

\usepackage{tkz-euclide}
\usetkzobj{all}

\usepackage{relsize}
\tikzset{fontscale/.style = {font=\relsize{#1}}
}

\title{
    \hfill{\small{\normalfont{DESY 19-133}}} \\~\\
Cutting massless four-loop propagators}

\author[a]{Vitaly Magerya,}
\emailAdd{vitaly.magerya@desy.de}
\affiliation[a]{II. Institut f\"ur Theoretische Physik, Universit\"at Hamburg,\\
Luruper Chaussee 149, 22761 Hamburg, Germany}

\author[b]{Andrey Pikelner}
\emailAdd{pikelner@theor.jinr.ru}
\affiliation[b]{Bogoliubov Laboratory of Theoretical Physics, Joint Institute\\
for Nuclear Research, 141980 Dubna, Russia}

\abstract{
Among the unitarity cuts of 4-loop massless propagators two kinds
are currently fully known: the 2-particle cuts with 3~loops corresponding
to form-factors, and the 5-particle phase-space integrals. In this
paper we calculate master integrals for the remaining ones: 3-particle
cuts with 2~loops, and 4-particle cuts with 1~loop.
The 4-particle cuts are calculated by solving dimensional
recurrence relations.
The 3-particle cuts are integrated directly using 1$\to$3
amplitudes with 2~loops, which we also re-derive here up to
transcendentality weight~7.
The results are verified both numerically, and by showing
consistency with previously known integrals using Cutkosky rules.
We provide the analytic results in the space-time dimension
$4-2\ep$ as series in~$\ep$ with coefficients being multiple
zeta values up to weight~12.
In the ancillary files we also provide dimensional recurrence
matrices and \noun{SummerTime} files suitable for numerical
evaluation of the series in arbitrary dimensions with any precision.
}
\begin{document}
\maketitle

\section{Introduction}
\label{sec:introduction}

\newcommand{\bbar}{\mkern 4.5mu\overline{\mkern-4.5mu B\mkern-1.0mu}\mkern 1.0mu}

Inclusive physical observables like total scattering cross
sections and related quantities are naturally defined within the
perturbation theory in terms of cut Feynman integrals.
Particularly, particle decay cross sections at the level of
N\textsuperscript{3}LO in massless QCD require the knowledge of
cuts of four-loop massless propagator-type integrals (two-point
functions).

While the optical theorem allows one to calculate massless total
cross sections~\cite{Gorishnii:1990vf,Baikov:2008jh} with only the knowledge of the discontinuities
of the propagators---without calculating each cut separately---the
knowledge of the cuts is necessary when they are used as a
building block in calculations of massive processes, exclusive
processes, or subtraction terms.
Examples of such cases include~\cite{CFHMSS15}, where a subset
of four-loop massless propagator cuts were used in the large-mass
expansion procedure needed for the boundary condition of
differential equations of $\bbar$ decay master integrals.
Also~\cite{GGG04}, where cuts of three-loop propagators were
used to develop infrared subtractions scheme for exclusive 2-jet
production at NNLO.
And finally~\cite{Mitov:2006wy,Gituliar15}, where cuts of three-loop massless
propagators were used in boundary conditions for differential
cross section master integrals.

The particular use case for cuts
of four-loop propagators we have in mind is the extraction of
NNLO time-like splitting functions from a semi-inclusive decay
cross section at N\textsuperscript{3}LO.
The time-like splitting functions are currently known from the
space-like case via an analytic continuation procedure, which leaves
one of the terms in quark-gluon and gluon-quark NNLO terms
undetermined~\cite{MMV06,MV07,AMV11}.
A direct calculation is needed to fix those terms.
As discussed in~\cite{GM15,Gituliar15} this direct calculation
requires the knowledge of decay cross sections differential in the
energy fraction of one of the outgoing partons---a calculation
for which the master integrals are not yet known, but can be
determined via the differential equations method~\cite{Kotikov:1990kg,Kotikov:1991pm}, as long as the
boundary conditions are known to fix the integration constants.
The cuts of four-loop propagators can be used for these boundary
conditions by noting that a differential cross section integrated
over its parameters must give precisely the fully inclusive
one, so integrating over a semi-inclusive master integral will
relate it to the cuts of four-loop propagators, providing enough
information to fix the integration constants.

These considerations motivate us to calculate all the cuts of
four-loop massless propagators.
Their 2-particle cuts (three-loop form factors) are already
known from~\cite{HHM07,HHKS09,LSS10}, 5-particle cuts are known
from~\cite{GMP18}, and in this article we shall complete this
knowledge by calculating the master integrals for the remaining
3- and 4-particle cuts.
To this end we shall use dimensional recurrence
relations~\cite{Tar96,Lee09} as well as direct phase-space
integration.
At the end we shall gather the values of all the cut master
integrals---old and new---and present all of them as $\ep$-series
in the space-time dimension of $4-2\ep$, with coefficients being
multiple zeta values (MZVs)~\cite{BBV09} up to transcendentality
weight~12.
As an intermediate step of our method we shall also present a
recalculation of 1$\to$3 two-loop amplitudes up to weight~7
(building upon the known weight-4 results from~\cite{GR00,GR01}).

\subsubsection*{What are cut integrals?}

To calculate a total cross section of an (off-shell or massive)
particle decay, one needs to integrate the probability density of
the final state over its phase space in all possible configurations,
\begin{equation}
\label{eq:sigma}
\sigma\sim\sum_{n}\int d\mathrm{PS}_{n}\big|\langle p_{1},\dots,p_{n}\vert S\vert q\rangle\big|^{2},
\end{equation}
where $S$ is the scattering matrix, and the phase-space element is defined as
\begin{equation}
d\mathrm{PS}_{n}=\left(2\pi\right)^{d}\delta^{d}\!\left(p_{1}+\dots+p_{n}-q\right)\prod_{i=1}^{n}\frac{d^{d}p_{i}}{\left(2\pi\right)^{d}}2\pi\delta\!\left(p_{i}^{2}\right)\Theta\!\left(p_{i}^{0}\right).\label{eq:dpsn-definition}
\end{equation}
Once the scattering amplitude $\langle p_{1},\dots,p_{n}\vert S\vert q\rangle$
is expanded in perturbation theory as a sum of Feynman diagrams,
\begin{equation}
\langle p_{1},\dots,p_{n}\vert S\vert q\rangle=A_{1\to n}^{\left(1\right)}+A_{1\to n}^{\left(2\right)}+\dots=\smallfig{amp1}+\smallfig{amp2}+\dots,
\end{equation}
expanding the modulus squared in eq.~\eqref{eq:sigma} gives rise to phase-space integrals of the form
\begin{equation}
\sigma\sim\int d\mathrm{PS}_{n}\,A_{1\to n}^{\left(1\right)}\left(A_{1\to n}^{\left(2\right)}\right)^{*}+\dots=\int d\mathrm{PS}_{n}\,\smallfig{amp1}\left(\smallfig{amp2}\right)^{*}+\dots.
\end{equation}
Each of the terms in this sum is a product of a decay amplitude,
a (different) conjugated amplitude, and a phase-space integration
operation.
Graphically, we denote these terms combined, with dashed likes
corresponding to the final-state momenta $p_i$,
\begin{equation}
\int d\mathrm{PS}_{3}\,\smallfig{amp1}\left(\smallfig{amp2}\right)^{*}=\int d\mathrm{PS}_{3}\,\smallfig{amp1}\;\smallfig{amp2conj}=\smallfig{amp1x2}.
\end{equation}

These are the cut integrals. One might view them as two-point Feynman
integrals partitioned into two parts---the ``left'' and the ``right''---with
all propagators between the two parts set on shell (or ``cut''), and every vertex
and loop integral in the ``right'' part complex-conjugated. This
view is useful for the optical theorem, which relates the discontinuity
of a virtual (uncut) diagram to its cuts.

Note that after cutting the integral into two parts, it is possible
to continue the process and cut each of the parts further, producing
``generalized cuts''~\cite{ABD14}, as opposed to the ``unitarity
cuts'', which we have here.

The goal of this paper is to calculate master integrals for all the
(unitarity) cuts of 4-loop massless two-point functions (``propagators'').

\subsubsection*{Notation for the integrals}

Throughout the paper we shall define our cut integrals in $d$ space-time dimensions as
\begin{equation}
    \label{eq:cut-definition}
    I = \int
        \underbrace{
            \left( \prod_i \frac{d^d l_i}{\left(2\pi\right)^d} \right)
            \left( \prod_j \frac{1}{k_j^2 + i0} \right)
        }_{\text{``left'' amplitude}}
        \underbrace{
            \left( \prod_{i'} \frac{d^d l_{i'}'}{\left(2\pi\right)^d} \right)
            \left( \prod_{j'} \frac{1}{k_{j'}'^2 - i0} \right)
        }_{\text{``right'' amplitude}}
        d\mathrm{PS}_n,
\end{equation}
where $d\mathrm{PS}_n$ is the same as eq.~\eqref{eq:dpsn-definition};
$l$ and $l'$ are the loop momenta; $k$ and $k'$ are the propagator momenta,
being linear combinations of $l$, $l'$, and the cut momenta $p_i$;
and the signs of the $i0$ prescription depend on whether the
propagator is to the left or to the right of the cut.
Note that the $i0$ prescription is only relevant for the propagators
involved in the loops, it does not matter for the rest of them.

In practical terms, it is often convenient to factor out a
common prefactor from this definition, and only work with the
normalized integrals without it.
For an integral in $d$ dimensions with $n$ cut lines, $m_L$
loops to the left of the cut, $m_R$ loops to the right, and $p$
propagators we shall factor out the full $n$-particle phase
space $\mathrm{PS}_n$, $m_L+m_R$ one-loop bubbles, and the full
$q^2$ dependence as follows:
\begin{equation}
    \label{eq:normalized-cut-definition}
    I = B^{m_L} \left(B^*\right)^{m_R} \mathrm{PS}_n \left( q^2 \right)^{\frac{d}{2}\left(n+m_L+m_R-1\right) -p -n} J,
\end{equation}
with the bubble $B$ given by
\begin{equation}
    \label{eq:normalization-b}
    B=\left. \smallfig{normG} \right|_{q^2=1}=
    \left(-1-i0\right)^{\frac{d-4}{2}}\frac{i\pi^{\frac{d}{2}}}{\left(2\pi\right)^{d}}\frac{\Gamma^{2}\!\left(\frac{d}{2}-1\right)\Gamma\!\left(2-\frac{d}{2}\right)}{\Gamma\!\left(d-2\right)},
\end{equation}
and the full $n$-particle phase space by
\begin{equation}
    \label{eq:normalization-psn}
    \mathrm{PS}_{n}=\left. \smallfig{normPS} \right|_{q^2=1}=
    \frac{2\pi}{\left(4\pi\right)^{\frac{d}{2}\left(n-1\right)}}\frac{\Gamma^{n}\!\left(\frac{d}{2}-1\right)}{\Gamma\!\left(\left(\frac{d}{2}-1\right)\left(n-1\right)\right)\Gamma\!\left(\left(\frac{d}{2}-1\right)n\right)}.
\end{equation}
Note that both normalization factors---$B$ and $\mathrm{PS}_n$---
are dimensionless, so the power of $q^2$ directly corresponds
to the dimensionality of the integral.

This normalization removes all $q^2$ dependence along with any
imaginary numbers from $J$, making the normalized integrals
real functions of $d$, and removing the distinction between the
``left'' and the ``right'' amplitudes (this distinction is fully
captured by the prefactors).
It also cancels the surface UV divergencies of the integrals;
this, for example, makes one-loop integrals finite when $d$ is
high enough to suppress the IR divergences.
Finally, it normalizes all of the following integrals to unity,
simplifying dimensional recurrence relations:
$\vvvv{1}$ from \autoref{tab:VVVV},
$\vvvr{1}$ from \autoref{tab:VVVR},
$\vvrv{3}$ from \autoref{tab:VVRV},
$\vrrv{1}$ from \autoref{tab:VRRV},
$\vvrr{4}$ from \autoref{tab:VVRR},
$\vrrr{16}$ from \autoref{tab:VRRR},
and $\rrrr{1}$ from \autoref{tab:RRRR}.

\subsubsection*{Classification of the cuts}

Following the natural structure of the cut integrals we will classify
them by the number of loops to the left of the cut (denoted as ``V'',
for ``virtual''), the number of propagators cut (``R'', for ``real''),
and the loop count to the right (``V'' again).

For 4-loop propagators there are 6 classes of cuts: two-particle
cuts VVVR and VVRV (see \autoref{tab:VVVR} and \autoref{tab:VVRV}),
three-particle cuts VVRR and VRRV (\autoref{tab:VVRR} and
\autoref{tab:VRRV}), four-particle cuts VRRR (\autoref{tab:VRRR}),
and five-particle cuts RRRR (\autoref{tab:RRRR}). The purely
virtual integrals with no cuts we shall denote as VVVV
(\autoref{tab:VVVV}).

\subsubsection*{State of the art}

The master integrals for 4-loop massless propagators (VVVV) have been
first calculated in~\cite{BC10} up to transcendentality weight
seven, and updated to weight twelve in~\cite{LSS11}.
Importantly, the latter provides the results in terms of quickly
convergent nested infinite sums, suitable for numerical evaluation
to arbitrary precision in arbitrary space-time dimension $d$ with
the \noun{SummerTime} package~\cite{LM15}.

The master integrals for 2-particle cuts of these propagators (VVVR
and VVRV) correspond to three-loop massless form-factor integrals;
these were calculated in~\cite{HHM07,HHKS09,LSS10}.
The corresponding \noun{SummerTime} files for these results are
distributed with the package itself.

The master integrals for 5-particle cuts (RRRR)---the five-particle
purely phase-space integrals---were calculated in~\cite{GMP18},
and \noun{SummerTime} files are available for these results too.

What still remains unfinished are the 3- and 4-particle cuts: VRRR,
VVRR, and VRRV.
A subset of these---11 integrals in total---was calculated
in~\cite{CFHMSS15} with either direct integration or through
Mellin-Barnes representation, with results provided in terms
of hypergeometric functions and/or series in $\ep$ up to transcendentality weight~6.
In this article we shall compute all these cuts completely.

\section{4-loop virtual integrals (VVVV)}
\label{sec:vvvv}

There are 28 master integrals for 4-loop massless propagators
in total.
These were calculated in~\cite{BC10}, and we have depicted them
in \autoref{tab:VVVV}.
Five topologies listed in \autoref{tab:topologies} are sufficient
to express all of these master integrals.
The first three of them directly correspond to the
master integrals with 11 propagators ($\vvvv{25}$, $\vvvv{26}$,
and $\vvvv{27}$ respectively), and the other two are sufficient
to cover all the remaining simpler integrals.

\begin{table}[H]
\begin{tabular}{>{\raggedleft}p{0.3\textwidth}>{\raggedright}m{0.6\textwidth}}
%
    \scalebox{0.75}{\input{fig/topo/h.tikz}}%
 & Topology H. Propagators in the indicated order: $q-p_{1}$; $q-p_{1}-p_{2}+p_{4}$; $q-p_{1}-p_{2}-p_{3}$;
$p_{1}+p_{2}+p_{3}$; $p_{1}+p_{2}$; $p_{1}$; $p_{4}-p_{2}$; $p_{2}$;
$p_{3}+p_{4}$; $p_{3}$; $p_{4}$.\tabularnewline
 & \tabularnewline
%
    \scalebox{0.75}{\input{fig/topo/n2.tikz}}%
 & Topology M. Propagators: $q-p_{1}$; $q-p_{1}+p_{2}$; $q-p_{1}+p_{2}-p_{3}$;
$q-p_{1}+p_{2}-p_{3}+p_{4}$; $p_{1}-p_{2}+p_{3}-p_{4}$; $p_{4}-p_{1}-p_{3}$;
$p_{1}-p_{4}$; $p_{1}$; $p_{2}$; $p_{3}$; $p_{4}$.\tabularnewline
 & \tabularnewline
%
    \scalebox{0.75}{\input{fig/topo/n1.tikz}}%
 & Topology N. Propagators: $q-p_{1}$; $q-p_{1}+p_{4}$; $q-p_{1}+p_{2}+p_{4}$;
$q-p_{1}+p_{2}+p_{3}+p_{4}$; $p_{1}-p_{2}-p_{3}-p_{4}$; $p_{1}-p_{2}-p_{3}$;
$p_{1}-p_{2}$; $p_{1}$; $p_{2}$; $p_{3}$; $p_{4}$.\tabularnewline
 & \tabularnewline
%
    \scalebox{0.75}{\input{fig/topo/l.tikz}}%
 & Topology L. Propagators: $q+p_{1}-p_{2}$;
$p_{2}-p_{1}$; $q-p_{2}$; $p_{1}$; $q-p_{3}$; $p_{3}-p_{2}$;
$p_{4}-p_{3}$; $q-p_{4}$; $p_{2}$; $p_{3}$; $p_{4}$.\tabularnewline
 & \tabularnewline
%
    \scalebox{0.75}{\input{fig/topo/j.tikz}}%
 & Topology J. Propagators: $q+p_{1}-p_{2}$; $p_{2}-p_{1}$; $q-p_{2}$;
$p_{1}$; $q-p_{2}+p_{3}-p_{4}$; $p_{4}-p_{3}$; $p_{3}-p_{2}$;
$q-p_{4}$; $p_{2}$; $p_{3}$; $p_{4}$.\tabularnewline
\end{tabular}
\caption{Generic topologies for 4-loop propagator master integrals.}
\label{tab:topologies}
\end{table}

\begin{table}[!t]
\begin{tabular}{>{\centering}p{2.6cm}>{\centering}p{2.6cm}>{\centering}p{2.6cm}>{\centering}p{2.6cm}>{\centering}p{2.6cm}}
%
    \scalebox{0.75}{\input{fig/vvvv/1.tikz}}%
 & %
    \scalebox{0.75}{\input{fig/vvvv/2.tikz}}%
 & %
    \scalebox{0.75}{\input{fig/vvvv/3.tikz}}%
 & %
    \scalebox{0.75}{\input{fig/vvvv/4.tikz}}%
 & %
    \scalebox{0.75}{\input{fig/vvvv/5.tikz}}%
\tabularnewline
$1,\mathrm{L}_{11101001111}$ & $2,\mathrm{L}_{01111001011}$ & $3,\mathrm{H}_{01101110010}$ & $4,\mathrm{H}_{01111110001}$ & $5,\mathrm{H}_{00100111110}$\tabularnewline
 &  &  &  & \tabularnewline
%
    \scalebox{0.75}{\input{fig/vvvv/6.tikz}}%
 & %
    \scalebox{0.75}{\input{fig/vvvv/7.tikz}}%
 & %
    \scalebox{0.75}{\input{fig/vvvv/8.tikz}}%
 & %
    \scalebox{0.75}{\input{fig/vvvv/9.tikz}}%
 & %
    \scalebox{0.75}{\input{fig/vvvv/10.tikz}}%
\tabularnewline
$6,\mathrm{H}_{10110111110}$ & $7,\mathrm{H}_{00100101101}$ & $8,\mathrm{H}_{10110101101}$ & $9,\mathrm{H}_{10101011110}$ & $10,\mathrm{H}_{01011110101}$\tabularnewline
 &  &  &  & \tabularnewline
%
    \scalebox{0.75}{\input{fig/vvvv/11.tikz}}%
 & %
    \scalebox{0.75}{\input{fig/vvvv/12.tikz}}%
 & %
    \scalebox{0.75}{\input{fig/vvvv/13.tikz}}%
 & %
    \scalebox{0.75}{\input{fig/vvvv/14.tikz}}%
 & %
    \scalebox{0.75}{\input{fig/vvvv/15.tikz}}%
\tabularnewline
$11,\mathrm{H}_{00101110110}$ & $12,\mathrm{M}_{01110111111}$ & $13,\mathrm{H}_{01110110011}$ & $14,\mathrm{H}_{01101111110}$ & $15,\mathrm{H}_{01110101001}$\tabularnewline
 &  &  &  & \tabularnewline
%
    \scalebox{0.75}{\input{fig/vvvv/16.tikz}}%
 & %
    \scalebox{0.75}{\input{fig/vvvv/17.tikz}}%
 & %
    \scalebox{0.75}{\input{fig/vvvv/18.tikz}}%
 & %
    \scalebox{0.75}{\input{fig/vvvv/19.tikz}}%
 & %
    \scalebox{0.75}{\input{fig/vvvv/20.tikz}}%
\tabularnewline
$16,\mathrm{L}_{11011101011}$ & $17,\mathrm{H}_{10111110101}$ & $18,\mathrm{N}_{11011101111}$ & $19,\mathrm{H}_{01110111011}$ & $20,\mathrm{N}_{11111111110}$\tabularnewline
 &  &  &  & \tabularnewline
%
    \scalebox{0.75}{\input{fig/vvvv/21.tikz}}%
 & %
    \scalebox{0.75}{\input{fig/vvvv/22.tikz}}%
 & %
    \scalebox{0.75}{\input{fig/vvvv/23.tikz}}%
 & %
    \scalebox{0.75}{\input{fig/vvvv/24.tikz}}%
 & %
    \scalebox{0.75}{\input{fig/vvvv/25.tikz}}%
\tabularnewline
$21,\mathrm{N}_{10111101111}$ & $22,\mathrm{H}_{01101110011}$ & $23,\mathrm{N}_{10111110111}$ & $24,\mathrm{H}_{11111101101}$ & $25,\mathrm{H}_{11111111111}$\tabularnewline
 &  &  &  & \tabularnewline
%
    \scalebox{0.75}{\input{fig/vvvv/26.tikz}}%
 & %
    \scalebox{0.75}{\input{fig/vvvv/27.tikz}}%
 & %
    \scalebox{0.75}{\input{fig/vvvv/28.tikz}}%
 &  & \tabularnewline
$26,\mathrm{M}_{11111111111}$ & $27,\mathrm{N}_{11111111111}$ & $28,\mathrm{J}_{11101111111}$ &  & \tabularnewline
\end{tabular}

\caption{\protect\strong{VVVV}, master integrals for 4-loop massless propagators.
The subscripts of the topology names are the indices, they denote
powers of the corresponding propagators.}
\label{tab:VVVV}
\end{table}

To identify the master integrals for the cuts one can try to
solve the integration-by-parts (IBP) relations~\cite{CT81}, but
a simpler procedure turns out to be sufficient: it is enough to
construct all possible 2-, 3-, 4-, and 5-particle cuts of these
28 VVVV integrals, and remove the symmetric duplicates from the
obtained set.
Further application of IBP reduction reveals no additional linear
relations between the obtained cut integrals---we have
checked this by using a combination of \noun{Fire}~\cite{Smi14}
and \noun{LiteRed}~\cite{Lee13} during the computation of
dimensional recurrence relations.
Moreover, we believe that the set of integrals obtained this way
constitutes a complete basis, because by the optical theorem any
full cross section can be expressed via the discontinuity of the
VVVV integrals, and any discontinuity of those can be expressed as
a linear combination of these cut integrals (we shall construct
these relations explicitly in \autoref{sec:Cutkosky-rules}).

\section{1-loop 4-particle cut integrals (VRRR)}
\label{sec:vrrr}

\begin{table}[H]
\begin{tabular}{>{\centering}p{2.6cm}>{\centering}p{2.6cm}>{\centering}p{2.6cm}>{\centering}p{2.6cm}>{\centering}p{2.6cm}}
%
    \scalebox{0.75}{\input{fig/vrrr/1.tikz}}%
 & %
    \scalebox{0.75}{\input{fig/vrrr/2.tikz}}%
 & %
    \scalebox{0.75}{\input{fig/vrrr/3.tikz}}%
 & %
    \scalebox{0.75}{\input{fig/vrrr/4.tikz}}%
 & %
    \scalebox{0.75}{\input{fig/vrrr/5.tikz}}%
\tabularnewline
$1,\mathrm{H}_{00*01**0*10}$ & $2,\mathrm{H}_{0*1*01100**}$ & $3,\mathrm{N}_{101*11*01**}$ & $4,\mathrm{H}_{00*00*1*1*0}$ & $5,\mathrm{M}_{011*01*1**1}$\tabularnewline
 &  &  &  & \tabularnewline
%
    \scalebox{0.75}{\input{fig/vrrr/6.tikz}}%
 & %
    \scalebox{0.75}{\input{fig/vrrr/7.tikz}}%
 & %
    \scalebox{0.75}{\input{fig/vrrr/8.tikz}}%
 & %
    \scalebox{0.75}{\input{fig/vrrr/9.tikz}}%
 & %
    \scalebox{0.75}{\input{fig/vrrr/10.tikz}}%
\tabularnewline
$6,\mathrm{N}_{10*11*101**}$ & $7,\mathrm{H}_{10*11**0*01}$ & $8,\mathrm{N}_{1*01**0*111}$ & $9,\mathrm{H}_{10*10*1*1*0}$ & $10,\mathrm{H}_{01*10*1*0*1}$\tabularnewline
 &  &  &  & \tabularnewline
%
    \scalebox{0.75}{\input{fig/vrrr/11.tikz}}%
 & %
    \scalebox{0.75}{\input{fig/vrrr/12.tikz}}%
 & %
    \scalebox{0.75}{\input{fig/vrrr/13.tikz}}%
 & %
    \scalebox{0.75}{\input{fig/vrrr/14.tikz}}%
 & %
    \scalebox{0.75}{\input{fig/vrrr/15.tikz}}%
\tabularnewline
$11,\mathrm{N}_{1*11**1*110}$ & $12,\mathrm{H}_{01*01*1*1*0}$ & $13,\mathrm{M}_{11*1*1**111}$ & $14,\mathrm{N}_{1*11**1*111}$ & $15,\mathrm{H}_{11*11*1*1*1}$\tabularnewline
 &  &  &  & \tabularnewline
%
    \scalebox{0.75}{\input{fig/vrrr/16.tikz}}%
 & %
    \scalebox{0.75}{\input{fig/vrrr/17.tikz}}%
 & %
    \scalebox{0.75}{\input{fig/vrrr/18.tikz}}%
 & %
    \scalebox{0.75}{\input{fig/vrrr/19.tikz}}%
 & %
    \scalebox{0.75}{\input{fig/vrrr/20.tikz}}%
\tabularnewline
$16,\mathrm{H}_{0*110*0*00*}$ & $17,\mathrm{N}_{*0111*10**1}$ & $18,\mathrm{N}_{10*1110**1*}$ & $19,\mathrm{L}_{1*0***01011}$ & $20,\mathrm{H}_{1*11*111**1}$\tabularnewline
 &  &  &  & \tabularnewline
%
    \scalebox{0.75}{\input{fig/vrrr/21.tikz}}%
 & %
    \scalebox{0.75}{\input{fig/vrrr/22.tikz}}%
 & %
    \scalebox{0.75}{\input{fig/vrrr/23.tikz}}%
 & %
    \scalebox{0.75}{\input{fig/vrrr/24.tikz}}%
 & %
    \scalebox{0.75}{\input{fig/vrrr/25.tikz}}%
\tabularnewline
$21,\mathrm{H}_{11*1*101*0*}$ & $22,\mathrm{N}_{111*11*11**}$ & $23,\mathrm{H}_{10*1*110*0*}$ & $24,\mathrm{N}_{1*01*101**1}$ & $25,\mathrm{H}_{0*110*1*01*}$\tabularnewline
 &  &  &  & \tabularnewline
%
    \scalebox{0.75}{\input{fig/vrrr/26.tikz}}%
 & %
    \scalebox{0.75}{\input{fig/vrrr/27.tikz}}%
 & %
    \scalebox{0.75}{\input{fig/vrrr/28.tikz}}%
 & %
    \scalebox{0.75}{\input{fig/vrrr/29.tikz}}%
 & %
    \scalebox{0.75}{\input{fig/vrrr/30.tikz}}%
\tabularnewline
$26,\mathrm{M}_{111*11*1**1}$ & $27,\mathrm{N}_{11*1***1110}$ & $28,\mathrm{N}_{1*11*111**1}$ & $29,\mathrm{N}_{11*1***1111}$ & $30,\mathrm{N}_{10*11*011**}$\tabularnewline
 &  &  &  & \tabularnewline
%
    \scalebox{0.75}{\input{fig/vrrr/31.tikz}}%
 & %
    \scalebox{0.75}{\input{fig/vrrr/32.tikz}}%
 & %
    \scalebox{0.75}{\input{fig/vrrr/33.tikz}}%
 & %
    \scalebox{0.75}{\input{fig/vrrr/34.tikz}}%
 & %
    \scalebox{0.75}{\input{fig/vrrr/35.tikz}}%
\tabularnewline
$31,\mathrm{M}_{11*11*11*1*}$ & $32,\mathrm{N}_{11*11*111**}$ & $33,\mathrm{H}_{11*1*111*1*}$ & $34,\mathrm{J}_{1110***11*1}$ & $35,\mathrm{J}_{11*0*11*1*1}$\tabularnewline
\end{tabular}

\caption{\protect\strong{VRRR}, 4-particle cut master integrals.
The index ``$*$'' denotes cut propagators.}
\label{tab:VRRR}
\end{table}

There are 35 master integrals for the VRRR cuts in total,
all depicted in \autoref{tab:VRRR}.
Out of these, integrals 1, 2, 3, 4, 5, 10, 12, and 18 have been
calculated in~\cite{CFHMSS15}.
Integral 16 is trivial, and normalizes to unity via eq.~\eqref{eq:normalized-cut-definition}.
Integrals 19, 34, and 35 have the one-loop amplitude factorized, and
thus can be expressed via the four-particle phase-space integrals
from~\cite{GGH03} (recomputed to weight 12 in~\cite{GMP18}).

\subsection{Direct integration over the phase space}

The four-particle phase-space is quite complicated~\cite{ERT80,KL86,GGH03}.
The parametrization of $d\mathrm{PS}_4$ from eq.~\eqref{eq:dpsn-definition} in terms of the scalar products
\begin{equation}
    s_{ij}=\frac{1}{q^{2}}\left(p_{i}+p_{j}\right)^{2},\qquad\text{where } 1\le i<j\le 4,
\end{equation}
has 5 degrees of freedom, and the following form:
\begin{equation}
    \label{eq:dps4-definition}
    d\mathrm{PS}_4=
        \left(q^{2}\right)^{\frac{3d-4}{2}} 
        \frac{ 2^{4-4d} \pi^{\frac{1}{2}-\frac{3}{2}d} }{ \Gamma\!\left(d-3\right) \Gamma\!\left(\frac{d-1}{2}\right) }
        \left(\Delta_{4}\right)^{\frac{d-5}{2}}
        \Theta\!\left(\Delta_{4}\right) \Theta\!\left(s_{ij}\right) \delta\!\left(1-\sum s_{ij}\right)\prod ds_{ij},
\end{equation}
where $\Delta_{4}$ is the Gram determinant,
\begin{equation}
\Delta_{4}=-\det\left|\begin{array}{cccc}
0 & s_{12} & s_{13} & s_{14}\\
s_{12} & 0 & s_{23} & s_{24}\\
s_{13} & s_{23} & 0 & s_{34}\\
s_{14} & s_{24} & s_{34} & 0
\end{array}\right|.
\end{equation}
Because the shape of the integration region is given by $\Theta\!
\left(\Delta_{4}\right) \Theta\!\left(s_{ij}\right)$, with $\Delta_{4}$ being
a polynomial of the third degree, integrating over this region
in the general case is a challenge.
Parametrizations such as the ``tripole parametrization''~\cite{GGH03,ERT80}
exist that remap this shape onto a hypercube, but they do so
at the expense of introducing square roots in the mapping from
$s_{ij}$ to the new parameters, which only allows one to complete the
integration analytically if the said roots drop out from
the integrand.

Still, for simpler integrals such an approach is viable.
Integrating $\vrrr{1}$ using the tripole parametrization, we find
\begin{equation}
    \label{eq:vrrr1}
    \vrrr{1}=\smallfig{vrrr/1}=B^* \mathrm{PS}_4 \left( q^2 \right)^{2d-6} \frac{
        \Gamma\!\left( 2d-4 \right) \Gamma\!\left( \frac{3}{2}d-4 \right)
    }{
        \Gamma\!\left( d-2 \right) \Gamma\!\left( \frac{5}{2}d-6 \right)
    }.
\end{equation}
Similarly for $\vrrr{4}$ we get
\begin{equation}
    \label{eq:vrrr4}
    \vrrr{4}=\smallfig{vrrr/4}=B\,\mathrm{PS}_4 \left( q^2 \right)^{2d-6} \frac{ 2d-5 }{ d-3 } \frac{
        \Gamma\!\left( \frac{3}{2}d-3 \right) \Gamma\!\left( \frac{3}{2}d-4 \right)
    }{
        \Gamma\!\left( \frac{1}{2}d-1 \right) \Gamma\!\left( \frac{5}{2}d-6 \right)
    }.
\end{equation}
In~\cite{CFHMSS15} a number of other VRRR integrals were
calculated in a similar fashion.
In the general case though, this approach does not apply, and
thus we shall turn to computing these integrals differently: by
solving the dimensional recurrence relations.

\subsection{An overview of dimensional recurrence relations}

Any Feynman integral $I$ can be cast into a parametric representation
(e.g. Feynman) in which the space-time dimension $d$
becomes a free parameter.
This is true for cut integrals as well, where a Baikov-like
representation can be used (we shall use it in \autoref{sec:drr-for-vrrr}).
With an insight gained from~\cite{Tar96}, it is then possible
to shift $d$ by $\pm2$, and express the resulting integral
$I\!\left(d\pm2\right)$ as a linear combination of integrals in
space-time dimension $d$.
Thus, a set of master integrals with shifted dimension
$J_i(d\pm2)$ can be expressed as a linear combination of the
same master integrals in space-time dimension~$d$ via the IBP
reduction, giving us dimensional recurrence
relations (DRR).
Using the terminology of~\cite{Lee13}, the ``lowering'' DRR take
the form
\begin{equation}
  \label{eq:ldrr}
  J_i(d+2) = M_{ij}(d)\,J_j(d).
\end{equation}
We have constructed such DRR systems for all the cut configurations
and VVVV using \noun{LiteRed} with \noun{Fire}.
For all of them the matrices $M$ turn out to be triangular, and
eq.~\eqref{eq:ldrr} can be easily split into the homogeneous and
inhomogeneous parts,
\begin{equation}
  \label{eq:ldrr-triangular}
    J_i(d+2) = M_{ii}(d)\,J_i(d) + \sum_{j<i}M_{ij}(d)\,J_j(d).
\end{equation}
The triangular form here is guaranteed by the fact that only a
single master integral is present in each sector, and thus no
coupled blocks form.
The general solution of this recurrence relation system can be
represented as
\begin{equation}
  \label{eq:drr-gen-sol}
  J_i(d) = \homH_i(d)\,\omega_i(d) + \partR_i(d),
\end{equation}
where
\begin{itemize}
    \item
        $\homH_i(d)$ is a homogeneous solution satisfying
        $\homH_i(d+2) = M_{ii}(d) \homH_i(d)$;
    \item
        $\partR_i(d)$ is a particular solution, determined by
        integrals from the lower sectors $J_{j<i}$;
    \item $ \omega_i(d)$ is an arbitrary periodic function, such that
        $\omega_i(d+2)=\omega_i(d)$; this function cannot be determined from
        the DRR relations alone, and needs to be fixed separately.
\end{itemize}
The triangular form greatly simplifies the construction of the
homogeneous solution compared to the case of coupled blocks
(explored for example in \cite{LS12,LM17b}).
For the diagonal entries $M_{ii}$ of the form
\begin{equation}
    \label{eq:drr-mii-form}
    M_{ii}(d)=\mathcal{C}\prod_{k}\left(\frac{d}{2}-a_{k}\right)^{n_{k}}\!,
\end{equation}
the homogeneous solution can be immediately constructed as
\begin{equation}
    \label{eq:drr-homogeneous-solution}
    \homH_i(d)=\mathcal{C}^{\frac{d}{2}}\prod_{k}\left\{\Gamma^{n_{k}}\!\left(\frac{d}{2}-a_{k}\right)
    \qquad\text{or}\qquad
    (-1)^{\frac{d}{2}n_k}\Gamma^{-n_{k}}\!\left(a_{k}-\frac{d}{2}+1\right)\right\}.
\end{equation}
Both forms of the factors are acceptable, and generally one
should choose one or the other depending on where it would be most
convenient to have the poles and zeros of $\homH(d)$ located.
The function \texttt{HomogeneousSolution} from the \noun{Dream}
package~\cite{LM17a} automates this construction.

The particular solution $\partR(d)$ can be constructed as an infinite sum,
\begin{equation}
    \label{eq:drr-particular-solution}
    \partR_i(d)=\homH_i(d)\left\{-\sum^{\infty}_{k=0}\quad\text{or}\quad\sum^{-1}_{k=-\infty}\right\}\homH^{-1}_i(d+2k+2)\sum_{j<i}M_{ij}(d+2k)\,J_j(d+2k),
\end{equation}
where the direction of the summation is chosen depending on which
one converges.
With the help of \noun{Dream} this sum can be evaluated
numerically as a series in $\ep$ with arbitrary precision---as
long as the integrals from lower sectors $J_{j<i}$ are known, of
course.
Normally this is a quickly converging geometric sum, and a precision
of thousands of digits can be easily achieved.

With $\homH$ and $\partR$ being well understood, the most
difficult part of solving eq.~\eqref{eq:ldrr-triangular} is
finding the periodic function~$\omega(d)$, which plays
the same role as integration constants play in the solution of
differential equations.

\subsection{Solving DRR for VRRR integrals}
\label{sec:drr-for-vrrr}

To fix $\fn{\omega}{d}$ we shall loosely follow the ``dimensional
recurrence and analyticity'' method (DRA) from~\cite{Lee09}.
As we shall soon see, only a simple incarnation of it is needed
for the VRRR integrals, but we shall return to the full method
in~\autoref{sec:drr-for-3cut}.

The overarching idea is to find restrictions on the possible
forms that $\fn{\omega}{d}$ is allowed to take by analyzing its
analytic properties in the whole complex plane.
Two essential sources of information are used for this: the
location of the poles in $d$ of $\fn{\omega}{d}$, and its
asymptotic behavior in the limit $\Im d\to\pm\infty$.

To perform such an analysis, let us first rewrite
eq.~\eqref{eq:drr-gen-sol} in the form
\begin{equation}
  \label{eq:omega-sigma}
    \fn{\omega_i}{d} = \fn{\homH_i^{-1}}{d} \fn{J_i}{d} - \fn{\homH_i^{-1}}{d} \fn{\partR_i}{d}.
\end{equation}
Because $\fn{\omega}{d}$ is periodic with a period of 2, we can
restrict the analysis to a stripe in the complex plane where
$\Re d \in (d_0, d_0+2]$ or $\Re d \in [d_0, d_0+2)$.
It is useful to choose this stripe such that $J$, $\homH^{-1}$,
and $\partR$ have as few poles on it as possible.
The poles of $\fn{J}{d}$ are particularly to be avoided.

Conveniently, all VRRR integrals normalized according to
eq.~\eqref{eq:normalized-cut-definition} are finite in the
stripe~$(6,8]$.
All $\homH_i$ can be chosen according to eq.~\eqref{eq:drr-homogeneous-solution}
to not have any zeros on this stripe too.
To analyze the behavior of $\partR_i$ constructed via eq.~\eqref{eq:drr-particular-solution},
the knowledge of the previous integrals $J_{j<i}$ is required,
having which we can evaluate $\fn{\partR_i}{d}$ numerically via
\noun{Dream}.
Proceeding with the solution steps for each $J_i$, and plotting
the numerical values of $\fn{\partR_i}{d}$ for $d\in(6,8]$,
every time we find that $\fn{\partR_i}{d}$ is smooth and finite---which
is expected, because in eq.~\eqref{eq:drr-particular-solution}
we have an infinite sum with finite coefficients that converges
geometrically.

Because none of the integrals $\fn{J}{d}$, $\fn{\homH^{-1}}{d}$,
or $\fn{\partR}{d}$ have poles when $\Re d\in(6,8]$, it follows
that $\fn{\omega}{d}$ is free from poles in $d$ in the whole
complex plane.

Next we shall turn to the investigation of the asymptotic behavior of
$\fn{\omega}{d}$ in the limit of $\Im d\to\pm\infty$.
For this we shall investigate each component of eq.~\eqref{eq:omega-sigma}
separately.

\subsubsection*{Asymptotic behavior of VRRR integrals}

To investigate the behaviour of VRRR integrals at $\fn{\Im}{d}\to \pm \infty$,
a parametric representation is needed.
To construct it, let us first rewrite eq.~\eqref{eq:cut-definition} in the form
\begin{equation}
    \label{eq:vrrr-parametrization}
    I = \int d\mathrm{PS}_4
        \underbrace{ \int \frac{d^d l}{(2\pi)^d} \prod_{i=1\vphantom{j}}^{n} \frac{1}{\left( l+ k_i \right)^2 + i0} }_{A_L}
        \,
        \underbrace{ \prod_{j} \frac{1}{ k_j'^2 } }_{A_T}
      \equiv B \, \mathrm{PS}_4 J,
\end{equation}
where $l$ is the loop momentum, $A_L$ is the loop part of the
amplitude (i.e. the propagators that contain $l$), $k_i$ are
some linear combinations of the cut momenta $p_j$, $A_T$ is the
tree part (i.e. the propagators that do not contain $l$), and
$J$ is the normalized integral.

The loop amplitude $A_L$ entering this form can be parametrized
with Feynman parameters, giving us
\begin{equation}
    \label{eq:vrrr-al}
    \frac{A_L}{B}
      = \frac{ \fn{\Gamma}{d-2} \fn{\Gamma}{n - \frac{d}{2}} }{  \fn{\Gamma^2}{\frac{d}{2}-1} \fn{\Gamma}{2-\frac{d}{2}} }
         \int \frac{
             \left( \sum_i x_i \right)^{n-d} \fn{\delta}{1 - \sum_i x_i}
         }{
             \left( \sum_{i<j} x_i x_j \left( k_i - k_j \right)^2 \right)^{n - \frac{d}{2}}
         } \prod_i d x_i .
\end{equation}
For the phase-space element $d\mathrm{PS}_4$ the direct parametrization
from eq.~\eqref{eq:dps4-definition} can be used.
Specifically, to analyze the normalized integral $J$, this form
with a factorized $\mathrm{PS}_4$ will be useful:
\begin{equation}
    \label{eq:vrrr-dps4}
    \frac{d \mathrm{PS}_4}{\mathrm{PS}_4} = 
        \left( q^2 \right)^{\frac{3d-4}{2}} \frac{2^{2d-5}}{\pi^{\frac{3}{2}}}
        \frac{ \fn{\Gamma}{d-\frac{3}{2}} \fn{\Gamma}{\frac{3}{2}d-3} }{ \fn{\Gamma}{d-3} \fn{\Gamma^3}{\frac{1}{2}d-1} }
        \left(\Delta_4\right)^{\frac{d-5}{2}} \fn{\Theta}{\Delta_4} \fn{\delta}{1-\textstyle\sum s_{ij}} \prod d s_{ij} .
\end{equation}
To analyze the asymptotic behaviour of $\abs{\fn{J}{d}}$ in the limit $\Im d\to\infty$,
it is enough to note that eqs.~\eqref{eq:vrrr-al} and~\eqref{eq:vrrr-dps4}
contain only two kinds of structures involving $d$: the Gamma functions
$\fn{\Gamma}{\alpha + \beta d}$ and the powers $x^{\alpha d}$.
The asymptotic behavior of $\fn{\Gamma}{z}$ follows from the Stirling formula,
\begin{equation}
    \label{eq:gamma-limit}
    \abs{\fn{\Gamma}{z}} = \sqrt{2 \pi} \, e^{-\frac{\pi}{2} \abs{\Im z}} \abs{\Im z}^{\Re z - \frac{1}{2}} \left( 1 + \fn{\mathcal{O}}{\abs{\Im z}^{-1}} \right).
\end{equation}
As long as the base of a power is positive, its
modulus does not depend on $\Im d$ at all,
\begin{equation}
    \abs{x^z} = x^{\Re z},\qquad\text{if }x\ge0.
\end{equation}
This suits our case, because the base of the power in
eqs.~\eqref{eq:vrrr-dps4} and~\eqref{eq:vrrr-al} are all in
fact always positive.
Combining all together, it follows that $\abs{\fn{J}{d}}$ is bounded in the limit by
\begin{equation}
    \abs{\fn{J}{d}}
      = \int \frac{d\mathrm{PS}_4}{\mathrm{PS}_4} \frac{A_L}{B} A_T
      \approx \fn{\mathcal{C}}{\Re d} \abs{\Im d}^{n + 1}.
\end{equation}

\subsubsection*{Asymptotic behavior of the homogeneous solutions}

All the diagonal elements of the DRR matrix for VRRR integrals
have the form of eq.~\eqref{eq:drr-mii-form}, with all $a_k\in[0,7/3]$.
For this reason it is possible to construct the homogeneous
solution via eq.~\eqref{eq:drr-homogeneous-solution}, and to avoid
both poles and zeros at $d>14/3$ by choosing the first form of the factors.
Then, applying the asymptotic of the Gamma functions to this
construction, for each VRRR master integral we find that
\begin{equation}
    \abs{\fn{\homH}{d}} \approx \fn{\mathcal{C}}{\Re d} \abs{\Im d}^\alpha,
\end{equation}
where $\alpha$ depends on the integral, but is always a small rational number, $\alpha\in[0,5/2]$.
It is important to note the absence of factors like $e^{\alpha\abs{\Im d}}$
in this asymptotic that could appear from eq.~\eqref{eq:gamma-limit}:
it turns out that they cancel each other.

\subsubsection*{Asymptotic behavior of the particular solutions}

Next, let us look at the particular solution $\fn{\partR}{d}$ in
the form of the infinite series from eq.~\eqref{eq:drr-particular-solution}.
The bounds established so far tell us that each term in that
series is asymptotically bounded by $\fn{\mathcal{C}}{\Re d}\abs{\Im d}^\beta$,
for some values of $\beta$.
To claim that the series as a whole is bounded similarly
too, it is enough to show that a series composed of these bounds
converges.
This is in fact the case, because the dependence of the series
terms on $\Re d$ approaches an exponential irrespective of the value
of $\Im d$, and the whole series converges geometrically.
Thus, there is such $\beta$ that
\begin{equation}
    \abs{\fn{\partR}{d}} \lesssim \fn{\mathcal{C}}{\Re d} \abs{\Im d}^\beta.
\end{equation}

\subsubsection*{Asymptotic behavior of the periodic function}

Because $J$, $\homH$, and $\partR$ all have the same form of
asymptotic behavior, from the definition of $\fn{\omega}{d}$ in
eq.~\eqref{eq:omega-sigma} it follows that $\fn{\omega}{d}$ is bounded similarly,
\begin{equation}
    \abs{\fn{\omega}{d}} \lesssim \fn{\mathcal{C}}{\Re d} \abs{\Im d}^\gamma,
\end{equation}
for some $\gamma$.
Furthermore, because $\fn{\omega}{d}$ is periodic, we can view it
as a function of $z$,
\begin{equation}
    \label{eq:z}
    z=e^{i \pi d}.
\end{equation}
In terms of $z$, the limit $\Im d \to +\infty$ corresponds to
$z\to0$, and $\Im d \to -\infty$ corresponds to $z\to\infty$.
Because $\fn{\omega}{d}$ has no poles in $d$, $\fn{\omega}{z}$
viewed as a function on the Riemann sphere can only have poles
at 0 and $\infty$.
Moreover, its growth at both of these limits is bounded by
$\abs{\Im d}^\gamma \approx \abs{\ln z}^\gamma$, which grows
slower than any non-zero power of $z$.
Then, representing $\fn{\omega}{z}$ as its Taylor series, these
constraints mean that only the $z^0$ term is allowed.
In other words, this means that $\fn{\omega}{z}$ can only be a
constant.

\subsubsection*{Fixing the constants}

Now that we have determined that all $\omega_i$ for VRRR master
integrals are constants, what is left is fixing them.
This is just one constant per integral, so it is sufficient to
for example calculate the leading term of $\ep$-expansion of
each integral.
To make this easy, one can use the following observation: for a
VRRR master integral with $n$ loop propagators, the superficial
degree of divergence of the loop part becomes zero when $d=2n$ (meaning that
the integral begins to diverge logarithmically in the UV region), and
importantly no IR divergences are present in this $d$ as well.
Changing $d$ to $2n-2\ep$ regulates the UV divergence via a
single $\ep$ pole, and being an UV pole, it does not depend on
any masses in the diagram.
Therefore, one can just as well insert some mass $m$ into the
loop without affecting the pole.
Then, applying the large mass expansion~\cite{Smirnov02} to the
massive diagram factorizes it into a massive one-loop vacuum
bubble (equal to the massive loop with external legs amputated)
and a 4-particle phase-space integral (equal to the original
integral with the loop shrinked into a dot), while still not
changing the pole.

For $\vrrr{31}$ this process can be illustrated as follows:
\begin{align}
    \label{eq:lmp}
    \smallfig{vrrr/31}
        &= \smallfig{vrrr/31-massive} + \fn{\mathcal{O}}{\ep^0}
        = \smallfig{lm-hard}\,\smallfig{lm-soft} + \fn{\mathcal{O}}{\ep^0,\frac{q^2}{m^2}} =\\
        &= -\frac{5500}{3} B\,\mathrm{PS}_4 + \fn{\mathcal{O}}{\ep^0,\frac{q^2}{m^2}},\nonumber
\end{align}
all with $d=10-2\ep$.
The vacuum bubble here was evaluated via
\begin{equation}
    \int \frac{d^d l}{\left(2\pi\right)^d} \frac{1}{\left( l^2-m^2+i0 \right)^k} = 
    \left(-1\right)^k \frac{i\pi^{\frac{d}{2}}}{\left(2\pi\right)^d} \frac{\fn{\Gamma}{k - \frac{d}{2}}}{\fn{\Gamma}{k}} \left(m^2\right)^{\frac{d}{2}-k},
\end{equation}
and the 4-particle phase-space integral was reduced to the
masters from~\cite{GGH03} using IBP and dimensional recurrence
relations.

From here it is possible to determine $\omega_{31}$ by inserting
eq.~\eqref{eq:lmp} into the definition of $\omega$ from eq.~\eqref{eq:omega-sigma}
along with a high-precision numerical series for $\fn{\partR_{31}}{d}$
calculated by \noun{Dream}.

Finally, the same procedure needs to be performed for all other
VRRR master integrals.
Interestingly, we have found that all $\omega_i$ are identically
zero, except for $\omega_1$, $\omega_4$, and $\omega_{16}$, which
all correspond to simple integrals.

This concludes the calculation of $\omega_i$ for VRRR master integrals.
Together with $\homH_i$ from eq.~\eqref{eq:drr-homogeneous-solution}
and $\partR_i$ from eq.~\eqref{eq:drr-particular-solution} this
gives us the full solution to the DRR eq.~\eqref{eq:ldrr} in
terms of nested infinite sums.
With the help of \noun{Dream} or \noun{SummerTime} these sums
can be evaluated as a series in $\ep$ around arbitrary $d$ to
any desired precision.
We have done so for $d=4-2\ep$ with 4000 digits of precision and
have restored the analytical answers via the PSLQ~\cite{FBA99}
algorithm in the basis of MZVs~\cite{BBV09} up
to weight~12 (see \autoref{sec:mzv} for the precise definition of this basis).
The results of this restoration up to weight~6 can be found in
\autoref{sec:results-vrrr}.
The full results up to weight 12 as well as the explicit
expressions for the sums in \noun{SummerTime} format can be
found in the ancillary files, as described in \autoref{sec:Ancillary-files}.

\subsection{Cross-checks}

One way to cross-check the obtained results is to recalculate
them numerically.
To this end the form of eq.~\eqref{eq:vrrr-parametrization}
can be used along with the Feynman parametrization of the
loop amplitude from eq.~\eqref{eq:vrrr-al} and the tripole
parameterization of the phase-space~\cite{GGH03,ERT80}.
Because VRRR integrals contain only one loop, their ultraviolet divergences
manifest themselves only in the prefactor of the Feynman parameterization.
The infrared divergences also disappear if we look at the series
around $d\ge6$.
For this reason this parameterization can be integrated directly
via the standard Monte-Carlo numerical integration methods (we use
the Vegas algorithm~\cite{Lepage78} implemented in \noun{Cuba}~\cite{Hah04})
in $d=6-2\ep$, and then lowered back to $d=4-2\ep$
via DRR---avoiding the need for methods like sector decomposition.

We have performed this numerical integration and can report that
for each VRRR integral the results for the first 3 orders in $
\ep$ match with the analytic answers within 1\% accuracy.

Additionally, we have compared our results for VRRR integrals
1, 2, 3, 4, 5, 10, 12, and 18 with the weight-6 series reported
in~\cite{CFHMSS15}, and found them to match fully.

Finally, a cross-check based on Cutkosky rules will be described
in \autoref{sec:Cutkosky-rules}.

\section{2-loop 3-particle cut integrals (VVRR, VRRV)}
\label{sec:VVRR-VRRV}

There are 22 VVRR and 9 VRRV master integrals; 27 in total, if one omits
the duplicates between these two sets.
Both of these sets are depicted in \autoref{tab:VVRR} and \autoref{tab:VRRV}.

\begin{table}[!th]
\begin{tabular}{>{\centering}p{2.6cm}>{\centering}p{2.6cm}>{\centering}p{2.6cm}>{\centering}p{2.6cm}>{\centering}p{2.6cm}}
%
    \scalebox{0.75}{\input{fig/vvrr/1.tikz}}%
 & %
    \scalebox{0.75}{\input{fig/vvrr/2.tikz}}%
 & %
    \scalebox{0.75}{\input{fig/vvrr/3.tikz}}%
 & %
    \scalebox{0.75}{\input{fig/vvrr/4.tikz}}%
 & %
    \scalebox{0.75}{\input{fig/vvrr/5.tikz}}%
\tabularnewline
$1,\mathrm{H}_{01*0*1100*0}$ & $2,\mathrm{H}_{01*0*1100*1}$ & $3,\mathrm{H}_{0*111**0001}$ & $4,\mathrm{L}_{0***1001011}$ & $5,\mathrm{H}_{10*0*0111*0}$\tabularnewline
 &  &  &  & \tabularnewline
%
    \scalebox{0.75}{\input{fig/vvrr/6.tikz}}%
 & %
    \scalebox{0.75}{\input{fig/vvrr/7.tikz}}%
 & %
    \scalebox{0.75}{\input{fig/vvrr/8.tikz}}%
 & %
    \scalebox{0.75}{\input{fig/vvrr/9.tikz}}%
 & %
    \scalebox{0.75}{\input{fig/vvrr/10.tikz}}%
\tabularnewline
$6,\mathrm{H}_{0*011**0101}$ & $7,\mathrm{H}_{00*0*1101*0}$ & $8,\mathrm{M}_{011*0*11*11}$ & $9,\mathrm{N}_{101*1*1011*}$ & $10,\mathrm{H}_{0*110**0011}$\tabularnewline
 &  &  &  & \tabularnewline
%
    \scalebox{0.75}{\input{fig/vvrr/11.tikz}}%
 & %
    \scalebox{0.75}{\input{fig/vvrr/12.tikz}}%
 & %
    \scalebox{0.75}{\input{fig/vvrr/13.tikz}}%
 & %
    \scalebox{0.75}{\input{fig/vvrr/14.tikz}}%
 & %
    \scalebox{0.75}{\input{fig/vvrr/15.tikz}}%
\tabularnewline
$11,\mathrm{N}_{*01111*0*11}$ & $12,\mathrm{N}_{110*1*0111*}$ & $13,\mathrm{H}_{1*1*1101*01}$ & $14,\mathrm{H}_{0*110**1011}$ & $15,\mathrm{N}_{101*1*0111*}$\tabularnewline
 &  &  &  & \tabularnewline
%
    \scalebox{0.75}{\input{fig/vvrr/16.tikz}}%
 & %
    \scalebox{0.75}{\input{fig/vvrr/17.tikz}}%
 & %
    \scalebox{0.75}{\input{fig/vvrr/18.tikz}}%
 & %
    \scalebox{0.75}{\input{fig/vvrr/19.tikz}}%
 & %
    \scalebox{0.75}{\input{fig/vvrr/20.tikz}}%
\tabularnewline
$16,\mathrm{H}_{01*0*1111*0}$ & $17,\mathrm{M}_{111*1*11*11}$ & $18,\mathrm{N}_{11*1*1111*0}$ & $19,\mathrm{N}_{11*1*1111*1}$ & $20,\mathrm{N}_{111*1*1111*}$\tabularnewline
 &  &  &  & \tabularnewline
%
    \scalebox{0.75}{\input{fig/vvrr/21.tikz}}%
 & %
    \scalebox{0.75}{\input{fig/vvrr/22.tikz}}%
 &  &  & \tabularnewline
$21,\mathrm{H}_{11*1*1111*1}$ & $22,\mathrm{J}_{11101*1*1*1}$ &  &  & \tabularnewline
 &  &  &  & \tabularnewline
\end{tabular}

\caption{\protect\strong{VVRR}, 3-particle cut master integrals with 2 loops
  on one side of the cut.}
\label{tab:VVRR}
\end{table}

\begin{table}[!th]
\begin{tabular}{>{\centering}p{2.6cm}>{\centering}p{2.6cm}>{\centering}p{2.6cm}>{\centering}p{2.6cm}>{\centering}p{2.6cm}}
%
    \scalebox{0.75}{\input{fig/vrrv/1.tikz}}%
 & %
    \scalebox{0.75}{\input{fig/vrrv/2.tikz}}%
 & %
    \scalebox{0.75}{\input{fig/vrrv/3.tikz}}%
 & %
    \scalebox{0.75}{\input{fig/vrrv/4.tikz}}%
 & %
    \scalebox{0.75}{\input{fig/vrrv/5.tikz}}%
\tabularnewline
$1,\mathrm{H}_{1*11*10000*}$ & $2,\mathrm{H}_{0*01*11010*}$ & $3,\mathrm{H}_{0*11*11000*}$ & $4,\mathrm{H}_{0*10*11001*}$ & $5,\mathrm{N}_{10*111*011*}$\tabularnewline
 &  &  &  & \tabularnewline
%
    \scalebox{0.75}{\input{fig/vrrv/6.tikz}}%
 & %
    \scalebox{0.75}{\input{fig/vrrv/7.tikz}}%
 & %
    \scalebox{0.75}{\input{fig/vrrv/8.tikz}}%
 & %
    \scalebox{0.75}{\input{fig/vrrv/9.tikz}}%
 & \tabularnewline
$6,\mathrm{H}_{1*11*10110*}$ & $7,\mathrm{H}_{1*11*11111*}$ & $8,\mathrm{N}_{11*111*111*}$ & $9,\mathrm{J}_{1110**11*11}$ & \tabularnewline
\end{tabular}

\caption{\protect\strong{VRRV}, 3-particle cut master integrals with 1 loop
  on each side of the cut.}
\label{tab:VRRV}
\end{table}

Ideally we would like to calculate VVRR master integrals by
solving the dimensional recurrence relations just as we did in
\autoref{sec:vrrr}.
The difficulty here lies in the fact that while for VRRR only one
constant per integral was needed to uniquely fix the periodic
function $\fn{\omega}{d}$, for VVRR integrals dozens might be
needed.
Therefore we shall postpone solving DRR---until \autoref{sec:drr-for-3cut}.
Fortunately the 3-particle phase space is considerably simpler
than the 4-particle one, and we can return to the idea of direct
integration over it.

In principle, the 3-particle phase-space volume element from eq.~\eqref{eq:dpsn-definition}
can be parameterized in terms of the kinematic invariants
\begin{equation}
s_{ij}=\frac{1}{q^{2}}\left(p_{i}+p_{j}\right)^{2}
\end{equation}
in the following way:
\begin{equation}
d\mathrm{PS}_{3}=\left(q^{2}\right)^{d-3}\frac{2^{4-3d}\pi^{\frac{3}{2}-d}}{\Gamma\!\left(\frac{d-2}{2}\right)\Gamma\!\left(\frac{d-1}{2}\right)}\left(s_{12}s_{13}s_{23}\right)^{\frac{d-4}{2}}\delta\!\left(1-s_{12}-s_{13}-s_{23}\right)ds_{12}ds_{13}ds_{23},\label{eq:dps3-definition}
\end{equation}
and the integration volume given by the $\delta$-function and
the condition $s_{ij}\geq0$ is simple enough that it is often
possible to integrate eq.~\eqref{eq:cut-definition} directly,
as long as both amplitudes entering the integral are known.
Specifically, one way to make the integration volume explicit
is
\begin{equation}
\int d\mathrm{PS}_{3}\,F\sim\int_{0}^{1}ds_{12}\int_{0}^{1-s_{12}}ds_{13}\,\left(s_{12}s_{13}s_{23}\right)^{\frac{d-4}{2}}F\Big|_{s_{23}=1-s_{12}-s_{13}}.
\end{equation}
In practice many of the amplitudes are only known as series in $\ep$,
and a series expansion operation does not necessarily commute with
integration. To illustrate this issue, let us consider the following
integral:
\begin{equation}
    \vvrr{8}=\smallfig{vvrr/8}=\int\smallfig{1to3/na19}d\mathrm{PS}_3.\label{eq:vvrr-example-integral}
\end{equation}
The amplitude in the integrand here is a single-scale integral (the
precise value is given by eq.~\eqref{eq:known-tricross}), and
just by dimensional analysis must be proportional to $\left(q^{2}s_{12}\right)^{d-6}$,
\begin{equation}
\smallfig{1to3/na19}=\mathcal{C}\,s_{12}^{d-6}.\label{eq:vvrr-example-amplitude-series}
\end{equation}
If we were to first integrate eq.~\eqref{eq:vvrr-example-integral}
directly and then expand the result in $\ep$, we would get
\begin{equation}
    \smallfig{vvrr/8}=\mathcal{C}2^{4-3d}\pi^{\frac{3}{2}-d}\frac{\Gamma\!\left(\frac{d}{2}-1\right)\Gamma\!\left(3\frac{d}{2}-7\right)}{\Gamma\!\left(\frac{d}{2}-\frac{1}{2}\right)\Gamma\!\left(5\frac{d}{2}-9\right)}=\frac{\mathcal{C}}{384\pi^{3}}\left(\frac{1}{\ep}+\left(2\ln\left(4\pi\right)-2\gamma-1\right)+\mathcal{O}\!\left(\ep\right)\right).\label{eq:vvrr-example-int-then-expand}
\end{equation}
In contrast, expanding both the amplitude and the phase-space element
in $\ep$ first using
\begin{equation}
    \smallfig{1to3/na19}=\frac{\mathcal{C}}{s_{12}^{2}}\left(1-2\ln s_{12}\ep+\mathcal{O}\!\left(\ep^{2}\right)\right),
\end{equation}
multiplying it by the expansion of the phase-space element from eq.~\eqref{eq:dps3-definition},
\begin{equation}
    \left(s_{12}s_{13}s_{23}\right)^{\frac{d-4}{2}}=\left(1+\left(2\ln\left(4\pi\right)-2\gamma+2-\ln\left(s_{12}s_{13}s_{23}\right)\right)\ep+\mathcal{O}\!\left(\ep^{2}\right)\right),
\end{equation}
\begin{align}
    d\mathrm{PS}_{3}=&\frac{1}{128\pi^{3}}\left(1+\left(2\ln\left(4\pi\right)-2\gamma+2-\ln\left(s_{12}s_{13}s_{23}\right)\right)\ep+\mathcal{O}\!\left(\ep^{2}\right)\right)\times \label{eq:dps3-series}\\
    &\times\delta\!\left(q^{2}-s_{12}-s_{13}-s_{23}\right)ds_{12}ds_{13}ds_{23},\nonumber
\end{align}
and then integrating order-by-order in $\ep$ would result in
divergences corresponding to the limit $s_{12}\to0$ in each order
of the series:
\begin{equation}
    \smallfig{vvrr/8}\overset{?}{=}\frac{\mathcal{C}}{128\pi^{3}}\int_{0}^{1}ds_{12}\int_{0}^{1-s_{12}}ds_{13}\left(\frac{1}{s_{12}^{2}}+\mathcal{O}\!\left(\ep\right)\right)=\frac{\mathcal{C}}{128\pi^{3}}\left(\frac{1}{0}+\mathcal{O}\!\left(\ep\right)\right).\label{eq:vvrr-example-expand-then-int}
\end{equation}
This should not be surprising, seeing that the true $\ep$-expansion
in eq.~\eqref{eq:vvrr-example-int-then-expand} starts with $\ep^{-1}$,
while the integrand in eq.~\eqref{eq:vvrr-example-expand-then-int}
starts with $\ep^{0}$, so integrating it order-by-order can
never give the expected series. In other words, the integral in eq.~\eqref{eq:vvrr-example-integral}
is not infrared finite, and the pole in $\ep$ regulating this
infinity cannot be obtained by integration in 4 dimensions.

Our solution to this problem stems from the fact that every Feynman
integral is free from infrared divergences if the dimension of space-time
is high enough. This can be seen already from eq.~\eqref{eq:dps3-definition}:
higher $d$ results in higher powers of the $\left(s_{12}s_{13}s_{23}\right)$
factor, which can eventually compensate any singularity of the integrand
in the infrared region. In particular, at $d=6$ all our 3-particle
master integrals are infrared-finite. Thus, we can overcome the divergence
in eq.~\eqref{eq:vvrr-example-expand-then-int} by integrating the
$\ep$-series not around $d=4-2\ep$, but rather around
$d=6-2\ep$. Once this is done, we can use dimensional recurrence
relations to restore the series in $d=4-2\ep$.

This procedure requires the knowledge of 1$\to$3 amplitudes at 1
and 2 loops as series in $\ep$, expanded around $d=6-2\ep$,
and DRR for the phase-space master integrals themselves.

Note that there is nothing magical about the basis at $d=6-2\ep$
specifically: any IR-finite basis would be sufficient for our
purposes.

\subsection{1-loop 1\texorpdfstring{$\to$}{->}3 amplitudes}
\label{sec:1to3-1l}

Up to relabeling of $p_{i}$, all the 1$\to$3 amplitudes at 1
loop fit into the box topology:
\begin{equation}
%
    \scalebox{0.75}{\input{fig/1to3/1l/topo.tikz}}%
.
\end{equation}
There are only four master integrals in this topology, with only two
being meaningfully distinct: the bubble and the box with one off-shell
leg.

\begin{table}[H]
\begin{longtable}[c]{>{\centering}p{2.6cm}>{\centering}p{2.6cm}>{\centering}p{2.6cm}>{\centering}p{2.6cm}}
%
    \scalebox{0.75}{\input{fig/1to3/1l/m1.tikz}}%
 & %
    \scalebox{0.75}{\input{fig/1to3/1l/m2.tikz}}%
 & %
    \scalebox{0.75}{\input{fig/1to3/1l/m3.tikz}}%
 & %
    \scalebox{0.75}{\input{fig/1to3/1l/m4.tikz}}%
\tabularnewline
\end{longtable}

\caption{Master integrals of the box topology.}
\end{table}

The values of these are known from the literature for arbitrary $d$.
See eq.~\eqref{eq:known-bubble} for the bubble, and eq.~\eqref{eq:known-box}
for the box.

In addition to the master integrals, a triangle with two off-shell
legs also appears in the VRRV integrals. It can be found via IBP as
\begin{equation}
\smallfig{1to3/1l/triangle}=\frac{d-3}{d-4}\frac{2}{s_{13}+s_{23}}\left(q^{2}\right)^{-1}\left(\smallfig{1to3/1l/m3}-\,\smallfig{1to3/1l/m1}\right).\label{eq:tri-1l-red}
\end{equation}

\subsection{2-loop 1\texorpdfstring{$\to$}{->}3 amplitudes}
\label{sec:1to3-2l}

The master integrals for these amplitudes were first calculated in~\cite{GR00,GR01}.
The results there are provided as series in $\ep$ with coefficients
in terms of ``2dHPLs'', a subclass of multiple polylogarithms,
up to transcendentality weight~4.
This turns out to be insufficient for our needs, because the $
\ep$-finite parts of the VVVV integrals are known to contain
$\zeta_7$, and thus we need the amplitudes to at least that
weight as well.
Thus, a re-derivation of these master integrals is required.

The overall idea of the method is to write down the differential equation
system for the master integrals in external kinematic invariants,
solve it, and determine the integration constants by enforcing regularity
conditions on the solution. Let us do this step by step.

\subsubsection*{Topologies and master integrals}

We start by determining IBP topologies that cover the master integrals.
Up to relabeling of $p_{i}$, three topologies are sufficient: one
planar (``PA''), and two non-planar (``NA'' and ``NB'').
These are depicted in \autoref{tab:1to3-2l-topologies}.

\begin{table}[!t]
\begin{longtable}[c]{ccc}
%
    \scalebox{0.75}{\input{fig/1to3/pb.tikz}}%
 & %
    \scalebox{0.75}{\input{fig/1to3/na.tikz}}%
 & %
    \scalebox{0.75}{\input{fig/1to3/nb.tikz}}%
\tabularnewline
PA & NA & NB\tabularnewline
\end{longtable}

\caption{Generic topologies for the 1$\to$3 master integrals at two loops.}
\label{tab:1to3-2l-topologies}
\end{table}

Solving the IBP relations in these topologies using \noun{LiteRed}~\cite{Lee13}
we arrive at 18 master integrals in PA, 22 in NA, and 29 in NB. Note
that there is an overlap between these three sets of master integrals,
but this will not bother us.

\begin{table}[!t]
\begin{tabular}{>{\centering}p{2.6cm}>{\centering}p{2.6cm}>{\centering}p{2.6cm}>{\centering}p{2.6cm}>{\centering}p{2.6cm}}
%
    \scalebox{0.75}{\input{fig/1to3/pb1.tikz}}%
 & %
    \scalebox{0.75}{\input{fig/1to3/pb2.tikz}}%
 & %
    \scalebox{0.75}{\input{fig/1to3/pb3.tikz}}%
 & %
    \scalebox{0.75}{\input{fig/1to3/pb4.tikz}}%
 & %
    \scalebox{0.75}{\input{fig/1to3/pb5.tikz}}%
\tabularnewline
$\mathrm{PA}_{1}$ & $\mathrm{PA}_{2}$ & $\mathrm{PA}_{3}$ & $\mathrm{PA}_{4}$ & $\mathrm{PA}_{5}$\tabularnewline
 &  &  &  & \tabularnewline
%
    \scalebox{0.75}{\input{fig/1to3/pb6.tikz}}%
 & %
    \scalebox{0.75}{\input{fig/1to3/pb7.tikz}}%
 & %
    \scalebox{0.75}{\input{fig/1to3/pb8.tikz}}%
 & %
    \scalebox{0.75}{\input{fig/1to3/pb9.tikz}}%
 & %
    \scalebox{0.75}{\input{fig/1to3/pb10.tikz}}%
\tabularnewline
$\mathrm{PA}_{6}$ & $\mathrm{PA}_{7}$ & $\mathrm{PA}_{8}$ & $\mathrm{PA}_{9}$ & $\mathrm{PA}_{10}$\tabularnewline
 &  &  &  & \tabularnewline
%
    \scalebox{0.75}{\input{fig/1to3/pb11.tikz}}%
 & %
    \scalebox{0.75}{\input{fig/1to3/pb12.tikz}}%
 & %
    \scalebox{0.75}{\input{fig/1to3/pb13.tikz}}%
 & %
    \scalebox{0.75}{\input{fig/1to3/pb14.tikz}}%
 & %
    \scalebox{0.75}{\input{fig/1to3/pb15.tikz}}%
\tabularnewline
$\mathrm{PA}_{11}$ & $\mathrm{PA}_{12}$ & $\mathrm{PA}_{13}$ & $\mathrm{PA}_{14}$ & $\mathrm{PA}_{15}$\tabularnewline
 &  &  &  & \tabularnewline
%
    \scalebox{0.75}{\input{fig/1to3/pb16.tikz}}%
 & %
    \scalebox{0.75}{\input{fig/1to3/pb17.tikz}}%
 & %
    \scalebox{0.75}{\input{fig/1to3/pb18.tikz}}%
 &  & \tabularnewline
$\mathrm{PA}_{16}$ & $\mathrm{PA}_{17}$ & $\mathrm{PA}_{18}$ &  & \tabularnewline
\end{tabular}

\caption{Master integrals for the 1$\to$3 amplitudes in the PA topology.}
\end{table}

\begin{table}[!t]
\begin{tabular}{>{\centering}p{2.6cm}>{\centering}p{2.6cm}>{\centering}p{2.6cm}>{\centering}p{2.6cm}>{\centering}p{2.6cm}}
%
    \scalebox{0.75}{\input{fig/1to3/na1.tikz}}%
 & %
    \scalebox{0.75}{\input{fig/1to3/na2.tikz}}%
 & %
    \scalebox{0.75}{\input{fig/1to3/na3.tikz}}%
 & %
    \scalebox{0.75}{\input{fig/1to3/na4.tikz}}%
 & %
    \scalebox{0.75}{\input{fig/1to3/na5.tikz}}%
\tabularnewline
$\mathrm{NA}_{1}$ & $\mathrm{NA}_{2}$ & $\mathrm{NA}_{3}$ & $\mathrm{NA}_{4}$ & $\mathrm{NA}_{5}$\tabularnewline
 &  &  &  & \tabularnewline
%
    \scalebox{0.75}{\input{fig/1to3/na6.tikz}}%
 & %
    \scalebox{0.75}{\input{fig/1to3/na7.tikz}}%
 & %
    \scalebox{0.75}{\input{fig/1to3/na8.tikz}}%
 & %
    \scalebox{0.75}{\input{fig/1to3/na9.tikz}}%
 & %
    \scalebox{0.75}{\input{fig/1to3/na10.tikz}}%
\tabularnewline
$\mathrm{NA}_{6}$ & $\mathrm{NA}_{7}$ & $\mathrm{NA}_{8}$ & $\mathrm{NA}_{9}$ & $\mathrm{NA}_{10}$\tabularnewline
 &  &  &  & \tabularnewline
%
    \scalebox{0.75}{\input{fig/1to3/na11.tikz}}%
 & %
    \scalebox{0.75}{\input{fig/1to3/na12.tikz}}%
 & %
    \scalebox{0.75}{\input{fig/1to3/na13.tikz}}%
 & %
    \scalebox{0.75}{\input{fig/1to3/na14.tikz}}%
 & %
    \scalebox{0.75}{\input{fig/1to3/na15.tikz}}%
\tabularnewline
$\mathrm{NA}_{11}$ & $\mathrm{NA}_{12}$ & $\mathrm{NA}_{13}$ & $\mathrm{NA}_{14}$ & $\mathrm{NA}_{15}$\tabularnewline
 &  &  &  & \tabularnewline
%
    \scalebox{0.75}{\input{fig/1to3/na16.tikz}}%
 & %
    \scalebox{0.75}{\input{fig/1to3/na17.tikz}}%
 & %
    \scalebox{0.75}{\input{fig/1to3/na18.tikz}}%
 & %
    \scalebox{0.75}{\input{fig/1to3/na19.tikz}}%
 & %
    \scalebox{0.75}{\input{fig/1to3/na20.tikz}}%
\tabularnewline
$\mathrm{NA}_{16}$ & $\mathrm{NA}_{17}$ & $\mathrm{NA}_{18}$ & $\mathrm{NA}_{19}$ & $\mathrm{NA}_{20}$\tabularnewline
 &  &  &  & \tabularnewline
%
    \scalebox{0.75}{\input{fig/1to3/na21.tikz}}%
 & %
    \scalebox{0.75}{\input{fig/1to3/na22.tikz}}%
 &  &  & \tabularnewline
$\mathrm{NA}_{21}$ & $\mathrm{NA}_{22}$ &  &  & \tabularnewline
\end{tabular}

\caption{Master integrals for the 1$\to$3 amplitudes in the NA topology.}
\end{table}

\begin{table}[!t]
\begin{tabular}{>{\centering}p{2.6cm}>{\centering}p{2.6cm}>{\centering}p{2.6cm}>{\centering}p{2.6cm}>{\centering}p{2.6cm}}
%
    \scalebox{0.75}{\input{fig/1to3/nb1.tikz}}%
 & %
    \scalebox{0.75}{\input{fig/1to3/nb2.tikz}}%
 & %
    \scalebox{0.75}{\input{fig/1to3/nb3.tikz}}%
 & %
    \scalebox{0.75}{\input{fig/1to3/nb4.tikz}}%
 & %
    \scalebox{0.75}{\input{fig/1to3/nb5.tikz}}%
\tabularnewline
$\mathrm{NB}_{1}$ & $\mathrm{NB}_{2}$ & $\mathrm{NB}_{3}$ & $\mathrm{NB}_{4}$ & $\mathrm{NB}_{5}$\tabularnewline
 &  &  &  & \tabularnewline
%
    \scalebox{0.75}{\input{fig/1to3/nb6.tikz}}%
 & %
    \scalebox{0.75}{\input{fig/1to3/nb7.tikz}}%
 & %
    \scalebox{0.75}{\input{fig/1to3/nb8.tikz}}%
 & %
    \scalebox{0.75}{\input{fig/1to3/nb9.tikz}}%
 & %
    \scalebox{0.75}{\input{fig/1to3/nb10.tikz}}%
\tabularnewline
$\mathrm{NB}_{6}$ & $\mathrm{NB}_{7}$ & $\mathrm{NB}_{8}$ & $\mathrm{NB}_{9}$ & $\mathrm{NB}_{10}$\tabularnewline
 &  &  &  & \tabularnewline
%
    \scalebox{0.75}{\input{fig/1to3/nb11.tikz}}%
 & %
    \scalebox{0.75}{\input{fig/1to3/nb12.tikz}}%
 & %
    \scalebox{0.75}{\input{fig/1to3/nb13.tikz}}%
 & %
    \scalebox{0.75}{\input{fig/1to3/nb14.tikz}}%
 & %
    \scalebox{0.75}{\input{fig/1to3/nb15.tikz}}%
\tabularnewline
$\mathrm{NB}_{11}$ & $\mathrm{NB}_{12}$ & $\mathrm{NB}_{13}$ & $\mathrm{NB}_{14}$ & $\mathrm{NB}_{15}$\tabularnewline
 &  &  &  & \tabularnewline
%
    \scalebox{0.75}{\input{fig/1to3/nb16.tikz}}%
 & %
    \scalebox{0.75}{\input{fig/1to3/nb17.tikz}}%
 & %
    \scalebox{0.75}{\input{fig/1to3/nb18.tikz}}%
 & %
    \scalebox{0.75}{\input{fig/1to3/nb19.tikz}}%
 & %
    \scalebox{0.75}{\input{fig/1to3/nb20.tikz}}%
\tabularnewline
$\mathrm{NB}_{16}$ & $\mathrm{NB}_{17}$ & $\mathrm{NB}_{18}$ & $\mathrm{NB}_{19}$ & $\mathrm{NB}_{20}$\tabularnewline
 &  &  &  & \tabularnewline
%
    \scalebox{0.75}{\input{fig/1to3/nb21.tikz}}%
 & %
    \scalebox{0.75}{\input{fig/1to3/nb22.tikz}}%
 & %
    \scalebox{0.75}{\input{fig/1to3/nb23.tikz}}%
 & %
    \scalebox{0.75}{\input{fig/1to3/nb24.tikz}}%
 & %
    \scalebox{0.75}{\input{fig/1to3/nb25.tikz}}%
\tabularnewline
$\mathrm{NB}_{21}$ & $\mathrm{NB}_{22}$ & $\mathrm{NB}_{23}$ & $\mathrm{NB}_{24}$ & $\mathrm{NB}_{25}$\tabularnewline
 &  &  &  & \tabularnewline
%
    \scalebox{0.75}{\input{fig/1to3/nb26.tikz}}%
 & %
    \scalebox{0.75}{\input{fig/1to3/nb27.tikz}}%
 & %
    \scalebox{0.75}{\input{fig/1to3/nb28.tikz}}%
 & %
    \scalebox{0.75}{\input{fig/1to3/nb29.tikz}}%
 & \tabularnewline
$\mathrm{NB}_{26}$ & $\mathrm{NB}_{27}$ & $\mathrm{NB}_{28}$ & $\mathrm{NB}_{29}$ & \tabularnewline
\end{tabular}

\caption{Master integrals for the 1$\to$3 amplitudes in the NB topology.}
\end{table}

There are only three independent external kinematic invariants for
the 1$\to$3 amplitudes: $p_{12}^{2}$, $p_{13}^{2}$, and $p_{23}^{2}$ ($p_{ij}\equiv p_i+p_j$),
with the incoming energy $q^{2}$ being the sum of all three. This
allows us to extract one of them as a dimensionful prefactor, with
a function of two dimensionless ratios remaining:
\begin{equation}
A\left(p_{12}^{2},p_{13}^{2},p_{23}^{2}\right)=\left(q^{2}\right)^{k}A\left(y,z\right),
\end{equation}
where $k$ is determined from dimensionality ($+\frac{d}{2}$ for
each loop, $-1$ for each propagator),
\begin{equation}
y=\frac{p_{13}^{2}}{q^{2}},\qquad\text{and}\qquad z=\frac{p_{23}^{2}}{q^{2}}.\label{eq:1to3-yz}
\end{equation}

\subsubsection*{Solution via differential equations in the $\ep$-form}

Next, for each topology $I=\left\{ \mathrm{PA},\mathrm{NA},\mathrm{NB}\right\} $
we can differentiate the integrand of each master integral $I_{i}$
by $y$ or $z$ and use IBP relations to express the derivatives
as a linear combination of $I_{i}$ themselves, forming linear differential
equation systems,
\begin{equation}
\frac{\partial}{\partial y}I_{i}=M_{ij}^{\left(I,y\right)}\!\left(d,y,z\right)I_{j},\qquad\text{and}\qquad\frac{\partial}{\partial z}I_{i}=M_{ij}^{\left(I,z\right)}\!\left(d,y,z\right)I_{j}.\label{eq:de-system}
\end{equation}
To solve these systems as series in $\ep$, it is convenient
to transform them into the combined $\ep$-form, where the dependence
on $d$ of both $M$ matrices is simultaneously factorized by a basis
change:
\begin{equation}
\frac{\partial}{\partial y}J_{i}=\ep S_{ij}^{\left(I,y\right)}\!\left(y,z\right)J_{j},\qquad\text{and}\qquad\frac{\partial}{\partial z}J_{i}=\ep S_{ij}^{\left(I,z\right)}\!\left(y,z\right)J_{j},\label{eq:epsilon-form}
\end{equation}
where $J_{i}$ is the set of master integrals in the $\ep$-basis,
related to the original basis $I_{i}$ via a transformation matrix
$T$,
\begin{equation}
I_{i}=T_{ij}^{\left(I\right)}J_{j},\label{eq:epsilon-form-transformation}
\end{equation}
and $S_{ij}$ are matrices of the form
\begin{equation}
S_{ij}^{\left(I,y\right)}\!\left(y,z\right)=\sum_{k}\frac{A_{ij}^{\left(I,y,k\right)}}{y-y_{k}\!\left(z\right)},\qquad\text{and}\qquad S_{ij}^{\left(I,z\right)}\!\left(y,z\right)=\sum_{k}\frac{A_{ij}^{\left(I,z,k\right)}}{z-z_{k}\!\left(y\right)}.\label{eq:epsilon-form-s}
\end{equation}
In particular our $S$ matrices only contain these six denominators
(the ``alphabet''):
\begin{equation}
\left\{ y,z,1-y-z,1-y,1-z,y+z\right\} ,\label{eq:alphabet}
\end{equation}
which correspond to singularities at limits $s_{ij}=0$, and $s_{ij}=1$.

Once the $\ep$-form is constructed, writing down the solution
as a series in $\ep$ becomes easy.

First, start at some arbitrary point, for example $\left(y,z\right)=\left(0,0\right)$,
and fix the value of $\mathbf{J}\equiv\{J_i\}$ at the point as a series of constants,
\begin{equation}
\mathbf{J}\!\left(0,0\right)=\mathbf{C}\!\left(\ep\right)=\ep^{-k_{0}}\sum_{k=0}^{\infty}\ep^{k}\mathbf{C}^{\left(k\right)},
\end{equation}
where $k_{0}$ is some arbitrary starting order of the series, chosen
high enough to cover the highest $\ep$-pole of the integrals.

Then, integrate along the $z$ axis using the right side of eq.~\eqref{eq:epsilon-form}
to obtain
\begin{equation}
\mathbf{J}\!\left(0,z\right)=\mathbb{W}^{\left(I\right)}\big|_{\left(0,0\right)\to\left(0,z\right)}\mathbf{C},
\end{equation}
where $\mathbb{W}$ is the fundamental solution matrix of the corresponding
differential equation system constructed as a path-ordered exponential
along the specified path,
\begin{align}
\mathbb{W}^{\left(I\right)}\big|_{\left(0,0\right)\to\left(0,z\right)} & =\mathrm{P}\exp\left(\ep\int_{0}^{z}dt\,\mathbb{S}^{\left(I,z\right)}\!\left(0,t\right)\right)=\\
 & =\mathds{1}+\ep\int_{0}^{z}dt\,\mathbb{S}^{\left(I,z\right)}\!\left(0,t\right)+\ep^{2}\int_{0}^{z}dt\,\mathbb{S}^{\left(I,z\right)}\!\left(0,t\right)\int_{0}^{t}dt'\,\mathbb{S}^{\left(I,z\right)}\!\left(0,t'\right)+\dots.\nonumber
\end{align}
Finally, use the left side of eq.~\eqref{eq:epsilon-form} to obtain
\begin{equation}
\mathbf{J}\!\left(y,z\right)=\mathbb{W}^{\left(I\right)}\big|_{\left(0,z\right)\to\left(y,z\right)}\mathbf{J}\!\left(0,z\right)=\mathbb{W}^{\left(I\right)}\big|_{\left(0,z\right)\to\left(y,z\right)}\mathbb{W}^{\left(I\right)}\big|_{\left(0,0\right)\to\left(0,z\right)}\mathbf{C},\label{eq:de-integration-order-zy}
\end{equation}
where
\begin{align}
\mathbb{W}^{\left(I\right)}\big|_{\left(0,z\right)\to\left(y,z\right)} & =\mathrm{P}\exp\left(\ep\int_{0}^{y}dt\,\mathbb{S}^{\left(I,y\right)}\!\left(t,z\right)\right)=\\
 & =\mathds{1}+\ep\int_{0}^{y}dt\,\mathbb{S}^{\left(I,y\right)}\!\left(t,z\right)+\ep^{2}\int_{0}^{y}dt\,\mathbb{S}^{\left(I,y\right)}\!\left(t,z\right)\int_{0}^{t}dt'\,\mathbb{S}^{\left(I,y\right)}\!\left(t',z\right)+\dots.\nonumber
\end{align}

\subsubsection*{Solution as multiple polylogarithms}

Because the $\mathbb{S}$ matrices have the structure of
eq.~\eqref{eq:epsilon-form-s}, the iterated integrals in the
general solution will result in terms of the form
\begin{equation}
    \label{eq:hlog-definition}
G\!\left(a,b,\dots;x\right)=\int_{0}^{x}\frac{dt}{t-a}\int_{0}^{t}\frac{dt'}{t'-b}\int_{0}^{t'}\dots=\int_{0}^{x}\frac{dt}{t-a}G\!\left(b,\dots;t\right).
\end{equation}
This is a class of functions known as ``multiple polylogarithms''~\cite{Goncharov98},
``Goncharov polylogarithms'' (GPLs), or ``hyperlogarithms''.
If the set of parameters is restricted to $\{0,1,-1\}$, these
correspond to a well known class of ``harmonic polylogarithms''~\cite{RV99}.
The set $\{0,1,1-x,-x\}$ corresponds to ``2d harmonic polylogarithms''~\cite{GR00}.

Note that for a consistent definition of GPLs, one needs to
introduce a special case for
\begin{equation}
G\!(\overbrace{0,\dots,0}^{n};x)=\frac{1}{n!}\ln\!\left(x\right).
\end{equation}
Because we have chosen to first integrate along the $\left(0,0\right)\to\left(0,z\right)$
direction, and our alphabet is limited to eq.~\eqref{eq:alphabet},
eq.~\eqref{eq:de-integration-order-zy} will result in each $J_{i}$
having the form
\begin{equation}
\sum\mathcal{C}\,G\!\left(\overrightarrow{w_{z}};y\right)G\!\left(\overrightarrow{w_{c}};z\right),\label{eq:hlogs-y-z}
\end{equation}
where $\mathcal{C}$ are some constants, and the parameters of $G$
are taken from sets
\begin{equation}
w_{z}\in\{0,1,1-z,-z\},\qquad\text{and}\qquad w_{c}\in\{0,1\}.
\end{equation}

\subsubsection*{Finding the combined $\ep$-form}

The possibility and usefulness of transforming the differential equations
into the form of eq.~\eqref{eq:epsilon-form} were first pointed
out in~\cite{Henn13}. Later in~\cite{Lee14} an algorithm was described
that constructs such transformations (in the form of the matrices
$T_{ij})$ for differential equation systems in one variable. The
algorithm was later updated in~\cite{LP17} and~\cite[section 8]{Blondel18}.
Two implementations of that algorithm followed, in the form of
\noun{Fuchsia}~\cite{GM17} and \noun{Epsilon}~\cite{Prausa17}
tools.
Other approaches include \noun{Canonica}~\cite{Meyer17} with its
own algorithm based on a rational ansatz for $T_{ij}$ explicitly designed for differential equations in
multiple variables, and an approach based on finding integrals
with constant leading singularities~\cite{Wasser18}.

While not being described in the original paper~\cite{Lee14},
the same algorithm treating single-variable case can be reused
for the case of multiple variables too.
Here is how:
\begin{enumerate}
\item Look at the first system from eq.~\eqref{eq:de-system}, the differential
equation system in $y$, and construct a transformation $I_{i}=T_{ij}^{\left(I,y\right)}J_{j}^{\left(y\right)}$
that reduces it to an $\ep$-form as described in~\cite{Lee14}.
The variable $z$ stands as a free parameter during this reduction.
\item Write down the differential equation system in the second variable,
$z$, for the new basis $J_{j}^{\left(y\right)}$, and reduce that
into the $\ep$-form as well, $J_{i}^{\left(y\right)}=T_{ij}^{\left(I,yz\right)}J_{j}$.
\item If $T_{ij}^{\left(I,yz\right)}$ is independent of $y$ and $d$---and
this is the case in practice---then it will not spoil the $\ep$-form
in $y$, so that the combined transformation $I_{i}=T_{ij}^{\left(I,y\right)}T_{ij}^{\left(I,yz\right)}J_{j}$
will result in an $\ep$-form in both $y$ and $z$.
\item Iterate from step 2 for each of the remaining variables, if any.
\end{enumerate}
This is an outline of the approach that \noun{Fuchsia} uses to find
$\ep$-form transformations for multi-variate systems,\footnote{We are using the new C++ version of \noun{Fuchsia} available at \url{https://github.com/magv/fuchsia.cpp}.} and how it was possible for us to compute the three $T_{ij}^{\left(I\right)}$
from eq.~\eqref{eq:epsilon-form-transformation}.

\subsubsection*{Fixing the integration constants}

To determine all the integration constants, it is sufficient to use
the following three conditions.
\begin{enumerate}
\item Some of the simpler integrals are known for arbitrary $d$ in terms
of Gamma functions $\Gamma$, and hypergeometric functions $\!_{2}F_{1}$
and $\!_{3}F_{2}$. In particular we fix the values of integrals 1,
2, 3, 4, 7, 8, 9, 10, and 11 from the PA topology; 1, 2, 3, 4, 5,
6, 7, 8, and 19 from NA; 1, 2, 3, 4, 5, 6, 7, 14, and 15 from NB.
We have collected the expressions for these integrals in \autoref{sec:Table-of-loop-integrals}.\\
Note that all 2-loop integrals have a common prefactor of the form
$\left(-q^{2}-i0\right)^{d}$. The $i0$ prescription is specified
here explicitly to fix the analytic continuation of the $\ln\!\left(-1\right)$
terms that will appear if one is to expand this prefactor in $\ep$.
Practically speaking, because all the integrals are proportional to
this factor, there is no need to expand it: it is better to separate
it out into a common prefactor, and only work with the remaining part
of the integral, which is free of imaginary numbers.
\item All of our integrals are massless, and therefore must only have discontinuities
at limits $s_{ij}\to0$ and nowhere else. On the other hand, the differential
equations we are solving also have poles at $s_{ij}\to1$, as the
list of denominators in eq.~\eqref{eq:alphabet} demonstrates. Thus,
requiring that the apparent discontinuities of the general solution
at $s_{ij}\to1$ vanish will generate nontrivial identities between
the integration constants. This requirement can be written down by
separating the terms proportional to $\ln\!\left(1-s_{ij}\right)$,
and enforcing that the coefficient in front of them vanishes in the
limit.
\item The planar integrals only have discontinuities at limits where adjacent
momenta go to zero. For the PA topology this means that it should
be regular at $s_{12}\to0$ (i.e. $y+z\to1$), as long as $q^{2}\neq0$.
Similarly for the planar integrals from other topologies. Here again
we are looking at the logarithmic terms like $\ln\!\left(s_{12}\right)$,
enforcing the cancellation of the coefficients in front of them in
the limit.
\end{enumerate}
To apply the regularity conditions above one needs to separate the
terms proportional to $\ln^{k}\left(1-s_{ij}\right)$, and require
that the coefficient of each is exactly zero in the limit $s_{ij}\to1$.
For the limit $y\to1$, to separate the divergent logarithms, it is
enough to employ the GPL shuffle relations to rewrite every
$G\!\left(1\dots,\overrightarrow{w_{y}};y\right)$ in eq.~\eqref{eq:hlogs-y-z}
into a product of the divergent factor $G\!\left(1\dots;y\right)$
and a part finite at $y\to1$. For the limit $z\to1$ the same cannot
be done directly on eq.~\eqref{eq:hlogs-y-z}, because $z$ appears
in the parameter list of $G\!\left(\dots;y\right)$. Instead, we can
rewrite eq.~\eqref{eq:hlogs-y-z} into the reverse form,
\begin{equation}
A=\sum C\,G\!\left(\overrightarrow{w_{y}};z\right)G\!\left(\overrightarrow{w_{c}};y\right),
\end{equation}
where
\begin{equation}
w_{y}\in\{0,1,1-y,-y\},
\end{equation}
and factor our logarithmic terms from that.

Such a rewrite can be achieved by the quite general technique of recursively
differentiating the GPL and then integrating it back:
\begin{equation}
G\!\left(\overrightarrow{w_{z}};y\right)=\left.G\!\left(\overrightarrow{w_{z}};y\right)\right|_{z=0}+\int_{0}^{z}dz\frac{\partial}{\partial z}G\!\left(\overrightarrow{w_{z}};y\right).\label{eq:hlogs-z-y}
\end{equation}
Practically speaking, a much more general incarnation of this technique
is implemented in the \code{fibrationBasis} routine from \noun{HyperInt}~\cite{Panzer14}.
In our experience it is powerful enough to perform these transformations
up to weight 9. At weight 10 the recursion results in too many terms
so that memory and performance optimizations become necessary. For
this reason we had to implement this transformation manually in C++
with the help of \noun{GiNaC}~\cite{BFK00}.

Note that applying the regularity conditions above to the weight-10
expansion of eq.~\eqref{eq:de-integration-order-zy} only
fixes the constants $\mathbf{C}$ up to weight 8.
These results are quite large (megabytes of text), and we are
providing them in machine-readable format in the ancillary files,
as described in \autoref{sec:Ancillary-files}.

Interestingly, weight~8 amplitudes are insufficient to compute all
of the VVRR integrals to weight~8 as well: because of an apparent
cancellation during DRR, some of the weight~8 information is
lost, and VRRR integrals 20-23 can only be obtained to weight~7.
Still, this is sufficient for practical needs, because weight~7
corresponds to terms of the order~$\ep^1$ or higher.
Moreover, after cross-checking these results we shall improve
upon them in \autoref{sec:drr-for-3cut}.

\subsection{Cross-checks}
\label{sec:num-check-VVRR-VRRV}

One way to cross-check the obtained 3-particle cut integrals is
to calculate them numerically.
This can be conveniently done by using sector decomposition as
implemented in \noun{Fiesta4}~\cite{Smirnov15}, all that is
required is a parametrization of the integrals suitable for
the \texttt{SDEvaluateDirect} function (\noun{Fiesta} does not
construct such parametrizations automationcaly for cut integrals).
To that end, the Feynman parametrization can be used for the loop
parts of the integrals, and the naive parameterization from
eq.~\eqref{eq:dps3-definition} can be used for the phase-space
parts.

Proceeding this way we were able to calculate the first 3
terms of the $\ep$-expansion for each VVRR and VRRV integral
numerically in $d=6-2\ep$ and $d=8-2\ep$.
Note that we were unable to do the same for the integrals in
$d=4-2\ep$: due to fairly high pole orders, \noun{Fiesta} has
identified thousands of sectors in some cases, and the whole
calculation effectively froze.
Still, a numerical match within 1\% of accuracy for both $6-2\ep$
and $8-2\ep$ gives us confidence in both the $4-2\ep$ results
and the DRR matrices.

Additionally, we have compared our results to the weight-6
series for $\vvrr{16}$, $\vvrr{8}$, and $\vvrr{9}$ reported
in~\cite{CFHMSS15}, and find them to match fully.

\section{3-loop 2-particle cut integrals (VVVR, VVRV)}

There are 22 distinct 2-particle cuts in total, all depicted in
\autoref{tab:VVVR} and \autoref{tab:VVRV}.
These integrals correspond to 3-loop form factors, and have been
calculated in~\cite{HHM07,HHKS09,LSS10}.
We list these integrals here for completeness, and as a reference
to the order in which their values are presented in the ancillary
files (as described in \autoref{sec:Ancillary-files}).

\begin{table}[H]
\begin{tabular}{>{\centering}p{2.6cm}>{\centering}p{2.6cm}>{\centering}p{2.6cm}>{\centering}p{2.6cm}>{\centering}p{2.6cm}}
%
    \scalebox{0.75}{\input{fig/vvvr/1.tikz}}%
 & %
    \scalebox{0.75}{\input{fig/vvvr/2.tikz}}%
 & %
    \scalebox{0.75}{\input{fig/vvvr/3.tikz}}%
 & %
    \scalebox{0.75}{\input{fig/vvvr/4.tikz}}%
 & %
    \scalebox{0.75}{\input{fig/vvvr/5.tikz}}%
\tabularnewline
$1,\mathrm{L}_{1110100*11*}$ & $2,\mathrm{H}_{10**0111110}$ & $3,\mathrm{L}_{0111100*01*}$ & $4,\mathrm{H}_{10**0101101}$ & $5,\mathrm{H}_{01**0101001}$\tabularnewline
 &  &  &  & \tabularnewline
%
    \scalebox{0.75}{\input{fig/vvvr/6.tikz}}%
 & %
    \scalebox{0.75}{\input{fig/vvvr/7.tikz}}%
 & %
    \scalebox{0.75}{\input{fig/vvvr/8.tikz}}%
 & %
    \scalebox{0.75}{\input{fig/vvvr/9.tikz}}%
 & %
    \scalebox{0.75}{\input{fig/vvvr/10.tikz}}%
\tabularnewline
$6,\mathrm{N}_{101**101111}$ & $7,\mathrm{L}_{1101110*01*}$ & $8,\mathrm{H}_{10**1110101}$ & $9,\mathrm{H}_{01**1110001}$ & $10,\mathrm{H}_{01**0110011}$\tabularnewline
 &  &  &  & \tabularnewline
%
    \scalebox{0.75}{\input{fig/vvvr/11.tikz}}%
 & %
    \scalebox{0.75}{\input{fig/vvvr/12.tikz}}%
 & %
    \scalebox{0.75}{\input{fig/vvvr/13.tikz}}%
 & %
    \scalebox{0.75}{\input{fig/vvvr/14.tikz}}%
 & %
    \scalebox{0.75}{\input{fig/vvvr/15.tikz}}%
\tabularnewline
$11,\mathrm{N}_{110**101111}$ & $12,\mathrm{N}_{101**110111}$ & $13,\mathrm{H}_{01**0111011}$ & $14,\mathrm{N}_{*011110*111}$ & $15,\mathrm{M}_{111**111111}$\tabularnewline
 &  &  &  & \tabularnewline
%
    \scalebox{0.75}{\input{fig/vvvr/16.tikz}}%
 & %
    \scalebox{0.75}{\input{fig/vvvr/17.tikz}}%
 & %
    \scalebox{0.75}{\input{fig/vvvr/18.tikz}}%
 & %
    \scalebox{0.75}{\input{fig/vvvr/19.tikz}}%
 & %
    \scalebox{0.75}{\input{fig/vvvr/20.tikz}}%
\tabularnewline
$16,\mathrm{N}_{111**111110}$ & $17,\mathrm{L}_{**011101011}$ & $18,\mathrm{H}_{11**1101101}$ & $19,\mathrm{N}_{111**111111}$ & $20,\mathrm{H}_{11**1111111}$\tabularnewline
 &  &  &  & \tabularnewline
%
    \scalebox{0.75}{\input{fig/vvvr/21.tikz}}%
 & %
    \scalebox{0.75}{\input{fig/vvvr/22.tikz}}%
 &  &  & \tabularnewline
$21,\mathrm{J}_{**101111111}$ & $22,\mathrm{J}_{1110111*11*}$ &  &  & \tabularnewline
\end{tabular}

\caption{\protect\strong{VVVR}, 2-particle cut master integrals with 3 loops
  on one side of the cut.}
\label{tab:VVVR}
\end{table}

\begin{table}[H]
\begin{tabular}{>{\centering}p{2.6cm}>{\centering}p{2.6cm}>{\centering}p{2.6cm}>{\centering}p{2.6cm}}
%
    \scalebox{0.75}{\input{fig/vvrv/1.tikz}}%
 & %
    \scalebox{0.75}{\input{fig/vvrv/2.tikz}}%
 & %
    \scalebox{0.75}{\input{fig/vvrv/3.tikz}}%
 & %
    \scalebox{0.75}{\input{fig/vvrv/4.tikz}}%
\tabularnewline
$1,\mathrm{L}_{0111*0010*1}$ & $2,\mathrm{L}_{1101*1010*1}$ & $3,\mathrm{L}_{1110*0011*1}$ & $4,\mathrm{J}_{11*01111*11}$\tabularnewline
\end{tabular}

\caption{\protect\strong{VVRV}, 2-particle cut master integrals with 2 loops
  on one side of the cut, and 1 loop on the other.}
\label{tab:VVRV}
\end{table}

\section{5-particle phase-space integrals (RRRR)}

There are 31 master integrals for the 5-particle cuts. All of these
have been calculated in~\cite{GMP18}, and we list them here for
completeness. These integrals will be important for the Cutkosky relations
in \autoref{sec:Cutkosky-rules}.

\begin{table}[H]
\begin{tabular}{>{\centering}p{2.6cm}>{\centering}p{2.6cm}>{\centering}p{2.6cm}>{\centering}p{2.6cm}>{\centering}p{2.6cm}}
%
    \scalebox{0.75}{\input{fig/rrrr/1.tikz}}%
 & %
    \scalebox{0.75}{\input{fig/rrrr/2.tikz}}%
 & %
    \scalebox{0.75}{\input{fig/rrrr/3.tikz}}%
 & %
    \scalebox{0.75}{\input{fig/rrrr/4.tikz}}%
 & %
    \scalebox{0.75}{\input{fig/rrrr/5.tikz}}%
\tabularnewline
$1,\mathrm{H}_{00*00*0**0*}$ & $2,\mathrm{M}_{01*101*1***}$ & $3,\mathrm{H}_{10*10*0**0*}$ & $4,\mathrm{N}_{*101*101***}$ & $5,\mathrm{H}_{11*11*0**0*}$\tabularnewline
 &  &  &  & \tabularnewline
%
    \scalebox{0.75}{\input{fig/rrrr/6.tikz}}%
 & %
    \scalebox{0.75}{\input{fig/rrrr/7.tikz}}%
 & %
    \scalebox{0.75}{\input{fig/rrrr/8.tikz}}%
 & %
    \scalebox{0.75}{\input{fig/rrrr/9.tikz}}%
 & %
    \scalebox{0.75}{\input{fig/rrrr/10.tikz}}%
\tabularnewline
$6,\mathrm{H}_{11*11*1**1*}$ & $7,\mathrm{H}_{10*0*0***10}$ & $8,\mathrm{M}_{011*011****}$ & $9,\mathrm{N}_{1*011*01***}$ & $10,\mathrm{H}_{01*10**00**}$\tabularnewline
 &  &  &  & \tabularnewline
%
    \scalebox{0.75}{\input{fig/rrrr/11.tikz}}%
 & %
    \scalebox{0.75}{\input{fig/rrrr/12.tikz}}%
 & %
    \scalebox{0.75}{\input{fig/rrrr/13.tikz}}%
 & %
    \scalebox{0.75}{\input{fig/rrrr/14.tikz}}%
 & %
    \scalebox{0.75}{\input{fig/rrrr/15.tikz}}%
\tabularnewline
$11,\mathrm{N}_{110*110****}$ & $12,\mathrm{H}_{01*10**10**}$ & $13,\mathrm{N}_{11*1**1**10}$ & $14,\mathrm{N}_{101*110****}$ & $15,\mathrm{M}_{111*111****}$\tabularnewline
 &  &  &  & \tabularnewline
%
    \scalebox{0.75}{\input{fig/rrrr/16.tikz}}%
 & %
    \scalebox{0.75}{\input{fig/rrrr/17.tikz}}%
 & %
    \scalebox{0.75}{\input{fig/rrrr/18.tikz}}%
 & %
    \scalebox{0.75}{\input{fig/rrrr/19.tikz}}%
 & %
    \scalebox{0.75}{\input{fig/rrrr/20.tikz}}%
\tabularnewline
$16,\mathrm{N}_{111*111****}$ & $17,\mathrm{H}_{01*0*1***10}$ & $18,\mathrm{M}_{11*1**1*1*1}$ & $19,\mathrm{M}_{11*11****11}$ & $20,\mathrm{M}_{11*111*1***}$\tabularnewline
 &  &  &  & \tabularnewline
%
    \scalebox{0.75}{\input{fig/rrrr/21.tikz}}%
 & %
    \scalebox{0.75}{\input{fig/rrrr/22.tikz}}%
 & %
    \scalebox{0.75}{\input{fig/rrrr/23.tikz}}%
 & %
    \scalebox{0.75}{\input{fig/rrrr/24.tikz}}%
 & %
    \scalebox{0.75}{\input{fig/rrrr/25.tikz}}%
\tabularnewline
$21,\mathrm{N}_{1*11***1*10}$ & $22,\mathrm{N}_{1*11***1*11}$ & $23,\mathrm{N}_{1*111*11***}$ & $24,\mathrm{N}_{11*1**1**11}$ & $25,\mathrm{H}_{11*1*1***11}$\tabularnewline
 &  &  &  & \tabularnewline
%
    \scalebox{0.75}{\input{fig/rrrr/26.tikz}}%
 & %
    \scalebox{0.75}{\input{fig/rrrr/27.tikz}}%
 & %
    \scalebox{0.75}{\input{fig/rrrr/28.tikz}}%
 & %
    \scalebox{0.75}{\input{fig/rrrr/29.tikz}}%
 & %
    \scalebox{0.75}{\input{fig/rrrr/30.tikz}}%
\tabularnewline
$26,\mathrm{H}_{01*01**00**}$ & $27,\mathrm{N}_{*011*110***}$ & $28,\mathrm{H}_{*11*110**0*}$ & $29,\mathrm{N}_{*111*111***}$ & $30,\mathrm{N}_{1*11*1**1*1}$\tabularnewline
 &  &  &  & \tabularnewline
%
    \scalebox{0.75}{\input{fig/rrrr/31.tikz}}%
 &  &  &  & \tabularnewline
$31,\mathrm{H}_{11*11**11**}$ &  &  &  & \tabularnewline
\end{tabular}

\caption{\protect\strong{RRRR}, 5-particle cut master integrals.}
\label{tab:RRRR}
\end{table}

\section{Relations from Cutkosky rules}
\label{sec:Cutkosky-rules}

By this point we already know all the cuts of VVVV integrals to
at least weight~7.
A good test for the consistency of all these results is the
optical theorem (Cutkosky rules, to be more precise), which
relates the discontinuity of the VVVV integrals to its cuts.

The general optical theorem comes from the requirement of unitarity
of the scattering matrix~$S$,
\begin{equation}
S^{\dagger}S=\mathds{1}.
\end{equation}
Introducing the transition matrix $T$ as
\begin{equation}
S=\mathds{1}+iT,
\end{equation}
it follows that
\begin{equation}
iT+\left(iT\right)^{\dagger}=-\left(iT\right)^{\dagger}iT.
\end{equation}
For a decay of a single particle with momentum $q$, rewriting this
relation in terms of the transition amplitudes produces
\begin{equation}
2\,\mathrm{Re}\left\langle q\right|iT\left|q\right\rangle =-\left\langle q\right|\left(iT\right)^{\dagger}\left(\sum_{x}\left|x\right\rangle \left\langle x\right|\right)iT\left|q\right\rangle =-\sum_{n}\int d\mathrm{PS}_{n}\,\big|\left\langle p_{1},\dots,p_{n}\right|iT\mid q\rangle\,\big|^{2}.
\end{equation}
This is the optical theorem. Note that once the transition amplitude
is expanded in terms of individual Feynman diagrams, the right-hand
side will consist precisely of our cut integrals as in eq.~\eqref{eq:cut-definition}.

Cutkosky rules~\cite{Cutkosky60,tHV74} provide a stronger form of
this relation that holds not only for the whole transition amplitude
$\left\langle q\right|iT\left|q\right\rangle $, but for each individual
Feynman diagram $F$ that comprise it too,
\begin{equation}
F+F^{*}=-\sum_{i}\mathrm{Cut}_{i}F,\label{eq:cutkosky}
\end{equation}
where the sum goes over all possible cuts of the diagram, each cut
being a partition into two sides, with the right-hand side complex-conjugated,
and the propagators between sides set on shell, exactly like in eq.~\eqref{eq:cut-definition}.

To write down these relations for the VVVV master integrals in this
convenient form, we'll augment our integrals with Feynman rules stemming
from a simple scalar field theory with the Lagrangian of the form
\begin{equation}
L=\frac{1}{2}\left(\partial\phi\right)^{2}+\frac{\lambda_{3}}{3!}\phi^{3}+\frac{\lambda_{4}}{4!}\phi^{4}+\dots.
\end{equation}

The momenta-space Feynman rules corresponding to this Lagrangian are
\[
%
    \scalebox{0.75}{\input{fig/3pt.tikz}}%
=i\lambda_{3}\hspace{4em}%
    \scalebox{0.75}{\input{fig/4pt.tikz}}%
=i\lambda_{4}\hspace{4em}%
    \scalebox{0.75}{\input{fig/5pt.tikz}}%
=\dots
\]
\[
%
    \scalebox{0.75}{\begin{tikzpicture}
	\begin{pgfonlayer}{nodelayer}
		\node [style=dot] (0) at (0, 0) {};
		\node [style=dot] (1) at (2, 0) {};
		\node [style=none] (2) at (1, 0.25) {p};
	\end{pgfonlayer}
	\begin{pgfonlayer}{edgelayer}
		\draw [style=edge] (0) to (1);
	\end{pgfonlayer}
\end{tikzpicture}
}%
=\frac{i}{p^{2}+i0}\hspace{4em}%
    \scalebox{0.75}{\begin{tikzpicture}
	\begin{pgfonlayer}{nodelayer}
		\node [style=dot] (0) at (0, 0) {};
		\node [style=dot] (1) at (2, 0) {};
		\node [style=none] (2) at (1, 0.25) {p};
	\end{pgfonlayer}
	\begin{pgfonlayer}{edgelayer}
		\draw [style=cut edge] (0) to (1);
	\end{pgfonlayer}
\end{tikzpicture}
}%
=2\pi\delta^{+}\!\left(p^{2}\right)
\]

An additional prescription for cut integrals is this: every vertex
and propagator to the right side of the cut needs to be conjugated.

Note that the values of $\lambda_{n}$ are not important for us here,
because a cut of a diagram will have the same overall $\lambda$ factor
as the initial diagram, which will thus factor out from eq.~\eqref{eq:cutkosky}.

With this in mind, writing down eq.~\eqref{eq:cutkosky} for each
Feynman diagram corresponding to a VVVV master integral, and mapping
the cuts onto our master integrals, we obtain the following relations:
\begin{align}
2\,\mathrm{Im}\,\smallfig{vvvv/1}= & +2\,\mathrm{Im}\,\smallfig{vvvr/1}-2\,\mathrm{Im}\,\smallfig{vvrv/3}\label{eq:ot-1}\\
2\,\mathrm{Im}\,\smallfig{vvvv/2}= & -I\,\smallfig{vvvr/3}+I\,\smallfig{vvrv/1}+\smallfig{vvrr/4}\\
2\,\mathrm{Im}\,\smallfig{vvvv/3}= & +2\,\mathrm{Re}\,\smallfig{vvrr/1}\\
2\,\mathrm{Im}\,\smallfig{vvvv/4}= & +I\,\smallfig{vvvr/9}+\smallfig{vvrr/3}-\smallfig{vrrv/3}\\
2\,\mathrm{Im}\,\smallfig{vvvv/5}= & +2\,\mathrm{Im}\,\smallfig{vrrr/4}\\
2\,\mathrm{Im}\,\smallfig{vvvv/6}= & -2\,\mathrm{Im}\,\smallfig{vvvr/2}+4\,\mathrm{Im}\,\smallfig{vrrr/9}\\
2\,\mathrm{Im}\,\smallfig{vvvv/7}= & -\smallfig{rrrr/1}\\
2\,\mathrm{Im}\,\smallfig{vvvv/8}= & -2\,\mathrm{Im}\,\smallfig{vvvr/4}-2\,\smallfig{rrrr/3}\\
2\,\mathrm{Im}\,\smallfig{vvvv/9}= & +2\,\mathrm{Re}\,\smallfig{vvrr/5}-2\,\smallfig{rrrr/7}\\
2\,\mathrm{Im}\,\smallfig{vvvv/10}= & +2\,\mathrm{Re}\,\smallfig{vvrr/6}-\smallfig{vrrv/2}\\
2\,\mathrm{Im}\,\smallfig{vvvv/11}= & +\smallfig{vvrr/7}+I\,\smallfig{vrrr/1}\\
2\,\mathrm{Im}\,\smallfig{vvvv/12}= & +2\,\mathrm{Re}\,\smallfig{vvrr/8}+4\,\mathrm{Im}\,\smallfig{vrrr/5}-4\,\smallfig{rrrr/8}-\\
 & -\smallfig{rrrr/2}\nonumber \\
2\,\mathrm{Im}\,\smallfig{vvvv/13}= & +I\,\smallfig{vvvr/10}+\smallfig{vvrr/10}-I\,\smallfig{vrrr/2}-\\
 & -\smallfig{rrrr/10}\nonumber \\
2\,\mathrm{Im}\,\smallfig{vvvv/14}= & +2\,\mathrm{Re}\,\smallfig{vvrr/16}+4\,\mathrm{Im}\,\smallfig{vrrr/12}-2\,\smallfig{rrrr/17}\\
2\,\mathrm{Im}\,\smallfig{vvvv/15}= & +I\,\smallfig{vvvr/5}+I\,\smallfig{vrrr/16}\\
2\,\mathrm{Im}\,\smallfig{vvvv/16}= & -I\,\smallfig{vvvr/7}+I\,\smallfig{vvrv/2}+I\,\smallfig{vvvr/17}-\\
 & -2I\,\smallfig{vrrr/19}\nonumber \\
2\,\mathrm{Im}\,\smallfig{vvvv/17}= & -2\,\mathrm{Im}\,\smallfig{vvvr/8}-2\,\mathrm{Im}\,\smallfig{vrrr/7}+2\,\mathrm{Im}\,\smallfig{vrrr/23}\\
2\,\mathrm{Im}\,\smallfig{vvvv/18}= & -2\,\mathrm{Im}\,\smallfig{vvvr/11}+2\,\mathrm{Re}\,\smallfig{vvrr/12}+2\,\mathrm{Im}\,\smallfig{vrrr/24}-\\
 & -2\,\mathrm{Im}\,\smallfig{vrrr/8}-\smallfig{rrrr/11}-\smallfig{rrrr/9}-\nonumber \\
 & -\smallfig{rrrr/4}\nonumber \\
2\,\mathrm{Im}\,\smallfig{vvvv/19}= & +I\,\smallfig{vvvr/13}+\smallfig{vvrr/14}-2I\,\smallfig{vrrr/10}+\\
 & +I\,\smallfig{vrrr/25}-2\,\smallfig{rrrr/12}\nonumber \\
2\,\mathrm{Im}\,\smallfig{vvvv/20}= & -2\,\mathrm{Im}\,\smallfig{vvvr/16}+2\,\mathrm{Re}\,\smallfig{vvrr/18}+4\,\mathrm{Im}\,\smallfig{vrrr/27}-\\
 & -2\,\mathrm{Im}\,\smallfig{vrrr/11}-4\,\smallfig{rrrr/13}-2\,\smallfig{rrrr/21}\nonumber \\
2\,\mathrm{Im}\,\smallfig{vvvv/21}= & +I\,\smallfig{vvvr/6}-I\,\smallfig{vvvr/14}+2\,\smallfig{vvrr/15}+\\
 & +2I\,\smallfig{vrrr/18}-I\,\smallfig{vrrr/30}-4\,\smallfig{rrrr/14}\nonumber \\
2\,\mathrm{Im}\,\smallfig{vvvv/22}= & +2\,\mathrm{Re}\,\smallfig{vvrr/2}-\smallfig{vrrv/4}-\smallfig{rrrr/26}\\
2\,\mathrm{Im}\,\smallfig{vvvv/23}= & +I\,\smallfig{vvvr/12}+2\,\smallfig{vvrr/9}-\smallfig{vrrv/5}+\label{eq:ot-23-coupled-VVRR}\\
 & +\smallfig{vvrr/11}-2I\,\smallfig{vrrr/3}-I\,\smallfig{vrrr/6}+\nonumber \\
 & +I\,\smallfig{vrrr/17}-2\,\smallfig{rrrr/27}\nonumber \\
2\,\mathrm{Im}\,\smallfig{vvvv/24}= & -2\,\mathrm{Im}\,\smallfig{vvvr/18}-\smallfig{vrrv/6}+2\,\mathrm{Re}\,\smallfig{vvrr/13}+\\
 & +2\,\mathrm{Im}\,\smallfig{vrrr/21}-\smallfig{rrrr/5}-\smallfig{rrrr/28}\nonumber \\
2\,\mathrm{Im}\,\smallfig{vvvv/25}= & -2\,\mathrm{Im}\,\smallfig{vvvr/20}+4\,\mathrm{Re}\,\smallfig{vvrr/21}-\smallfig{vrrv/7}+\\
 & +4\,\mathrm{Im}\,\smallfig{vrrr/15}+4\,\mathrm{Im}\,\smallfig{vrrr/33}+2\,\mathrm{Im}\,\smallfig{vrrr/20}-\nonumber \\
 & -2\,\smallfig{rrrr/6}-2\,\smallfig{rrrr/31}-4\,\smallfig{rrrr/25}\nonumber \\
2\,\mathrm{Im}\,\smallfig{vvvv/26}= & -2\,\mathrm{Im}\,\smallfig{vvvr/15}+4\,\mathrm{Re}\,\smallfig{vvrr/17}+4\,\mathrm{Im}\,\smallfig{vrrr/26}+\\
 & +2\,\mathrm{Im}\,\smallfig{vrrr/31}+4\,\mathrm{Im}\,\smallfig{vrrr/13}-2\,\smallfig{rrrr/15}-\nonumber \\
 & -2\,\smallfig{rrrr/20}-4\,\smallfig{rrrr/19}-4\,\smallfig{rrrr/18}\nonumber \\
2\,\mathrm{Im}\,\smallfig{vvvv/27}= & -2\,\mathrm{Im}\,\smallfig{vvvr/19}+2\,\mathrm{Re}\,\smallfig{vvrr/20}-\smallfig{vrrv/8}+\label{eq:ot-27-coupled-VVRR}\\
 & +2\,\mathrm{Re}\,\smallfig{vvrr/19}+2\,\mathrm{Im}\,\smallfig{vrrr/22}+2\,\mathrm{Im}\,\smallfig{vrrr/32}+\nonumber \\
 & +2\,\mathrm{Im}\,\smallfig{vrrr/29}+2\,\mathrm{Im}\,\smallfig{vrrr/28}-2\,\mathrm{Im}\,\smallfig{vrrr/14}-\nonumber \\
 & -\smallfig{rrrr/16}-2\,\smallfig{rrrr/24}-\smallfig{rrrr/23}-\nonumber \\
 & -2\,\smallfig{rrrr/30}-2\,\smallfig{rrrr/22}-\smallfig{rrrr/29}\nonumber \\
2\,\mathrm{Im}\,\smallfig{vvvv/28}= & -I\,\smallfig{vvvr/22}-I\,\smallfig{vvrv/4}+I\,\smallfig{vvvr/21}+\label{eq:ot-28}\\
 & +2\,\smallfig{vvrr/22}-2\,\smallfig{vrrv/9}+I\,\smallfig{vrrr/34}+\nonumber \\
 & +4I\,\smallfig{vrrr/35}\nonumber
\end{align}

Inserting the values of our cut integrals into these relations, we
find that they all hold precisely. This concludes our cross-check.

\section{Dimensional recurrence relations for 3-particle cut integrals}
\label{sec:drr-for-3cut}

In \autoref{sec:VVRR-VRRV} we have postponed solving the dimensional
recurrence relations for 3-particle cut master integrals because
we did not have enough information to fix all the $\omega_i$.
Since that time we have obtained multiple terms of the $\ep$
expansion, as well as the Cutkosky relations.
This information combined will be enough to fix $\omega_i$ and
solve the DRR.
Such a solution will upgrade our knowledge of 3-particle cut
integrals from weight-7 series to weight-12, and more generally
will provide expressions that can be evaluated in arbitrary $d$
to arbitrary $\ep$ order numerically with any desired precision.

\subsection{DRA method by example: VRRV}
\label{sec:drr-for-vrrv}

We have already gone through the process of solving DRR in the
simpler case of VRRR integrals in \autoref{sec:vrrr}.
Let us now consider the more complicated cases.
To summarize, following the ``dimensional recurrence and
analyticity'' (DRA) method described in~\cite{Lee09}, to solve
DRR in the form of eq.~\eqref{eq:ldrr-triangular} for a given
integral, one needs to:

\begin{enumerate}
\item Solve all integrals in subsectors.
\item Choose a semi-open stripe of width 2 in the complex plane, such that $\Re d \in (d_0,d_0+2]$ or $\Re d \in [d_0,d_0+2)$, and restrict the analysis to this stripe.
\item Construct the homogeneous solution $\fn{\homH}{d}$ via eq.~\eqref{eq:drr-homogeneous-solution}, trying to minimize the number of its zeros on the chosen stripe.
\item Construct the particular solution $\fn{\partR}{d}$ via eq.~\eqref{eq:drr-particular-solution}.
\item Determine the set of poles and their multiplicities of $\fn{J}{d}$, $\fn{\homH^{-1}}{d}$, and $\fn{\partR}{d}$ on the chosen stripe.
\item Construct an ansatz for the periodic function $\fn{\omega}{d}$ by looking at the poles and the behaviour at $\Im d\to\pm\infty$ of $\fn{J}{d}$, $\fn{\homH^{-1}}{d}$, and $\fn{\partR}{d}$.
\item Fix the constants in the ansatz by expanding eq.~\eqref{eq:omega-sigma} near the poles, and/or other considerations.
\end{enumerate}

Let us illustrate these steps using a simple, but non trivial
case of $\vrrv{8}$ from \autoref{tab:VRRV}.
The VRRV family of integrals is considerably simpler than the
VVRR, because the loop part of these integrals factorizes into
a product of two one-loop integrals, for which arbitrary-$d$
expressions are known (see \autoref{sec:1to3-1l}), and which
simplifies the pole analysis and makes a numerical evaluation of
the integrals in arbitrary~$d$ easy via Monte-Carlo integration.

We shall restrict our analysis to the strip of $\Re d\in(6,8]$,
because all normalized VRRV integrals are finite on it.
The reason is that IR divergences are suppressed at $d\ge6$, and
surface UV divergences are cancelled by the normalization of
eq.~\eqref{eq:normalized-cut-definition}.
We can verify this finiteness numerically via Monte-Carlo
integration, which can be performed by parametrizing the three-particle
phase-space element entering eq.~\eqref{eq:cut-definition}
via eq.~\eqref{eq:dps3-definition}, and using the analytical
expressions for the remaining two one-loop amplitudes found in
\autoref{sec:1to3-1l}.
The result for $\vrrv{8}$ is presented on \autoref{fig:drr-I8-J};
it is finite as expected.

\subsubsection*{The homogeneous solution $\fn{\homH}{d}$}

\begin{figure}
    \centering
    \subfloat[][]{\includegraphics[width=0.5\textwidth]{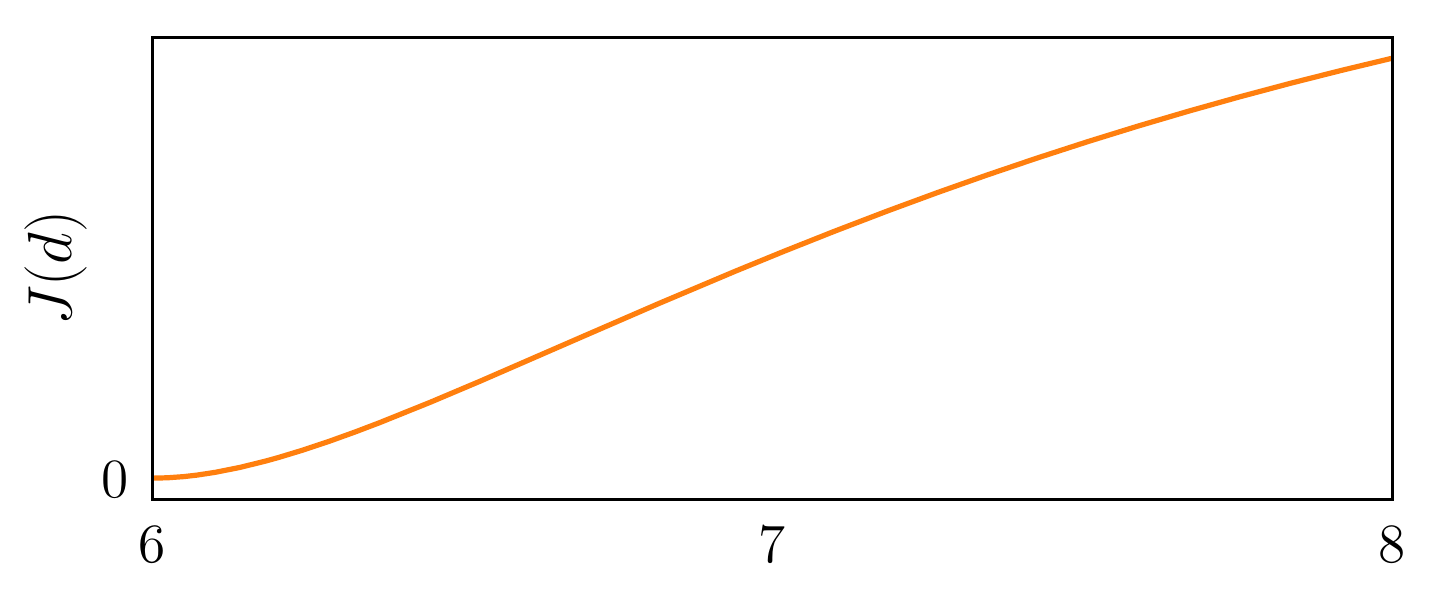}\label{fig:drr-I8-J}}
    \subfloat[][]{\includegraphics[width=0.5\textwidth]{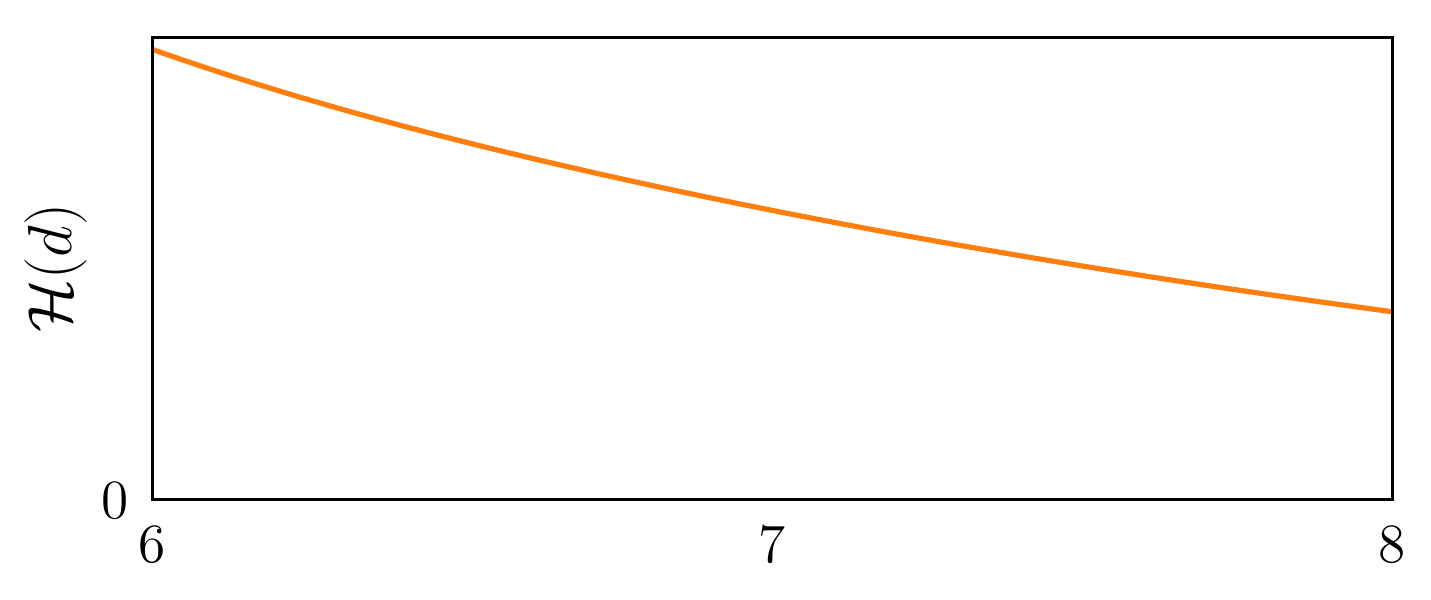}\label{fig:drr-I8-H}}
    \\
    \subfloat[][]{\includegraphics[width=0.5\textwidth]{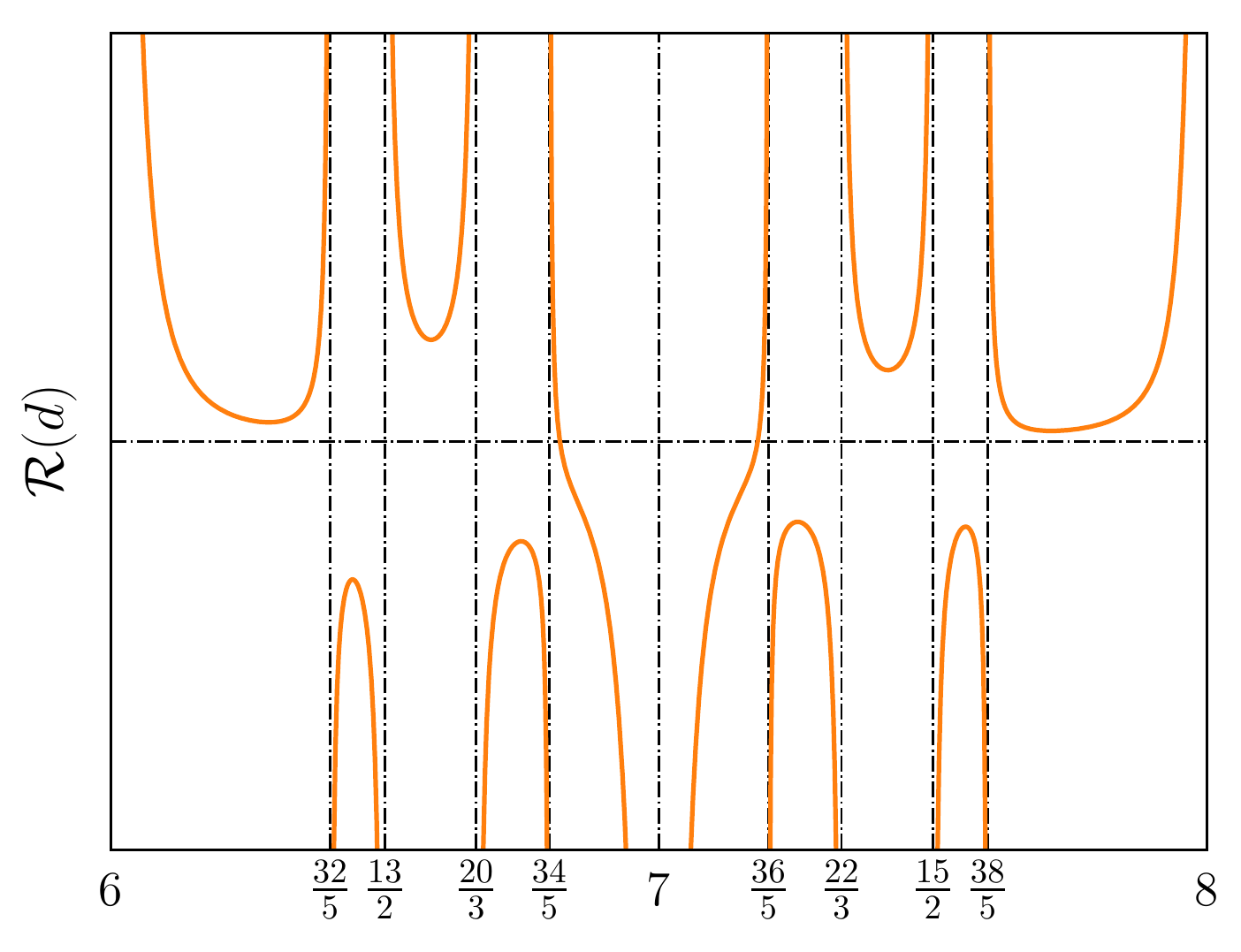}\label{fig:drr-I8-R}}
    \subfloat[][]{\includegraphics[width=0.5\textwidth]{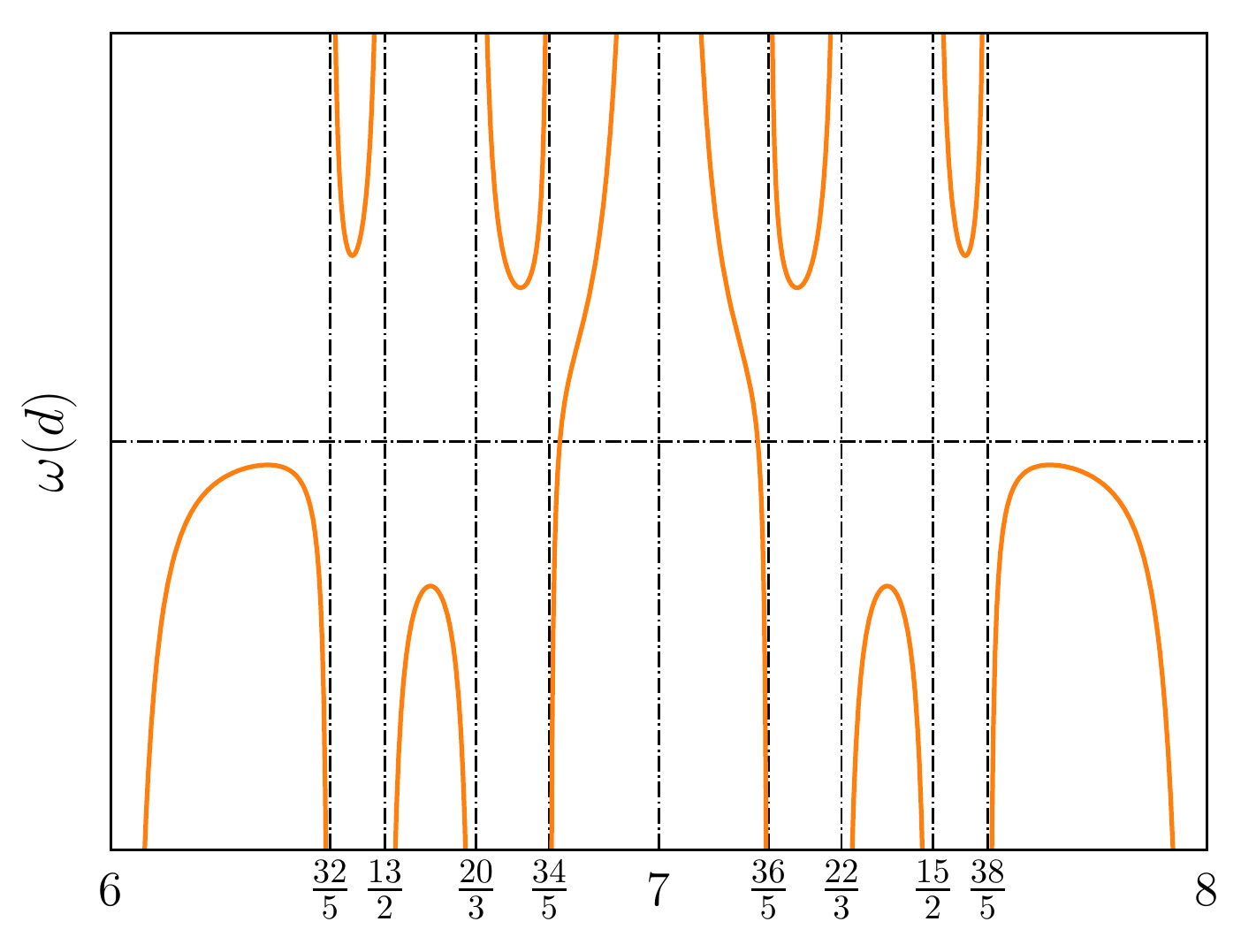}\label{fig:drr-I8-W}}
    \caption{Components of the DRR solution for $\vrrv{8}$ on the stripe $d \in (6,8]$.
      The plots show:
      \protect\subref{fig:drr-I8-J} the normalized integral itself;
      \protect\subref{fig:drr-I8-H} the homogeneous solution;
      \protect\subref{fig:drr-I8-R} the particular solution; and \protect\subref{fig:drr-I8-W} the periodic function.
      These plots lack vertical scale indicators because they
      only aim at illustrating the divergences of the functions.
    }
    \label{fig:drr-I8-JHRW}
\end{figure}

Next, to construct the homogeneous solution to eq.~\eqref{eq:ldrr-triangular}, we need the diagonal element of the DRR matrix, $M_{ii}$.
For $\vrrv{8}$ it has the form of
\begin{equation}
    \fn{M_{ii}}{d} = \frac{3^6}{5^5} 
        \frac{ \left(\frac{d}{2}-\frac{7}{3}\right) \left(\frac{d}{2}-2\right)^2 \left(\frac{d}{2}-\frac{5}{3}\right) \left(\frac{d}{2}-\frac{2}{3}\right) \left(\frac{d}{2}-\frac{1}{2}\right)^2 \left(\frac{d}{2}-\frac{1}{3}\right)}{\left(\frac{d}{2}-\frac{11}{5}\right) \left(\frac{d}{2}-\frac{9}{5}\right) \left(\frac{d}{2}-\frac{8}{5}\right) \left(\frac{d}{2}-\frac{3}{2}\right)^2 \left(\frac{d}{2}-\frac{7}{5}\right) \left(\frac{d}{2}-1\right)^2}.
\end{equation}
Note that the constant subtracted from $d/2$ in the factors is
always $\le7/3$.
This means that it is possible to construct such a $\fn{\homH}{d}$
that is free of poles and zeros at $d>14/3$.
Using eq.~\eqref{eq:drr-homogeneous-solution} and choosing the
first form of factors does precisely that, resulting in
\begin{equation}
    \label{eq:vrrv8-h}
    \fn{\homH}{d}=
    \left( \frac{3^6}{5^5} \right)^{\frac{d}{2}}
        \frac{ \fn{\Gamma}{\frac{d}{2}-\frac{7}{3}} \fn{\Gamma^2}{\frac{d}{2}-2} \fn{\Gamma}{\frac{d}{2}-\frac{5}{3}} \fn{\Gamma}{\frac{d}{2}-\frac{2}{3}} \fn{\Gamma^2}{\frac{d}{2}-\frac{1}{2}} \fn{\Gamma}{\frac{d}{2}-\frac{1}{3}} }{ \fn{\Gamma}{\frac{d}{2}-\frac{11}{5}} \fn{\Gamma}{\frac{d}{2}-\frac{9}{5}} \fn{\Gamma}{\frac{d}{2}-\frac{8}{5}} \fn{\Gamma^2}{\frac{d}{2}-\frac{3}{2}} \fn{\Gamma}{\frac{d}{2}-\frac{7}{5}} \fn{\Gamma^2}{\frac{d}{2}-1} }.
\end{equation}
This answer can also be obtained via the \texttt{HomogeneousSolution}
function from the \noun{Dream} package.
This solution is indeed finite and free from zeroes on the stripe
$(6,8]$ as can be seen on \autoref{fig:drr-I8-H}.

\subsubsection*{The particular solution $\fn{\partR}{d}$}

With $\fn{\homH}{d}$ ready, the particular solution $\fn{\partR}{d}$
can be constructed via eq.~\eqref{eq:drr-particular-solution}.
The numerical evaluation of the nested infinite sums in that
formula is the main functionality of the \noun{Dream} package,
and with its help we can plot $\fn{\partR}{d}$ on $d\in(6,8]$ with
arbitrary precision in many points.
The result can be seen on \autoref{fig:drr-I8-R}.
From this plot it is then easy to find the position of $\fn{\partR}{d}$ poles,
\begin{equation}
  \label{eq:vrrv8-poles}
  d_i = \left\{ \frac{32}{5},\, \frac{13}{2},\, \frac{20}{3},\, \frac{34}{5},\, 7,\, \frac{36}{5},\, \frac{22}{3},\, \frac{15}{2},\, \frac{38}{5},\, 8\right\}.
\end{equation}
Using \noun{Dream} again, we can now evaluate the series for
$\fn{\partR}{d}$ at $d=d_i-2\ep$, and find out the multiplicity
of the poles.
Only $d=7$ and $d=8$ turn out to be double poles, other poles
are single.

Because $\fn{J}{d}$ has no poles, and $\fn{\homH^{-1}}{d}$
has neither poles nor zeros on the stripe $(6,8]$, from
eq.~\eqref{eq:omega-sigma} it follows that $\fn{\omega}{d}$
must have the same pole structure as $\fn{\partR}{d}$.

\subsubsection*{The ansatz for the periodic function $\fn{\omega}{d}$}

With this information it is now possible to construct an ansatz
for $\fn{\omega}{d}$.
Note that just as it was for VRRR integrals, applying
eq.~\eqref{eq:gamma-limit} to eq.~\eqref{eq:vrrv8-h} the
asymptotic of $\fn{\homH}{d}$ at $\Im d\to\pm\infty$ can be found
as
\begin{equation}
    \abs{\fn{\homH}{d}} \approx
    \frac{1}{4} \left( \frac{3^6}{5^5} \right)^{\Re \frac{d}{2}} \abs{\Im d}^2.
\end{equation}
With similar arguments as in \autoref{sec:vrrr}, one can also
show that $\abs{\fn{J}{d}}$ and $\abs{\fn{\partR}{d}}$ asymptotically
behave as $\abs{\Im d}^\alpha$, for some $\alpha$, and thus, so
does $\abs{\fn{\omega}{d}}$.
For VRRR integrals $\fn{\omega}{d}$ was free from poles, and
together with this asymptotic that meant that $\fn{\omega}{d}$ was a constant.
The difference with the case of $\vrrv{8}$ is that $\fn{\omega}{d}$
does have poles to compensate for the poles of $\fn{\partR}{d}$.
To deal with them observe that if one is able to subtract from $\fn{\omega}{d}$
an expression that would cancel its poles while simultaneously not
altering the form of its asymptotic behavior, then the resulting
difference will have no poles, and by the same argument could
only be a constant.

For a function that cancels a pole at $d_i$, but does not spoil
the asymptotic behavior at $\Im d\to\pm\infty$, one convenient
choice is a cotangent function,
\begin{equation}
    \label{eq:cotangent}
    \fn{C_{d_i}}{d} \equiv \fn{\cot}{\frac{\pi}{2}\left(d-d_i\right)},
\end{equation}
which in terms of eq.~\eqref{eq:z} is just
\begin{equation}
    \fn{C_{d_i}}{z}=i\frac{z+z_i}{z-z_i},
\end{equation}
one of the simplest functions with a single pole that is constant
at both $z\to0$ and $z\to\infty$.

The whole ansatz can then be constructed as a sum of a constant
term and $\fn{C^n_{d_i}}{d}$ for each pole of multiplicity $n$,
\begin{equation}
    \label{eq:omega-ansatz}
    \fn{\omega}{d} = a_0 + \sum_{i,k} a_{i,k} \, \fn{C_{d_i}^k}{d},
\end{equation}
where $a_{i,k}$ are unknown are constants that need to be determined.
Specifically for $\vrrv{8}$ we might use
\begin{align}
  \label{eq:vrrv8-omega-ansatz}
  \omega(d)  =
    a_0 \Big[ 1& + a_1 \, \fn{C_{\frac{32}{5}}}{d}
                 + a_2 \, \fn{C_{\frac{13}{2}}}{d}
                 + a_3 \, \fn{C_{\frac{20}{3}}}{d}
                 + a_4 \, \fn{C_{\frac{34}{5}}}{d}
                 +\\
               &
                 + a_{5,1} \, \fn{C_{7}}{d}
                 + a_{5,2} \, \fn{C^2_{7}}{d}
                 + a_6 \, \fn{C_{\frac{36}{5}}}{d}
                 + a_7 \, \fn{C_{\frac{22}{3}}}{d}
                 +\nonumber\\
               &
                 + a_8 \, \fn{C_{\frac{15}{2}}}{d}
                 + a_9 \, \fn{C_{\frac{38}{5}}}{d}
                 + a_{10,1} \, \fn{C_{8}}{d}
                 + a_{10,2} \, \fn{C^2_{8}}{d}
                 \Big].\nonumber
\end{align}

\subsubsection*{Fixing the constants in the $\fn{\omega}{d}$ ansatz}

To fix these constants numerically it is enough to insert the
series for $\fn{\partR}{d_i-2\ep}$ obtained via \noun{Dream}
into eq.~\eqref{eq:drr-gen-sol}, and for each $d_i$ from
eq.~\eqref{eq:vrrv8-poles} demand the cancellation of poles,
\begin{equation}
    \label{eq:drr-cancel-poles}
    \lim_{\ep \to 0} \ep\left(\fn{\omega}{d_i-2\ep} \fn{\homH}{d_i-2\ep} + \fn{\partR}{d_i-2\ep}\right) = 0.
\end{equation}
Note that because $\fn{J}{d}$ is finite, it drops out from these
equations.

After $a_i$ have all been fixed numerically with high enough
precision, we can reconstruct their analytic values, obtaining
the following:
\begin{align}
    \label{eq:vrrv8-ansatz-constants}
    &a_0 =\frac{2^6 5^{10} \sqrt{5} \pi^2}{3^{12}},\quad
    a_1 = -a_9 = \frac{3}{\sqrt{2}}\sqrt{1+\frac{1}{\sqrt{5}}},\quad
    -a_2 = a_8 = \frac{45}{8},\\
    &a_3 = -a_7 = \frac{10}{\sqrt{3}},\quad
    -a_4 = a_6 = \frac{3}{\sqrt{2}}\sqrt{1-\frac{1}{\sqrt{5}}},\quad
    a_{5,1} = a_{10,1} = 0,\nonumber\\
    &a_{5,2} = \frac{45}{16},\quad
    a_{10,2} = -\frac{29}{16}.\nonumber
\end{align}
Reconstructing the rational constants here is easy enough;
for the irrational ones we had to resort to using an educated guess
and the Inverse Symbolic Calculator\footnote{\url{http://wayback.cecm.sfu.ca/projects/ISC/ISCmain.html}}.
The result of inserting these values into the ansatz is plotted
on \autoref{fig:drr-I8-W}.

With the constants from eq.~\eqref{eq:vrrv8-omega-ansatz} now fixed,
all that is left is to use \noun{Dream} to calculate the $\ep$-series
for $\fn{J}{4-2\ep}$, and restore them in terms of MZVs.
The results of this calculation for all VRRV integrals up to
weight~6 are listed in \autoref{sec:results-vrrv}, and weight-12
results along with the full expressions in \noun{SummerTime}
format are available in the ancillary files, as described in
\autoref{sec:Ancillary-files}.
All of these results of course match what we have calculated via
direct integration in \autoref{sec:VVRR-VRRV}.

\subsection{Solving DRR for VVRR integrals}

The case of the VVRR integrals brings another complication on
top of what VRRV had: not only the periodic function $\fn{\omega}{d}$,
but also the full normalized integral $\fn{J}{d}$ has poles now.
These are ultraviolet poles coming from subdivergences.
For VRRR and VRRV integrals UV poles only came from surface
divergences, and thus were suppressed by the normalization
factors from eq.~\eqref{eq:normalized-cut-definition}; this is
not the case for VVRR integrals.

Knowing the location and multiplicity of these poles is important
because the ansatz for $\fn{\omega}{d}$ from eq.~\eqref{eq:omega-ansatz}
must now include them too.
To determine these locations one can apply \noun{Fiesta} to the same
parametrization as in \autoref{sec:num-check-VVRR-VRRV}.

The way poles of $\fn{J}{d}$ complicate the calculations is
by preventing the usage of eq.~\eqref{eq:drr-cancel-poles} when
$\fn{J}{d_i}$ is divergent, because $\fn{J}{d_i-2\ep}$ no longer
drops out.
In these cases one needs to find additional sources of information
to fix the corresponding constants in the ansatz.

\subsubsection*{VVRR integrals entering Cutkosky relations alone}

One source of additional information are the Cutkosky
relations from \autoref{sec:Cutkosky-rules}.
Looking closely at relations in eq.~\eqref{eq:ot-1} through
eq.~\eqref{eq:ot-28}, one can see that most of them contain
no more that one VVRR integral.
Only $\vvrr{9}$ with $\vvrr{11}$, and $\vvrr{19}$ with
$\vvrr{20}$ enter in pairs in eqs.~\eqref{eq:ot-23-coupled-VVRR}
and~\eqref{eq:ot-27-coupled-VVRR} respectively.
The rest of VVRR integrals can be directly determined from those
relations, since we already have DRR solutions for all the other
cuts.

Practically speaking, to obtain VVRR integrals in the same
format as the other DRR solution, we only use Cutkosky relations
numerically: to evaluate $\fn{\omega}{d}$ in 1000 points using
eq.~\eqref{eq:omega-sigma} with 100 digits of precision.
From this plot we determine the poles of $\fn{\omega}{d}$,
construct an ansatz in terms of the cotangent functions from
eq.~\eqref{eq:cotangent}, fix its constants numerically via the
\noun{Mathematica} function \texttt{FindFit}, and reconstruct
them analytically (more on this later).

The DRR solutions obtained this way can be double-checked by
inserting their numerical values into the Cutkosky relations at
multiple $d$ with high precision, by inserting their reconstructed
analytical expressions into the same relations, and also by
comparing them with the series at $d=4-2\ep$ calculated in
\autoref{sec:VVRR-VRRV}.
All of these checks hold.

\subsubsection*{VVRR\textsubscript{9} and VVRR\textsubscript{11}}

Integrals $\vvrr{9}$ and $\vvrr{11}$ enter the Cutkosky relation
eq.~\eqref{eq:ot-23-coupled-VVRR} in a pair.
Being different cuts of the same VVVV integral both have identical
diagonal DRR matrix elements, $M_{9,9} = M_{11,11}$, and thus
identical homogeneous solutions.
It might seem like untangling them is impossible, but because
they enter with different prefactors---one being $2B$, and the
other $B^*$---and because each Cutkosky relation can be split
into the real and imaginary parts, this turns out to be enough to
uniquely fix both $\fn{\omega_9}{d}$ and $\fn{\omega_{11}}{d}$
just from eq.~\eqref{eq:ot-23-coupled-VVRR}, and then proceed
as explained above.

\subsubsection*{VVRR\textsubscript{19} and VVRR\textsubscript{20}}

In the case of $\vvrr{19}$ and $\vvrr{20}$ the Cutkosky relation
from eq.~\eqref{eq:ot-27-coupled-VVRR} only constrains the sum
of $\fn{\omega_{19}}{\nu} + \fn{\omega_{20}}{\nu}$.
For these integrals we proceed in the same way as in \autoref{sec:drr-for-vrrv},
using eq.~\eqref{eq:drr-cancel-poles} with $d_i$ for which $\fn{J}{d_i}$
is finite---this is enough to fix most of the constants.
For the rest, we compare the $\ep$-series for $\fn{J}{4-2\ep}$
with the ones calculated in \autoref{sec:VVRR-VRRV}: using the
first few terms of those series is enough to fix the remaining
constants.
The rest of the terms can be used to cross-check the results,
together with eq.~\eqref{eq:ot-27-coupled-VVRR}, which provides
another independent check.

\subsubsection*{Restoring the analytic expressions for the $\omega$ ansatz constants}

Once the constants $a_{i,k}$ in the $\fn{\omega}{d}$ ansatz from
eq.~\eqref{eq:omega-ansatz} are fixed numerically with high
precision, it is desired---even though not strictly speaking
necessary---to reconstruct their analytical forms.

For most integrals an educated guess and the Inverse Symbolic
Calculator do the job, but $\vvrr{17}$, $\vvrr{19}$, and
$\vvrr{20}$ proved to be a challenge.

The basic difficulty is that as we have seen in eq.~\eqref{eq:vrrv8-ansatz-constants}
these constants contain algebraic numbers, and for example an
MZV basis with rational coefficients is insufficient to
reconstruct them.
In the simple cases when $a_{i,k}$ is a product of an algebraic
number and some power of~$\pi$, one can try dividing its value
by consecutive powers of~$\pi$, and reconstructing the rest in
terms of radicals.
If however $a_{i,k}$ is a sum of terms with different $\pi$
powers, or even general MZVs, one would need to split these terms
beforehand.

Fortunately we have experimentally observed that the particular
solutions $\fn{\partR}{d}$ for each of these integrals can in
fact be reconstructed in terms of an MZV basis with rational
coefficients.
Keeping in mind that the full solution $\fn{J}{d}$ should be
representable in terms of MZVs as well, from eq.~\eqref{eq:drr-gen-sol}
it follows that $\fn{\homH}{d} \fn{\omega}{d}$ should be too.
Then, expanding around $d=4-2\ep$,
\begin{equation}
  \label{eq:omegaH-MZV}
    \fn{\homH}{4-2\ep} \fn{\omega}{4-2\ep} = \sum\limits_{k=-4}^{\infty} c_k\,\ep^k,
    \qquad{c_k=\sum_{j_1, j_2, \dots} c_{k,j}\,\zeta_{j_1}\zeta_{j_2}\dots},
\end{equation}
where constants $c_{k,j}$ are rational numbers, and can be
reconstructed analytically via PSLQ.

From here it is possible to evaluate the left hand side numerically
using eq.~\eqref{eq:drr-homogeneous-solution} for $\homH$ and
eq.~\eqref{eq:omega-ansatz} for $\omega$, reconstruct $c_{k,j}$
via PSLQ in the basis of MZVs, then evaluate the left hand side
again symbolically, and finally solve for the constants $a_{i,k}$ from the $\omega$ ansatz.
The radicals we have seen in eq.~\eqref{eq:vrrv8-ansatz-constants}
will appear here as the result of the expansion of $\fn{\homH}{4-2\ep}$
in an $\ep$-series.

The practical complication here is that fixing $\mathcal{O}(20)$ of
$a_{i,k}$ constants this way means working with about the same
number of terms in the $\ep$-series in eq.~\eqref{eq:omegaH-MZV},
and thus with higher and higher transcendentality weights, and
larger and larger MZV bases, all requiring increasingly high
numerical precision.
This quickly becomes computationally expensive.

The trick is to exploit the observation that $a_{i,k}$ seem to
only contain powers of $\pi$, and are free from MZVs otherwise.
With this conjecture in mind, we can evaluate the left hand side
of eq.~\eqref{eq:omegaH-MZV} symbolically, drop all terms
multiplied by MZVs that are not powers of $\pi$, then evaluate
the resulting series numerically, reconstruct them via PSLQ in
the basis of even powers of $\pi$, and finally return to the
symbolical series, solving for $a_{i,k}$ in the form of
\begin{equation}
    a_{i,k}=\sum_p a_{i,k,p} \, \pi^{2p},
\end{equation}
where $a_{i,k,p}$ are algebraic numbers.
Because the basis of $\pi$ powers is much smaller than the general
MZV basis, this procedure becomes computationally tractable.

With all $\fn{\omega}{d}$ ansatz constants finally fixed, we proceed
to evaluate the VVRR master integrals as $\ep$-series around
$d=4-2\ep$ with 4000 digits of precision, and restore the results
in terms of MZVs up to weight~12 using the basis from \autoref{sec:mzv}.
These series truncated after weight~6 are listed in \autoref{sec:results-vvrr}.
The full weight-12 results along with the expressions in
\noun{SummerTime} format are available in the ancillary files,
as described in \autoref{sec:Ancillary-files}.
As a final check, these reconstructed series match what we have
calculated up to weight~7 via direct integration in \autoref{sec:VVRR-VRRV}.

\section{Conclusions}

We have presented the calculation of the previously unknown
master integrals for 3- and 4-particle cuts of massless four-loop
propagators.
Together with the already known 2- and 5-particle cuts this
completes the knowledge of all such cuts.
Both direct integration over the phase space and the solution of
dimensional recurrence relations were used in the calculation,
with the latter finally resulting in expressions that allow the
numerical evaluation of the integrals as $\ep$-series to arbitrary
order with arbitrary precision via the \noun{SummerTime} package.

In \autoref{sec:Results} we have provided analytic expressions
for 3- and 4-particle cut master integrals restored via PSLQ
in the basis of MZVs up to weight~6.
The ancillary files described in \autoref{sec:Ancillary-files}
contain analytic results for all cut structures (including 2-
and 5-particle cuts, as well as the uncut virtual loop integrals)
up to weight~12, as well as the corresponding \noun{SummerTime}
files.
We hope that these results gathered in one place will serve
as a complete reference.

All the results have been cross-checked via numerical integration,
by comparing values obtained via different methods, by comparing
with the known results from the literature, as well as by showing
consistency with Cutkosky rules.

The same methods used here are applicable to the cuts of five-loop
propagators as well, although we expect that calculation to be
harder for two reasons: firstly, solving IBP relations for five-loop
problems is computationally much more challenging; secondly,
five-loop propagators will have multiple master integrals per
sector, so we expect the appearance of coupled blocks in the DRR
equations, as well as the appearance of elliptic integrals in
the amplitudes. Still, this is one viable direction for further
research.

Another direction to go is investigating massive propagators and
their cuts, with the massless ones serving as boundary conditions
for the differential equations.

Finally, our original motivation of calculating semi-inclusive
cut integrals and time-like splitting functions through them
now becomes feasible, with the presented cut integrals allowing us to
fix the integration constants in the differential equations for
the semi-inclusive master integrals.

\subsubsection*{Acknowledgments}

The work of V.M. is supported by the DFG with the grant MO~1801/4-1
within research unit FOR~2926 "Next Generation Perturbative QCD
for Hadron Structure: Preparing for the Electron-Ion Collider".
The work of A.P. is supported by the Foundation for the Advancement
of Theoretical Physics and Mathematics ``BASIS'' and by the RFBR
grant No.~17-02-00872-a.
This article is based upon work from COST Action CA16201
PARTICLEFACE supported by COST (European Cooperation in Science
and Technology).

We would like to thank Sven-Olaf Moch for support and for providing
the original motivation for this work; Oleksandr Gituliar for
initiating this project, and helping during the early stages of
the calculations; and Vsevolod Chestnov for discussions about
the properties of Goncharov polylogarithms.

\appendix

\section{Results}
\label{sec:Results}

Here we provide the values of the master integrals for 3- and 4-particle
cuts of 4-loop propagators as $\ep$-series around $d=4-2\ep$.
The normalization of the integrals is as discussed in
\autoref{sec:introduction}, with prefactors $B$ and
$\mathrm{PS}_n$ defined in eqs.~\eqref{eq:normalization-b} and~\eqref{eq:normalization-psn}
respectively.

For brevity, the series here are truncated after transcendentality weight~6,
which is enough to cover all the $\ep$-finite parts of the integrals.
The full results up to weight 12 are available in the ancillary
files on arXiv, see \autoref{sec:Ancillary-files} for a description
of those.

Note that in all our results MZVs are defined as
\begin{equation}
    \label{eq:mzv}
    \zeta_{a,b,\dots} = \sum_{n_1>n_2>\ldots>0}\frac{{\mathrm{sgn}(a)}^{n_1}}{n_1^{|a|}}\frac{{\mathrm{sgn}(b)}^{n_2}}{n_2^{|b|}}\ldots,
\end{equation}
which is a common ``physicist'' notation adopted by
e.g.~\cite{BBV09,HM07}, but which has the opposite order of
parameters compared to the ``mathematician'' notation used
in~\cite{Goncharov98,Panzer14}.

\subsection{VRRR}
\label{sec:results-vrrr}

\begin{align}
\mathrm{VRRR}_{1} & =\smallfig{vrrr/1}=B^{*}\mathrm{PS}_{4}\left(q^{2}\right)^{2-4\ep}\Big[1+\frac{5}{6}\,\ep+\Big(\frac{143}{36}-2\zeta_{2}\Big)\,\ep^{2}+\Big(\frac{4409}{216}-\\
 & -\frac{5}{3}\zeta_{2}-14\zeta_{3}\Big)\,\ep^{3}+\Big(\frac{136295}{1296}-\frac{143}{18}\zeta_{2}-\frac{35}{3}\zeta_{3}-\frac{142}{5}\zeta_{2}^{2}\Big)\,\ep^{4}+\Big(\frac{4171625}{7776}-\nonumber \\
 & -\frac{4409}{108}\zeta_{2}-\frac{1001}{18}\zeta_{3}-\frac{71}{3}\zeta_{2}^{2}+28\zeta_{2}\zeta_{3}-378\zeta_{5}\Big)\,\ep^{5}+\Big(\frac{126614375}{46656}-\nonumber \\
 & -\frac{136295}{648}\zeta_{2}-\frac{30863}{108}\zeta_{3}-\frac{10153}{90}\zeta_{2}^{2}+\frac{70}{3}\zeta_{2}\zeta_{3}-315\zeta_{5}-\frac{1772}{5}\zeta_{2}^{3}+\nonumber \\
 & +98\zeta_{3}^{2}\Big)\,\ep^{6}+\mathcal{O}\!\left(\ep^{7}\right)\Big]\nonumber \\
\mathrm{VRRR}_{2} & =\smallfig{vrrr/2}=B\,\mathrm{PS}_{4}\left(q^{2}\right)^{1-4\ep}\Big[3+\Big(14-6\zeta_{2}\Big)\,\ep+\Big(\frac{141}{2}+5\zeta_{2}-\\
 & -60\zeta_{3}\Big)\,\ep^{2}+\Big(\frac{1401}{4}-23\zeta_{2}+68\zeta_{3}-141\zeta_{2}^{2}\Big)\,\ep^{3}+\Big(\frac{13901}{8}-129\zeta_{2}-\nonumber \\
 & -146\zeta_{3}+\frac{1733}{10}\zeta_{2}^{2}+84\zeta_{2}\zeta_{3}-1890\zeta_{5}\Big)\,\ep^{4}+\Big(\frac{138177}{16}-\frac{1345}{2}\zeta_{2}-867\zeta_{3}-\nonumber \\
 & -\frac{2801}{10}\zeta_{2}^{2}-70\zeta_{2}\zeta_{3}+2331\zeta_{5}-\frac{65307}{35}\zeta_{2}^{3}+240\zeta_{3}^{2}\Big)\,\ep^{5}+\mathcal{O}\!\left(\ep^{6}\right)\Big]\nonumber \\
\mathrm{VRRR}_{3} & =\smallfig{vrrr/3}=B\,\mathrm{PS}_{4}\left(q^{2}\right)^{-1-4\ep}\Big[3\,\frac{1}{\ep^{4}}-\frac{59}{2}\,\frac{1}{\ep^{3}}+\Big(\frac{203}{2}-12\zeta_{2}\Big)\,\frac{1}{\ep^{2}}+\\
 & +\Big(-144+118\zeta_{2}-108\zeta_{3}\Big)\,\frac{1}{\ep}+\Big(72-406\zeta_{2}+1062\zeta_{3}-246\zeta_{2}^{2}\Big)+\nonumber \\
 & +\Big(576\zeta_{2}-3654\zeta_{3}+2419\zeta_{2}^{2}+216\zeta_{2}\zeta_{3}-3348\zeta_{5}\Big)\,\ep+\Big(-288\zeta_{2}+\nonumber \\
 & +5184\zeta_{3}-8323\zeta_{2}^{2}-2124\zeta_{2}\zeta_{3}+32922\zeta_{5}-\frac{111702}{35}\zeta_{2}^{3}+792\zeta_{3}^{2}\Big)\,\ep^{2}+\nonumber \\
 & +\mathcal{O}\!\left(\ep^{3}\right)\Big]\nonumber \\
\mathrm{VRRR}_{4} & =\smallfig{vrrr/4}=B\,\mathrm{PS}_{4}\left(q^{2}\right)^{2-4\ep}\Big[1+\frac{7}{3}\,\ep+\Big(\frac{193}{18}-4\zeta_{2}\Big)\,\ep^{2}+\Big(\frac{5611}{108}-\\
 & -\frac{28}{3}\zeta_{2}-24\zeta_{3}\Big)\,\ep^{3}+\Big(\frac{166621}{648}-\frac{386}{9}\zeta_{2}-56\zeta_{3}-\frac{192}{5}\zeta_{2}^{2}\Big)\,\ep^{4}+\Big(\frac{4985347}{3888}-\nonumber \\
 & -\frac{5611}{27}\zeta_{2}-\frac{772}{3}\zeta_{3}-\frac{448}{5}\zeta_{2}^{2}+96\zeta_{2}\zeta_{3}-528\zeta_{5}\Big)\,\ep^{5}+\Big(\frac{149515789}{23328}-\nonumber \\
 & -\frac{166621}{162}\zeta_{2}-\frac{11222}{9}\zeta_{3}-\frac{6176}{15}\zeta_{2}^{2}+224\zeta_{2}\zeta_{3}-1232\zeta_{5}-\frac{1824}{5}\zeta_{2}^{3}+\nonumber \\
 & +288\zeta_{3}^{2}\Big)\,\ep^{6}+\mathcal{O}\!\left(\ep^{7}\right)\Big]\nonumber \\
\mathrm{VRRR}_{5} & =\smallfig{vrrr/5}=B\,\mathrm{PS}_{4}\left(q^{2}\right)^{-1-4\ep}\Big[10\,\frac{1}{\ep^{4}}-\frac{445}{3}\,\frac{1}{\ep^{3}}+\Big(980-68\zeta_{2}\Big)\,\frac{1}{\ep^{2}}+\\
 & +\Big(-\frac{12290}{3}+\frac{3026}{3}\zeta_{2}-396\zeta_{3}\Big)\,\frac{1}{\ep}+\Big(13490-6664\zeta_{2}+5874\zeta_{3}-\nonumber \\
 & -\frac{2732}{5}\zeta_{2}^{2}\Big)+\Big(-40950+\frac{83572}{3}\zeta_{2}-38808\zeta_{3}+\frac{121574}{15}\zeta_{2}^{2}+2104\zeta_{2}\zeta_{3}-\nonumber \\
 & -7780\zeta_{5}\Big)\,\ep+\Big(122850-91732\zeta_{2}+162228\zeta_{3}-\frac{267736}{5}\zeta_{2}^{2}-\frac{93628}{3}\zeta_{2}\zeta_{3}+\nonumber \\
 & +\frac{346210}{3}\zeta_{5}-\frac{58376}{15}\zeta_{2}^{3}+6228\zeta_{3}^{2}\Big)\,\ep^{2}+\mathcal{O}\!\left(\ep^{3}\right)\Big]\nonumber \\
\mathrm{VRRR}_{6} & =\smallfig{vrrr/6}=B\,\mathrm{PS}_{4}\left(q^{2}\right)^{-1-4\ep}\Big[9\,\frac{1}{\ep^{4}}-\frac{177}{2}\,\frac{1}{\ep^{3}}+\Big(\frac{609}{2}-36\zeta_{2}\Big)\,\frac{1}{\ep^{2}}+\\
 & +\Big(-432+354\zeta_{2}-204\zeta_{3}\Big)\,\frac{1}{\ep}+\Big(216-1218\zeta_{2}+2006\zeta_{3}-\frac{1512}{5}\zeta_{2}^{2}\Big)+\nonumber \\
 & +\Big(1728\zeta_{2}-6902\zeta_{3}+\frac{14868}{5}\zeta_{2}^{2}+816\zeta_{2}\zeta_{3}-4116\zeta_{5}\Big)\,\ep+\Big(-864\zeta_{2}+\nonumber \\
 & +9792\zeta_{3}-\frac{51156}{5}\zeta_{2}^{2}-8024\zeta_{2}\zeta_{3}+40474\zeta_{5}-\frac{19296}{7}\zeta_{2}^{3}+2376\zeta_{3}^{2}\Big)\,\ep^{2}+\nonumber \\
 & +\mathcal{O}\!\left(\ep^{3}\right)\Big]\nonumber \\
\mathrm{VRRR}_{7} & =\smallfig{vrrr/7}=B^{*}\mathrm{PS}_{4}\left(q^{2}\right)^{-4\ep}\Big[\Big(-12+12\zeta_{2}\Big)+\Big(-74-34\zeta_{2}+\\
 & +132\zeta_{3}\Big)\,\ep+\Big(-450+72\zeta_{2}-374\zeta_{3}+\frac{1608}{5}\zeta_{2}^{2}\Big)\,\ep^{2}+\Big(-2570+296\zeta_{2}+\nonumber \\
 & +552\zeta_{3}-\frac{4556}{5}\zeta_{2}^{2}-432\zeta_{2}\zeta_{3}+4836\zeta_{5}\Big)\,\ep^{3}+\Big(-14130+1800\zeta_{2}+\nonumber \\
 & +1776\zeta_{3}+1104\zeta_{2}^{2}+1224\zeta_{2}\zeta_{3}-13702\zeta_{5}+\frac{158912}{35}\zeta_{2}^{3}-1056\zeta_{3}^{2}\Big)\,\ep^{4}+\nonumber \\
 & +\mathcal{O}\!\left(\ep^{5}\right)\Big]\nonumber \\
\mathrm{VRRR}_{8} & =\smallfig{vrrr/8}=B^{*}\mathrm{PS}_{4}\left(q^{2}\right)^{-1-4\ep}\Big[6\zeta_{2}\,\frac{1}{\ep^{2}}+\Big(-59\zeta_{2}+72\zeta_{3}\Big)\,\frac{1}{\ep}+\\
 & +\Big(203\zeta_{2}-708\zeta_{3}+\frac{927}{5}\zeta_{2}^{2}\Big)+\Big(-288\zeta_{2}+2436\zeta_{3}-\frac{18231}{10}\zeta_{2}^{2}-\nonumber \\
 & -264\zeta_{2}\zeta_{3}+2862\zeta_{5}\Big)\,\ep+\Big(144\zeta_{2}-3456\zeta_{3}+\frac{62727}{10}\zeta_{2}^{2}+2596\zeta_{2}\zeta_{3}-\nonumber \\
 & -28143\zeta_{5}+\frac{94067}{35}\zeta_{2}^{3}-768\zeta_{3}^{2}\Big)\,\ep^{2}+\mathcal{O}\!\left(\ep^{3}\right)\Big]\nonumber \\
\mathrm{VRRR}_{9} & =\smallfig{vrrr/9}=B\,\mathrm{PS}_{4}\left(q^{2}\right)^{-4\ep}\Big[\Big(-12+12\zeta_{2}\Big)+\Big(-62-46\zeta_{2}+\\
 & +144\zeta_{3}\Big)\,\ep+\Big(-352+82\zeta_{2}-552\zeta_{3}+\frac{1872}{5}\zeta_{2}^{2}\Big)\,\ep^{2}+\Big(-1924+244\zeta_{2}+\nonumber \\
 & +696\zeta_{3}-\frac{7176}{5}\zeta_{2}^{2}-432\zeta_{2}\zeta_{3}+5688\zeta_{5}\Big)\,\ep^{3}+\Big(-10268+1400\zeta_{2}+\nonumber \\
 & +1440\zeta_{3}+\frac{7608}{5}\zeta_{2}^{2}+1656\zeta_{2}\zeta_{3}-21804\zeta_{5}+5472\zeta_{2}^{3}-864\zeta_{3}^{2}\Big)\,\ep^{4}+\nonumber \\
 & +\mathcal{O}\!\left(\ep^{5}\right)\Big]\nonumber \\
\mathrm{VRRR}_{10} & =\smallfig{vrrr/10}=B\,\mathrm{PS}_{4}\left(q^{2}\right)^{-4\ep}\Big[-12\zeta_{3}+\Big(34\zeta_{3}-\frac{114}{5}\zeta_{2}^{2}\Big)\,\ep+\Big(-96\zeta_{3}+\\
 & +\frac{323}{5}\zeta_{2}^{2}+168\zeta_{2}\zeta_{3}-432\zeta_{5}\Big)\,\ep^{2}+\Big(-156\zeta_{3}-\frac{912}{5}\zeta_{2}^{2}-476\zeta_{2}\zeta_{3}+\nonumber \\
 & +1224\zeta_{5}-\frac{4062}{35}\zeta_{2}^{3}+480\zeta_{3}^{2}\Big)\,\ep^{3}+\mathcal{O}\!\left(\ep^{4}\right)\Big]\nonumber \\
\mathrm{VRRR}_{11} & =\smallfig{vrrr/11}=B^{*}\mathrm{PS}_{4}\left(q^{2}\right)^{-2-4\ep}\Big[2\,\frac{1}{\ep^{4}}-\frac{89}{3}\,\frac{1}{\ep^{3}}+\Big(196+20\zeta_{2}\Big)\,\frac{1}{\ep^{2}}+\\
 & +\Big(-\frac{2458}{3}-\frac{890}{3}\zeta_{2}+316\zeta_{3}\Big)\,\frac{1}{\ep}+\Big(2698+1960\zeta_{2}-\frac{14062}{3}\zeta_{3}+\nonumber \\
 & +\frac{4348}{5}\zeta_{2}^{2}\Big)+\Big(-8190-\frac{24580}{3}\zeta_{2}+30968\zeta_{3}-\frac{193486}{15}\zeta_{2}^{2}-1848\zeta_{2}\zeta_{3}+\nonumber \\
 & +15028\zeta_{5}\Big)\,\ep+\Big(24570+26980\zeta_{2}-\frac{388364}{3}\zeta_{3}+\frac{426104}{5}\zeta_{2}^{2}+27412\zeta_{2}\zeta_{3}-\nonumber \\
 & -\frac{668746}{3}\zeta_{5}+\frac{498824}{35}\zeta_{2}^{3}-5380\zeta_{3}^{2}\Big)\,\ep^{2}+\mathcal{O}\!\left(\ep^{3}\right)\Big]\nonumber \\
\mathrm{VRRR}_{12} & =\smallfig{vrrr/12}=B\,\mathrm{PS}_{4}\left(q^{2}\right)^{-4\ep}\Big[24\zeta_{3}\,\frac{1}{\ep}+\Big(-164\zeta_{3}+\frac{372}{5}\zeta_{2}^{2}\Big)+\Big(464\zeta_{3}-\\
 & -\frac{2542}{5}\zeta_{2}^{2}-264\zeta_{2}\zeta_{3}+1368\zeta_{5}\Big)\,\ep+\Big(-456\zeta_{3}+\frac{7192}{5}\zeta_{2}^{2}+1804\zeta_{2}\zeta_{3}-\nonumber \\
 & -9348\zeta_{5}+\frac{33024}{35}\zeta_{2}^{3}-780\zeta_{3}^{2}\Big)\,\ep^{2}+\mathcal{O}\!\left(\ep^{3}\right)\Big]\nonumber \\
\mathrm{VRRR}_{13} & =\smallfig{vrrr/13}=B\,\mathrm{PS}_{4}\left(q^{2}\right)^{-3-4\ep}\Big[-3\,\frac{1}{\ep^{5}}+\frac{191}{6}\,\frac{1}{\ep^{4}}+\Big(-\frac{73}{9}+\\
 & +42\zeta_{2}\Big)\,\frac{1}{\ep^{3}}+\Big(-1704-\frac{1285}{3}\zeta_{2}+\frac{1394}{3}\zeta_{3}\Big)\,\frac{1}{\ep^{2}}+\Big(\frac{132922}{9}+\frac{6268}{9}\zeta_{2}-\nonumber \\
 & -\frac{42197}{9}\zeta_{3}+\frac{6338}{5}\zeta_{2}^{2}\Big)\,\frac{1}{\ep}+\Big(-\frac{685640}{9}+\frac{33116}{3}\zeta_{2}+10222\zeta_{3}-\nonumber \\
 & -\frac{189191}{15}\zeta_{2}^{2}-\frac{7124}{3}\zeta_{2}\zeta_{3}+\frac{66670}{3}\zeta_{5}\Big)+\Big(\frac{2702030}{9}-\frac{870556}{9}\zeta_{2}+\nonumber \\
 & +\frac{637816}{9}\zeta_{3}+\frac{88733}{3}\zeta_{2}^{2}+\frac{209378}{9}\zeta_{2}\zeta_{3}-\frac{1962631}{9}\zeta_{5}+\frac{2268184}{105}\zeta_{2}^{3}-\nonumber \\
 & -4818\zeta_{3}^{2}\Big)\,\ep+\mathcal{O}\!\left(\ep^{2}\right)\Big]\nonumber \\
\mathrm{VRRR}_{14} & =\smallfig{vrrr/14}=B^{*}\mathrm{PS}_{4}\left(q^{2}\right)^{-3-4\ep}\Big[\frac{16}{3}\,\frac{1}{\ep^{5}}-\frac{238}{9}\,\frac{1}{\ep^{4}}+\Big(-\frac{2327}{9}-\\
 & -\frac{196}{3}\zeta_{2}\Big)\,\frac{1}{\ep^{3}}+\Big(\frac{26788}{9}+\frac{4378}{9}\zeta_{2}-600\zeta_{3}\Big)\,\frac{1}{\ep^{2}}+\Big(-\frac{129430}{9}+\frac{6812}{9}\zeta_{2}+\nonumber \\
 & +5080\zeta_{3}-1334\zeta_{2}^{2}\Big)\,\frac{1}{\ep}+\Big(\frac{147622}{3}-\frac{184828}{9}\zeta_{2}-\frac{6410}{3}\zeta_{3}+\frac{36509}{3}\zeta_{2}^{2}+\nonumber \\
 & +3592\zeta_{2}\zeta_{3}-22188\zeta_{5}\Big)+\Big(-150150+\frac{986380}{9}\zeta_{2}-128560\zeta_{3}-\nonumber \\
 & -\frac{53195}{3}\zeta_{2}^{2}-\frac{95476}{3}\zeta_{2}\zeta_{3}+207498\zeta_{5}-\frac{624186}{35}\zeta_{2}^{3}+10044\zeta_{3}^{2}\Big)\,\ep+\mathcal{O}\!\left(\ep^{2}\right)\Big]\nonumber \\
\mathrm{VRRR}_{15} & =\smallfig{vrrr/15}=B\,\mathrm{PS}_{4}\left(q^{2}\right)^{-3-4\ep}\Big[14\,\frac{1}{\ep^{4}}-\frac{427}{3}\,\frac{1}{\ep^{3}}+\Big(\frac{1838}{9}-\\
 & -\frac{380}{3}\zeta_{2}\Big)\,\frac{1}{\ep^{2}}+\Big(\frac{37964}{9}+\frac{12998}{9}\zeta_{2}-\frac{2924}{3}\zeta_{3}\Big)\,\frac{1}{\ep}+\Big(-\frac{343292}{9}-\nonumber \\
 & -\frac{41716}{9}\zeta_{2}+\frac{104942}{9}\zeta_{3}-\frac{5872}{3}\zeta_{2}^{2}\Big)+\Big(\frac{1837780}{9}-\frac{130648}{9}\zeta_{2}-\nonumber \\
 & -\frac{410540}{9}\zeta_{3}+\frac{1094576}{45}\zeta_{2}^{2}+\frac{16304}{3}\zeta_{2}\zeta_{3}-\frac{95596}{3}\zeta_{5}\Big)\,\ep+\Big(-890644+\nonumber \\
 & +\frac{1894120}{9}\zeta_{2}-\frac{223568}{9}\zeta_{3}-\frac{972364}{9}\zeta_{2}^{2}-\frac{578600}{9}\zeta_{2}\zeta_{3}+\frac{3582094}{9}\zeta_{5}-\nonumber \\
 & -\frac{2594504}{105}\zeta_{2}^{3}+13656\zeta_{3}^{2}\Big)\,\ep^{2}+\mathcal{O}\!\left(\ep^{3}\right)\Big]\nonumber \\
\mathrm{VRRR}_{16} & =\smallfig{vrrr/16}=B^{*}\mathrm{PS}_{4}\left(q^{2}\right)^{2-4\ep}\\
\mathrm{VRRR}_{17} & =\smallfig{vrrr/17}=B^{*}\mathrm{PS}_{4}\left(q^{2}\right)^{-1-4\ep}\Big[48\zeta_{3}\,\frac{1}{\ep}+\Big(-472\zeta_{3}+\frac{984}{5}\zeta_{2}^{2}\Big)+\\
 & +\Big(1624\zeta_{3}-\frac{9676}{5}\zeta_{2}^{2}+192\zeta_{2}\zeta_{3}+2712\zeta_{5}\Big)\,\ep+\Big(-2304\zeta_{3}+\frac{33292}{5}\zeta_{2}^{2}-\nonumber \\
 & -1888\zeta_{2}\zeta_{3}-26668\zeta_{5}+\frac{124784}{35}\zeta_{2}^{3}+192\zeta_{3}^{2}\Big)\,\ep^{2}+\mathcal{O}\!\left(\ep^{3}\right)\Big]\nonumber \\
\mathrm{VRRR}_{18} & =\smallfig{vrrr/18}=B^{*}\mathrm{PS}_{4}\left(q^{2}\right)^{-1-4\ep}\Big[-\frac{96}{5}\zeta_{2}^{2}+\Big(112\zeta_{2}^{2}+144\zeta_{2}\zeta_{3}-\\
 & -696\zeta_{5}\Big)\,\ep+\Big(-\frac{1776}{5}\zeta_{2}^{2}-840\zeta_{2}\zeta_{3}+4060\zeta_{5}-\frac{26672}{35}\zeta_{2}^{3}+1008\zeta_{3}^{2}\Big)\,\ep^{2}+\nonumber \\
 & +\mathcal{O}\!\left(\ep^{3}\right)\Big]\nonumber \\
\mathrm{VRRR}_{19} & =\smallfig{vrrr/19}=B\,\mathrm{PS}_{4}\left(q^{2}\right)^{-4\ep}\Big[\Big(-12+12\zeta_{2}\Big)+\Big(-26-58\zeta_{2}+\\
 & +108\zeta_{3}\Big)\,\ep+\Big(-82+44\zeta_{2}-522\zeta_{3}+\frac{1212}{5}\zeta_{2}^{2}\Big)\,\ep^{2}+\Big(-254-8\zeta_{2}+\nonumber \\
 & +396\zeta_{3}-\frac{5858}{5}\zeta_{2}^{2}+216\zeta_{2}\zeta_{3}+2484\zeta_{5}\Big)\,\ep^{3}+\Big(-778-16\zeta_{2}-72\zeta_{3}+\nonumber \\
 & +\frac{4444}{5}\zeta_{2}^{2}-1044\zeta_{2}\zeta_{3}-12006\zeta_{5}+\frac{13128}{5}\zeta_{2}^{3}+972\zeta_{3}^{2}\Big)\,\ep^{4}+\mathcal{O}\!\left(\ep^{5}\right)\Big]\nonumber \\
\mathrm{VRRR}_{20} & =\smallfig{vrrr/20}=B\,\mathrm{PS}_{4}\left(q^{2}\right)^{-3-4\ep}\Big[\frac{50}{3}\,\frac{1}{\ep^{4}}-\frac{1325}{9}\,\frac{1}{\ep^{3}}+\Big(-\frac{482}{3}-\\
 & -\frac{440}{3}\zeta_{2}\Big)\,\frac{1}{\ep^{2}}+\Big(\frac{76900}{9}+\frac{17420}{9}\zeta_{2}-\frac{4040}{3}\zeta_{3}\Big)\,\frac{1}{\ep}+\Big(-66216-\nonumber \\
 & -\frac{29960}{3}\zeta_{2}+\frac{154100}{9}\zeta_{3}-3488\zeta_{2}^{2}\Big)+\Big(\frac{3077692}{9}+\frac{193520}{9}\zeta_{2}-\nonumber \\
 & -\frac{732680}{9}\zeta_{3}+\frac{129920}{3}\zeta_{2}^{2}-\frac{640}{3}\zeta_{2}\zeta_{3}-\frac{138040}{3}\zeta_{5}\Big)\,\ep+\Big(-1477440+\nonumber \\
 & +\frac{97120}{3}\zeta_{2}+\frac{1134640}{9}\zeta_{3}-\frac{583096}{3}\zeta_{2}^{2}+\frac{160}{9}\zeta_{2}\zeta_{3}+\frac{5131660}{9}\zeta_{5}-\nonumber \\
 & -\frac{1085488}{21}\zeta_{2}^{3}+2720\zeta_{3}^{2}\Big)\,\ep^{2}+\mathcal{O}\!\left(\ep^{3}\right)\Big]\nonumber \\
\mathrm{VRRR}_{21} & =\smallfig{vrrr/21}=B\,\mathrm{PS}_{4}\left(q^{2}\right)^{-1-4\ep}\Big[\frac{42}{5}\zeta_{2}^{2}\,\frac{1}{\ep}+\Big(-\frac{497}{5}\zeta_{2}^{2}-168\zeta_{2}\zeta_{3}+\\
 & +552\zeta_{5}\Big)+\Big(\frac{2247}{5}\zeta_{2}^{2}+1988\zeta_{2}\zeta_{3}-6532\zeta_{5}+\frac{21874}{35}\zeta_{2}^{3}-768\zeta_{3}^{2}\Big)\,\ep+\nonumber \\
 & +\mathcal{O}\!\left(\ep^{2}\right)\Big]\nonumber \\
\mathrm{VRRR}_{22} & =\smallfig{vrrr/22}=B\,\mathrm{PS}_{4}\left(q^{2}\right)^{-3-4\ep}\Big[\frac{16}{3}\,\frac{1}{\ep^{5}}-\frac{298}{9}\,\frac{1}{\ep^{4}}+\Big(-151-\\
 & -\frac{160}{3}\zeta_{2}\Big)\,\frac{1}{\ep^{3}}+\Big(\frac{19439}{9}+\frac{3376}{9}\zeta_{2}-592\zeta_{3}\Big)\,\frac{1}{\ep^{2}}+\Big(-\frac{91330}{9}+\frac{2008}{3}\zeta_{2}+\nonumber \\
 & +\frac{12928}{3}\zeta_{3}-\frac{7648}{5}\zeta_{2}^{2}\Big)\,\frac{1}{\ep}+\Big(\frac{91964}{3}-\frac{123716}{9}\zeta_{2}+\frac{17420}{3}\zeta_{3}+\frac{170392}{15}\zeta_{2}^{2}+\nonumber \\
 & +2112\zeta_{2}\zeta_{3}-23496\zeta_{5}\Big)+\Big(-\frac{218764}{3}+\frac{457792}{9}\zeta_{2}-148424\zeta_{3}+\nonumber \\
 & +\frac{186116}{15}\zeta_{2}^{2}-17440\zeta_{2}\zeta_{3}+179108\zeta_{5}-\frac{828648}{35}\zeta_{2}^{3}+11432\zeta_{3}^{2}\Big)\,\ep+\mathcal{O}\!\left(\ep^{2}\right)\Big]\nonumber \\
\mathrm{VRRR}_{23} & =\smallfig{vrrr/23}=B\,\mathrm{PS}_{4}\left(q^{2}\right)^{-4\ep}\Big[24\zeta_{3}\,\ep+\Big(-68\zeta_{3}+\frac{312}{5}\zeta_{2}^{2}\Big)\,\ep^{2}+\Big(48\zeta_{3}-\\
 & -\frac{884}{5}\zeta_{2}^{2}-432\zeta_{2}\zeta_{3}+1488\zeta_{5}\Big)\,\ep^{3}+\Big(\frac{624}{5}\zeta_{2}^{2}+1224\zeta_{2}\zeta_{3}-4216\zeta_{5}+\nonumber \\
 & +\frac{32864}{35}\zeta_{2}^{3}-1200\zeta_{3}^{2}\Big)\,\ep^{4}+\mathcal{O}\!\left(\ep^{5}\right)\Big]\nonumber \\
\mathrm{VRRR}_{24} & =\smallfig{vrrr/24}=B\,\mathrm{PS}_{4}\left(q^{2}\right)^{-1-4\ep}\Big[-6\zeta_{2}\,\frac{1}{\ep^{2}}+\Big(59\zeta_{2}-60\zeta_{3}\Big)\,\frac{1}{\ep}+\\
 & +\Big(-203\zeta_{2}+590\zeta_{3}-129\zeta_{2}^{2}\Big)+\Big(288\zeta_{2}-2030\zeta_{3}+\frac{2537}{2}\zeta_{2}^{2}+192\zeta_{2}\zeta_{3}-\nonumber \\
 & -1806\zeta_{5}\Big)\,\ep+\Big(-144\zeta_{2}+2880\zeta_{3}-\frac{8729}{2}\zeta_{2}^{2}-1888\zeta_{2}\zeta_{3}+17759\zeta_{5}-\nonumber \\
 & -\frac{58089}{35}\zeta_{2}^{3}+828\zeta_{3}^{2}\Big)\,\ep^{2}+\mathcal{O}\!\left(\ep^{3}\right)\Big]\nonumber \\
\mathrm{VRRR}_{25} & =\smallfig{vrrr/25}=B^{*}\mathrm{PS}_{4}\left(q^{2}\right)^{-4\ep}\Big[-48\zeta_{3}+\Big(136\zeta_{3}-96\zeta_{2}^{2}\Big)\,\ep+\\
 & +\Big(-384\zeta_{3}+272\zeta_{2}^{2}+672\zeta_{2}\zeta_{3}-1872\zeta_{5}\Big)\,\ep^{2}+\Big(-624\zeta_{3}-768\zeta_{2}^{2}-\nonumber \\
 & -1904\zeta_{2}\zeta_{3}+5304\zeta_{5}-\frac{5568}{7}\zeta_{2}^{3}+2400\zeta_{3}^{2}\Big)\,\ep^{3}+\mathcal{O}\!\left(\ep^{4}\right)\Big]\nonumber \\
\mathrm{VRRR}_{26} & =\smallfig{vrrr/26}=B\,\mathrm{PS}_{4}\left(q^{2}\right)^{-3-4\ep}\Big[\frac{7}{3}\,\frac{1}{\ep^{5}}+\frac{217}{18}\,\frac{1}{\ep^{4}}+\Big(-\frac{3014}{9}-\\
 & -30\zeta_{2}\Big)\,\frac{1}{\ep^{3}}+\Big(\frac{12386}{9}+\frac{163}{3}\zeta_{2}-\frac{422}{3}\zeta_{3}\Big)\,\frac{1}{\ep^{2}}+\Big(\frac{21226}{9}+\frac{18530}{9}\zeta_{2}+\nonumber \\
 & +\frac{635}{9}\zeta_{3}+\frac{214}{5}\zeta_{2}^{2}\Big)\,\frac{1}{\ep}+\Big(-\frac{135160}{3}-\frac{36280}{3}\zeta_{2}+\frac{36016}{3}\zeta_{3}-\frac{5801}{3}\zeta_{2}^{2}+\nonumber \\
 & +\frac{4612}{3}\zeta_{2}\zeta_{3}+\frac{5174}{3}\zeta_{5}\Big)+\Big(\frac{742994}{3}+\frac{48452}{9}\zeta_{2}-\frac{612964}{9}\zeta_{3}+\frac{281747}{15}\zeta_{2}^{2}-\nonumber \\
 & -\frac{61234}{9}\zeta_{2}\zeta_{3}-\frac{343535}{9}\zeta_{5}+\frac{2138552}{315}\zeta_{2}^{3}+\frac{15190}{3}\zeta_{3}^{2}\Big)\,\ep+\mathcal{O}\!\left(\ep^{2}\right)\Big]\nonumber \\
\mathrm{VRRR}_{27} & =\smallfig{vrrr/27}=B\,\mathrm{PS}_{4}\left(q^{2}\right)^{-2-4\ep}\Big[2\,\frac{1}{\ep^{4}}-\frac{89}{3}\,\frac{1}{\ep^{3}}+\Big(196+24\zeta_{2}\Big)\,\frac{1}{\ep^{2}}+\\
 & +\Big(-\frac{2458}{3}-356\zeta_{2}+280\zeta_{3}\Big)\,\frac{1}{\ep}+\Big(2698+2352\zeta_{2}-\frac{12460}{3}\zeta_{3}+\nonumber \\
 & +\frac{3292}{5}\zeta_{2}^{2}\Big)+\Big(-8190-9832\zeta_{2}+27440\zeta_{3}-\frac{146494}{15}\zeta_{2}^{2}-600\zeta_{2}\zeta_{3}+\nonumber \\
 & +9240\zeta_{5}\Big)\,\ep+\Big(24570+32376\zeta_{2}-\frac{344120}{3}\zeta_{3}+\frac{322616}{5}\zeta_{2}^{2}+8900\zeta_{2}\zeta_{3}-\nonumber \\
 & -137060\zeta_{5}+\frac{966272}{105}\zeta_{2}^{3}-1620\zeta_{3}^{2}\Big)\,\ep^{2}+\mathcal{O}\!\left(\ep^{3}\right)\Big]\nonumber \\
\mathrm{VRRR}_{28} & =\smallfig{vrrr/28}=B\,\mathrm{PS}_{4}\left(q^{2}\right)^{-3-4\ep}\Big[\frac{58}{3}\,\frac{1}{\ep^{5}}-\frac{1633}{9}\,\frac{1}{\ep^{4}}+\Big(\frac{2990}{9}-\\
 & -\frac{340}{3}\zeta_{2}\Big)\,\frac{1}{\ep^{3}}+\Big(\frac{21622}{9}+\frac{9838}{9}\zeta_{2}-480\zeta_{3}\Big)\,\frac{1}{\ep^{2}}+\Big(-\frac{153638}{9}-\frac{7010}{3}\zeta_{2}+\nonumber \\
 & +\frac{15136}{3}\zeta_{3}-\frac{1294}{5}\zeta_{2}^{2}\Big)\,\frac{1}{\ep}+\Big(\frac{188774}{3}-\frac{108892}{9}\zeta_{2}-\frac{139048}{9}\zeta_{3}+\nonumber \\
 & +\frac{61961}{15}\zeta_{2}^{2}+3512\zeta_{2}\zeta_{3}-3908\zeta_{5}\Big)+\Big(-193830+\frac{288952}{3}\zeta_{2}-\frac{66136}{3}\zeta_{3}-\nonumber \\
 & -\frac{1228993}{45}\zeta_{2}^{2}-\frac{96748}{3}\zeta_{2}\zeta_{3}+\frac{179942}{3}\zeta_{5}+\frac{126054}{35}\zeta_{2}^{3}+8276\zeta_{3}^{2}\Big)\,\ep+\nonumber \\
 & +\mathcal{O}\!\left(\ep^{2}\right)\Big]\nonumber \\
\mathrm{VRRR}_{29} & =\smallfig{vrrr/29}=B\,\mathrm{PS}_{4}\left(q^{2}\right)^{-3-4\ep}\Big[\frac{26}{3}\,\frac{1}{\ep^{5}}-\frac{443}{9}\,\frac{1}{\ep^{4}}+\Big(-\frac{2893}{9}-\\
 & -96\zeta_{2}\Big)\,\frac{1}{\ep^{3}}+\Big(\frac{37001}{9}+\frac{2320}{3}\zeta_{2}-\frac{2584}{3}\zeta_{3}\Big)\,\frac{1}{\ep^{2}}+\Big(-\frac{177920}{9}+\frac{680}{9}\zeta_{2}+\nonumber \\
 & +\frac{65836}{9}\zeta_{3}-\frac{28048}{15}\zeta_{2}^{2}\Big)\,\frac{1}{\ep}+\Big(\frac{194738}{3}-\frac{63820}{3}\zeta_{2}-\frac{43136}{9}\zeta_{3}+\nonumber \\
 & +\frac{732388}{45}\zeta_{2}^{2}+\frac{10928}{3}\zeta_{2}\zeta_{3}-28032\zeta_{5}\Big)+\Big(-183834+\frac{971392}{9}\zeta_{2}-\nonumber \\
 & -\frac{1405216}{9}\zeta_{3}-\frac{732362}{45}\zeta_{2}^{2}-\frac{293168}{9}\zeta_{2}\zeta_{3}+248032\zeta_{5}-\frac{2545792}{105}\zeta_{2}^{3}+\nonumber \\
 & +\frac{39608}{3}\zeta_{3}^{2}\Big)\,\ep+\mathcal{O}\!\left(\ep^{2}\right)\Big]\nonumber \\
\mathrm{VRRR}_{30} & =\smallfig{vrrr/30}=B\,\mathrm{PS}_{4}\left(q^{2}\right)^{-1-4\ep}\Big[72\zeta_{3}\,\frac{1}{\ep}+\Big(-420\zeta_{3}+\frac{456}{5}\zeta_{2}^{2}\Big)+\\
 & +\Big(1332\zeta_{3}-532\zeta_{2}^{2}-1152\zeta_{2}\zeta_{3}+1824\zeta_{5}\Big)\,\ep+\Big(-336\zeta_{3}+\frac{8436}{5}\zeta_{2}^{2}+\nonumber \\
 & +6720\zeta_{2}\zeta_{3}-10640\zeta_{5}-\frac{2656}{35}\zeta_{2}^{3}-4752\zeta_{3}^{2}\Big)\,\ep^{2}+\mathcal{O}\!\left(\ep^{3}\right)\Big]\nonumber \\
\mathrm{VRRR}_{31} & =\smallfig{vrrr/31}=B\,\mathrm{PS}_{4}\left(q^{2}\right)^{-3-4\ep}\Big[\frac{38}{3}\,\frac{1}{\ep^{5}}-\frac{239}{9}\,\frac{1}{\ep^{4}}+\Big(-\frac{7270}{9}-\\
 & -124\zeta_{2}\Big)\,\frac{1}{\ep^{3}}+\Big(\frac{41438}{9}+\frac{1630}{3}\zeta_{2}-564\zeta_{3}\Big)\,\frac{1}{\ep^{2}}+\Big(\frac{1678}{3}+\frac{14984}{3}\zeta_{2}+\nonumber \\
 & +\frac{5530}{3}\zeta_{3}-\frac{172}{15}\zeta_{2}^{2}\Big)\,\frac{1}{\ep}+\Big(-\frac{933718}{9}-\frac{358712}{9}\zeta_{2}+\frac{278552}{9}\zeta_{3}-\nonumber \\
 & -\frac{38146}{9}\zeta_{2}^{2}+\frac{16952}{3}\zeta_{2}\zeta_{3}+\frac{4300}{3}\zeta_{5}\Big)+\Big(\frac{5766122}{9}+\frac{575704}{9}\zeta_{2}-\nonumber \\
 & -\frac{2007832}{9}\zeta_{3}+\frac{2367992}{45}\zeta_{2}^{2}-\frac{299420}{9}\zeta_{2}\zeta_{3}-\frac{678790}{9}\zeta_{5}+\frac{82072}{5}\zeta_{2}^{3}+\nonumber \\
 & +\frac{52276}{3}\zeta_{3}^{2}\Big)\,\ep+\mathcal{O}\!\left(\ep^{2}\right)\Big]\nonumber \\
\mathrm{VRRR}_{32} & =\smallfig{vrrr/32}=B\,\mathrm{PS}_{4}\left(q^{2}\right)^{-3-4\ep}\Big[36\,\frac{1}{\ep^{5}}-\frac{916}{3}\,\frac{1}{\ep^{4}}+\Big(\frac{1339}{9}-\\
 & -208\zeta_{2}\Big)\,\frac{1}{\ep^{3}}+\Big(\frac{66805}{9}+1920\zeta_{2}-1368\zeta_{3}\Big)\,\frac{1}{\ep^{2}}+\Big(-\frac{386258}{9}-\nonumber \\
 & -\frac{29780}{9}\zeta_{2}+\frac{38620}{3}\zeta_{3}-\frac{12688}{5}\zeta_{2}^{2}\Big)\,\frac{1}{\ep}+\Big(147080-\frac{225220}{9}\zeta_{2}-\nonumber \\
 & -\frac{223600}{9}\zeta_{3}+\frac{72232}{3}\zeta_{2}^{2}+3728\zeta_{2}\zeta_{3}-34640\zeta_{5}\Big)+\Big(-\frac{1263472}{3}+\nonumber \\
 & +\frac{480872}{3}\zeta_{2}-\frac{453260}{3}\zeta_{3}-\frac{439760}{9}\zeta_{2}^{2}-\frac{106360}{3}\zeta_{2}\zeta_{3}+\frac{985760}{3}\zeta_{5}-\nonumber \\
 & -\frac{3208376}{105}\zeta_{2}^{3}+11624\zeta_{3}^{2}\Big)\,\ep+\mathcal{O}\!\left(\ep^{2}\right)\Big]\nonumber \\
\mathrm{VRRR}_{33} & =\smallfig{vrrr/33}=B\,\mathrm{PS}_{4}\left(q^{2}\right)^{-3-4\ep}\Big[18\,\frac{1}{\ep^{4}}-175\,\frac{1}{\ep^{3}}+\Big(\frac{370}{3}-\frac{488}{3}\zeta_{2}\Big)\,\frac{1}{\ep^{2}}+\\
 & +\Big(6564+\frac{17564}{9}\zeta_{2}-\frac{3952}{3}\zeta_{3}\Big)\,\frac{1}{\ep}+\Big(-\frac{165664}{3}-\frac{69428}{9}\zeta_{2}+\frac{145456}{9}\zeta_{3}-\nonumber \\
 & -2900\zeta_{2}^{2}\Big)+\Big(288908-\frac{29276}{9}\zeta_{2}-\frac{619612}{9}\zeta_{3}+\frac{540178}{15}\zeta_{2}^{2}+\frac{10856}{3}\zeta_{2}\zeta_{3}-\nonumber \\
 & -\frac{125008}{3}\zeta_{5}\Big)\,\ep+\Big(-1243464+\frac{1613084}{9}\zeta_{2}+\frac{250196}{9}\zeta_{3}-\frac{2389216}{15}\zeta_{2}^{2}-\nonumber \\
 & -\frac{416660}{9}\zeta_{2}\zeta_{3}+\frac{4705888}{9}\zeta_{5}-\frac{2463464}{63}\zeta_{2}^{3}+\frac{32948}{3}\zeta_{3}^{2}\Big)\,\ep^{2}+\mathcal{O}\!\left(\ep^{3}\right)\Big]\nonumber\\
\mathrm{VRRR}_{34} & =\smallfig{vrrr/34}=B^*\,\mathrm{PS}_{4}\left(q^{2}\right)^{-2-4\ep}\Big[60\,\frac{1}{\ep^4}-590\,\frac{1}{\ep^3}+2030\,\frac{1}{\ep^2}+\\
 & +\Big(-2880+48\zeta_{3}\Big)\,\frac{1}{\ep}+\Big(1440-472\zeta_{3}+144\zeta_{2}^{2}\Big)+\Big(1624\zeta_{3}-1416\zeta_{2}^{2}+\nonumber \\
 & +1728\zeta_{5}\Big)\,\ep+\Big(-2304\zeta_{3}+4872\zeta_{2}^{2}-16992\zeta_{5}+\frac{10560}{7}\zeta_{2}^{3}+192\zeta_{3}^{2}\Big)\,\ep^2+\nonumber \\
 & +\mathcal{O}\!\left(\ep^3\right)\Big] \nonumber \\
\mathrm{VRRR}_{35} & =\smallfig{vrrr/35}=B^*\,\mathrm{PS}_{4}\left(q^{2}\right)^{-2-4\ep}\Big[9\,\frac{1}{\ep^4}-\frac{177}{2}\,\frac{1}{\ep^3}+\Big(\frac{609}{2}-30\zeta_{2}\Big)\,\frac{1}{\ep^2}+\\
 & +\Big(-432+295\zeta_{2}-294\zeta_{3}\Big)\,\frac{1}{\ep}+\Big(216-1015\zeta_{2}+2891\zeta_{3}-\frac{3414}{5}\zeta_{2}^{2}\Big)+\nonumber \\
 & +\Big(1440\zeta_{2}-9947\zeta_{3}+\frac{33571}{5}\zeta_{2}^{2}-492\zeta_{2}\zeta_{3}-7266\zeta_{5}\Big)\,\ep+\Big(-720\zeta_{2}+\nonumber \\
 & +14112\zeta_{3}-\frac{115507}{5}\zeta_{2}^{2}+4838\zeta_{2}\zeta_{3}+71449\zeta_{5}-\frac{262796}{35}\zeta_{2}^{3}-2310\zeta_{3}^{2}\Big)\,\ep^2+\nonumber \\
 & +\mathcal{O}\!\left(\ep^3\right)\Big]\nonumber
\end{align}

\subsection{VVRR}
\label{sec:results-vvrr}

\begin{align}
\mathrm{VVRR}_{1} & =\smallfig{vvrr/1}=B^{2}\mathrm{PS}_{3}\left(q^{2}\right)^{2-4\ep}\Big[-\frac{1}{4}\,\ep-\frac{5}{8}\,\ep^{2}-\frac{27}{16}\,\ep^{3}+\Big(-\frac{153}{32}+\\
 & +\frac{3}{2}\zeta_{3}\Big)\,\ep^{4}+\Big(-\frac{891}{64}+\frac{15}{4}\zeta_{3}+\frac{9}{10}\zeta_{2}^{2}\Big)\,\ep^{5}+\Big(-\frac{5265}{128}+\frac{81}{8}\zeta_{3}+\frac{9}{4}\zeta_{2}^{2}+\nonumber \\
 & +\frac{21}{2}\zeta_{5}\Big)\,\ep^{6}+\Big(-\frac{31347}{256}+\frac{459}{16}\zeta_{3}+\frac{243}{40}\zeta_{2}^{2}+\frac{105}{4}\zeta_{5}+\frac{36}{7}\zeta_{2}^{3}-\frac{9}{2}\zeta_{3}^{2}\Big)\,\ep^{7}+\nonumber \\
 & +\mathcal{O}\!\left(\ep^{8}\right)\Big]\nonumber \\
\mathrm{VVRR}_{2} & =\smallfig{vvrr/2}=B^{2}\mathrm{PS}_{3}\left(q^{2}\right)^{1-4\ep}\Big[\frac{1}{2}+2\,\ep+\Big(\frac{37}{4}-3\zeta_{2}\Big)\,\ep^{2}+\Big(\frac{343}{8}-\\
 & -9\zeta_{2}-15\zeta_{3}\Big)\,\ep^{3}+\Big(\frac{3223}{16}-\frac{87}{2}\zeta_{2}-48\zeta_{3}-\frac{189}{10}\zeta_{2}^{2}\Big)\,\ep^{4}+\Big(\frac{30763}{32}-\nonumber \\
 & -\frac{855}{4}\zeta_{2}-\frac{459}{2}\zeta_{3}-\frac{117}{2}\zeta_{2}^{2}+84\zeta_{2}\zeta_{3}-282\zeta_{5}\Big)\,\ep^{5}+\Big(\frac{297703}{64}-\frac{8475}{8}\zeta_{2}-\nonumber \\
 & -\frac{4449}{4}\zeta_{3}-\frac{1125}{4}\zeta_{2}^{2}+252\zeta_{2}\zeta_{3}-867\zeta_{5}-\frac{9657}{70}\zeta_{2}^{3}+213\zeta_{3}^{2}\Big)\,\ep^{6}+\mathcal{O}\!\left(\ep^{7}\right)\Big]\nonumber \\
\mathrm{VVRR}_{3} & =\smallfig{vvrr/3}=\left(B^{*}\right)^{2}\mathrm{PS}_{3}\left(q^{2}\right)^{1-4\ep}\Big[1+\frac{3}{2}\,\ep+\Big(\frac{11}{2}-2\zeta_{2}\Big)\,\ep^{2}+\Big(21-\\
 & -3\zeta_{2}-10\zeta_{3}\Big)\,\ep^{3}+\Big(82-11\zeta_{2}-15\zeta_{3}-14\zeta_{2}^{2}\Big)\,\ep^{4}+\Big(324-42\zeta_{2}-55\zeta_{3}-\nonumber \\
 & -21\zeta_{2}^{2}+20\zeta_{2}\zeta_{3}-150\zeta_{5}\Big)\,\ep^{5}+\Big(1288-164\zeta_{2}-210\zeta_{3}-77\zeta_{2}^{2}+30\zeta_{2}\zeta_{3}-\nonumber \\
 & -225\zeta_{5}-\frac{476}{5}\zeta_{2}^{3}+50\zeta_{3}^{2}\Big)\,\ep^{6}+\mathcal{O}\!\left(\ep^{7}\right)\Big]\nonumber \\
\mathrm{VVRR}_{4} & =\smallfig{vvrr/4}=B^{2}\mathrm{PS}_{3}\left(q^{2}\right)^{1-4\ep}\\
\mathrm{VVRR}_{5} & =\smallfig{vvrr/5}=B^{2}\mathrm{PS}_{3}\left(q^{2}\right)^{1-4\ep}\Big[\frac{1}{2}+2\,\ep+\Big(\frac{41}{4}-\zeta_{2}\Big)\,\ep^{2}+\Big(\frac{415}{8}-4\zeta_{2}-\\
 & -14\zeta_{3}\Big)\,\ep^{3}+\Big(\frac{4175}{16}-\frac{41}{2}\zeta_{2}-56\zeta_{3}-\frac{116}{5}\zeta_{2}^{2}\Big)\,\ep^{4}+\Big(\frac{41875}{32}-\frac{415}{4}\zeta_{2}-\nonumber \\
 & -287\zeta_{3}-\frac{464}{5}\zeta_{2}^{2}+28\zeta_{2}\zeta_{3}-282\zeta_{5}\Big)\,\ep^{5}+\Big(\frac{419375}{64}-\frac{4175}{8}\zeta_{2}-\frac{2905}{2}\zeta_{3}-\nonumber \\
 & -\frac{2378}{5}\zeta_{2}^{2}+112\zeta_{2}\zeta_{3}-1128\zeta_{5}-\frac{1156}{5}\zeta_{2}^{3}+196\zeta_{3}^{2}\Big)\,\ep^{6}+\mathcal{O}\!\left(\ep^{7}\right)\Big]\nonumber \\
\mathrm{VVRR}_{6} & =\smallfig{vvrr/6}=\left(B^{*}\right)^{2}\mathrm{PS}_{3}\left(q^{2}\right)^{1-4\ep}\Big[1+3\,\ep+\Big(\frac{29}{2}-4\zeta_{2}\Big)\,\ep^{2}+\Big(\frac{285}{4}-\\
 & -12\zeta_{2}-24\zeta_{3}\Big)\,\ep^{3}+\Big(\frac{2825}{8}-58\zeta_{2}-72\zeta_{3}-\frac{192}{5}\zeta_{2}^{2}\Big)\,\ep^{4}+\Big(\frac{28125}{16}-\nonumber \\
 & -285\zeta_{2}-348\zeta_{3}-\frac{576}{5}\zeta_{2}^{2}+96\zeta_{2}\zeta_{3}-528\zeta_{5}\Big)\,\ep^{5}+\Big(\frac{280625}{32}-\frac{2825}{2}\zeta_{2}-\nonumber \\
 & -1710\zeta_{3}-\frac{2784}{5}\zeta_{2}^{2}+288\zeta_{2}\zeta_{3}-1584\zeta_{5}-\frac{1824}{5}\zeta_{2}^{3}+288\zeta_{3}^{2}\Big)\,\ep^{6}+\mathcal{O}\!\left(\ep^{7}\right)\Big]\nonumber \\
\mathrm{VVRR}_{7} & =\smallfig{vvrr/7}=B^{2}\mathrm{PS}_{3}\left(q^{2}\right)^{2-4\ep}\Big[-\frac{1}{12}\,\ep-\frac{25}{72}\,\ep^{2}+\Big(-\frac{691}{432}+\frac{1}{3}\zeta_{2}\Big)\,\ep^{3}+\\
 & +\Big(-\frac{20005}{2592}+\frac{25}{18}\zeta_{2}+\frac{5}{2}\zeta_{3}\Big)\,\ep^{4}+\Big(-\frac{590875}{15552}+\frac{691}{108}\zeta_{2}+\frac{125}{12}\zeta_{3}+\nonumber \\
 & +\frac{7}{2}\zeta_{2}^{2}\Big)\,\ep^{5}+\Big(-\frac{17603125}{93312}+\frac{20005}{648}\zeta_{2}+\frac{3455}{72}\zeta_{3}+\frac{175}{12}\zeta_{2}^{2}-10\zeta_{2}\zeta_{3}+\nonumber \\
 & +\frac{95}{2}\zeta_{5}\Big)\,\ep^{6}+\Big(-\frac{526406875}{559872}+\frac{590875}{3888}\zeta_{2}+\frac{100025}{432}\zeta_{3}+\frac{4837}{72}\zeta_{2}^{2}-\nonumber \\
 & -\frac{125}{3}\zeta_{2}\zeta_{3}+\frac{2375}{12}\zeta_{5}+\frac{1082}{35}\zeta_{2}^{3}-\frac{75}{2}\zeta_{3}^{2}\Big)\,\ep^{7}+\mathcal{O}\!\left(\ep^{8}\right)\Big]\nonumber \\
\mathrm{VVRR}_{8} & =\smallfig{vvrr/8}=B^{2}\mathrm{PS}_{3}\left(q^{2}\right)^{-1-4\ep}\Big[\frac{2}{3}\,\frac{1}{\ep^{3}}-\frac{23}{3}\,\frac{1}{\ep^{2}}+\Big(\frac{122}{3}-6\zeta_{2}\Big)\,\frac{1}{\ep}+\\
 & +\Big(-146+69\zeta_{2}-\frac{94}{3}\zeta_{3}\Big)+\Big(450-366\zeta_{2}+\frac{1081}{3}\zeta_{3}-26\zeta_{2}^{2}\Big)\,\ep+\nonumber \\
 & +\Big(-1350+1314\zeta_{2}-\frac{5734}{3}\zeta_{3}+299\zeta_{2}^{2}+224\zeta_{2}\zeta_{3}-422\zeta_{5}\Big)\,\ep^{2}+\Big(4050-\nonumber \\
 & -4050\zeta_{2}+6862\zeta_{3}-1586\zeta_{2}^{2}-2576\zeta_{2}\zeta_{3}+4853\zeta_{5}-\frac{1552}{21}\zeta_{2}^{3}+\nonumber \\
 & +\frac{2014}{3}\zeta_{3}^{2}\Big)\,\ep^{3}+\mathcal{O}\!\left(\ep^{4}\right)\Big]\nonumber \\
\mathrm{VVRR}_{9} & =\smallfig{vvrr/9}=B^{2}\mathrm{PS}_{3}\left(q^{2}\right)^{-1-4\ep}\Big[\frac{1}{\ep^{3}}-\frac{13}{2}\,\frac{1}{\ep^{2}}+\Big(\frac{27}{2}-6\zeta_{2}\Big)\,\frac{1}{\ep}+\Big(-9+\\
 & +39\zeta_{2}-38\zeta_{3}\Big)+\Big(-81\zeta_{2}+247\zeta_{3}-\frac{259}{5}\zeta_{2}^{2}\Big)\,\ep+\Big(54\zeta_{2}-513\zeta_{3}+\nonumber \\
 & +\frac{3367}{10}\zeta_{2}^{2}+202\zeta_{2}\zeta_{3}-755\zeta_{5}\Big)\,\ep^{2}+\Big(342\zeta_{3}-\frac{6993}{10}\zeta_{2}^{2}-1313\zeta_{2}\zeta_{3}+\nonumber \\
 & +\frac{9815}{2}\zeta_{5}-\frac{44701}{105}\zeta_{2}^{3}+646\zeta_{3}^{2}\Big)\,\ep^{3}+\mathcal{O}\!\left(\ep^{4}\right)\Big]\nonumber \\
\mathrm{VVRR}_{10} & =\smallfig{vvrr/10}=\left(B^{*}\right)^{2}\mathrm{PS}_{3}\left(q^{2}\right)^{1-4\ep}\Big[\frac{1}{2}+\frac{1}{2}\,\ep+\Big(\frac{3}{4}+\zeta_{2}\Big)\,\ep^{2}+\Big(-\frac{19}{8}+\\
 & +\frac{5}{2}\zeta_{2}\Big)\,\ep^{3}+\Big(-\frac{551}{16}+\frac{27}{2}\zeta_{2}+\frac{15}{2}\zeta_{3}+\frac{9}{2}\zeta_{2}^{2}\Big)\,\ep^{4}+\Big(-\frac{8259}{32}+\frac{145}{2}\zeta_{2}+\nonumber \\
 & +60\zeta_{3}+\frac{69}{4}\zeta_{2}^{2}-30\zeta_{2}\zeta_{3}+\frac{165}{2}\zeta_{5}\Big)\,\ep^{5}+\Big(-\frac{103951}{64}+\frac{1535}{4}\zeta_{2}+\frac{1545}{4}\zeta_{3}+\nonumber \\
 & +\frac{435}{4}\zeta_{2}^{2}-75\zeta_{2}\zeta_{3}+\frac{1065}{4}\zeta_{5}+\frac{3937}{70}\zeta_{2}^{3}-75\zeta_{3}^{2}\Big)\,\ep^{6}+\mathcal{O}\!\left(\ep^{7}\right)\Big]\nonumber \\
\mathrm{VVRR}_{11} & =\smallfig{vvrr/11}=\left(B^{*}\right)^{2}\mathrm{PS}_{3}\left(q^{2}\right)^{-1-4\ep}\Big[4\zeta_{3}+\Big(-26\zeta_{3}+12\zeta_{2}^{2}\Big)\,\ep+\\
 & +\Big(54\zeta_{3}-78\zeta_{2}^{2}+28\zeta_{2}\zeta_{3}+76\zeta_{5}\Big)\,\ep^{2}+\Big(-36\zeta_{3}+162\zeta_{2}^{2}-182\zeta_{2}\zeta_{3}-\nonumber \\
 & -494\zeta_{5}+\frac{4412}{35}\zeta_{2}^{3}-88\zeta_{3}^{2}\Big)\,\ep^{3}+\mathcal{O}\!\left(\ep^{4}\right)\Big]\nonumber \\
\mathrm{VVRR}_{12} & =\smallfig{vvrr/12}=B^{2}\mathrm{PS}_{3}\left(q^{2}\right)^{-1-4\ep}\Big[2\zeta_{2}\,\frac{1}{\ep}+\Big(-13\zeta_{2}+16\zeta_{3}\Big)+\Big(27\zeta_{2}-\\
 & -104\zeta_{3}+\frac{156}{5}\zeta_{2}^{2}\Big)\,\ep+\Big(-18\zeta_{2}+216\zeta_{3}-\frac{1014}{5}\zeta_{2}^{2}-90\zeta_{2}\zeta_{3}+448\zeta_{5}\Big)\,\ep^{2}+\nonumber \\
 & +\Big(-144\zeta_{3}+\frac{2106}{5}\zeta_{2}^{2}+585\zeta_{2}\zeta_{3}-2912\zeta_{5}+\frac{34304}{105}\zeta_{2}^{3}-401\zeta_{3}^{2}\Big)\,\ep^{3}+\nonumber \\
 & +\mathcal{O}\!\left(\ep^{4}\right)\Big]\nonumber \\
\mathrm{VVRR}_{13} & =\smallfig{vvrr/13}=B^{2}\mathrm{PS}_{3}\left(q^{2}\right)^{-1-4\ep}\Big[\Big(-2\zeta_{2}\zeta_{3}-20\zeta_{5}\Big)\,\ep+\Big(17\zeta_{2}\zeta_{3}+\\
 & +170\zeta_{5}-\frac{448}{15}\zeta_{2}^{3}-42\zeta_{3}^{2}\Big)\,\ep^{2}+\mathcal{O}\!\left(\ep^{3}\right)\Big]\nonumber \\
\mathrm{VVRR}_{14} & =\smallfig{vvrr/14}=\left(B^{*}\right)^{2}\mathrm{PS}_{3}\left(q^{2}\right)^{-4\ep}\Big[-8\zeta_{3}\,\ep+\Big(-4\zeta_{3}-\frac{42}{5}\zeta_{2}^{2}\Big)\,\ep^{2}+\\
 & +\Big(-88\zeta_{3}-\frac{21}{5}\zeta_{2}^{2}+56\zeta_{2}\zeta_{3}-114\zeta_{5}\Big)\,\ep^{3}+\Big(-424\zeta_{3}-\frac{462}{5}\zeta_{2}^{2}+28\zeta_{2}\zeta_{3}-\nonumber \\
 & -57\zeta_{5}-\frac{566}{35}\zeta_{2}^{3}+308\zeta_{3}^{2}\Big)\,\ep^{4}+\mathcal{O}\!\left(\ep^{5}\right)\Big]\nonumber \\
\mathrm{VVRR}_{15} & =\smallfig{vvrr/15}=B^{2}\mathrm{PS}_{3}\left(q^{2}\right)^{-1-4\ep}\Big[12\zeta_{3}+\Big(-30\zeta_{3}+12\zeta_{2}^{2}\Big)\,\ep+\Big(138\zeta_{3}-\\
 & -30\zeta_{2}^{2}-120\zeta_{2}\zeta_{3}+188\zeta_{5}\Big)\,\ep^{2}+\Big(396\zeta_{3}+138\zeta_{2}^{2}+300\zeta_{2}\zeta_{3}-470\zeta_{5}-\nonumber \\
 & -\frac{296}{35}\zeta_{2}^{3}-672\zeta_{3}^{2}\Big)\,\ep^{3}+\mathcal{O}\!\left(\ep^{4}\right)\Big]\nonumber \\
\mathrm{VVRR}_{16} & =\smallfig{vvrr/16}=B^{2}\mathrm{PS}_{3}\left(q^{2}\right)^{-4\ep}\Big[4\zeta_{3}+\Big(-14\zeta_{3}+8\zeta_{2}^{2}\Big)\,\ep+\Big(36\zeta_{3}-\\
 & -28\zeta_{2}^{2}-56\zeta_{2}\zeta_{3}+156\zeta_{5}\Big)\,\ep^{2}+\Big(36\zeta_{3}+72\zeta_{2}^{2}+196\zeta_{2}\zeta_{3}-546\zeta_{5}+\nonumber \\
 & +\frac{464}{7}\zeta_{2}^{3}-224\zeta_{3}^{2}\Big)\,\ep^{3}+\mathcal{O}\!\left(\ep^{4}\right)\Big]\nonumber \\
\mathrm{VVRR}_{17} & =\smallfig{vvrr/17}=B^{2}\mathrm{PS}_{3}\left(q^{2}\right)^{-3-4\ep}\Big[\frac{19}{36}\,\frac{1}{\ep^{4}}+\frac{29}{24}\,\frac{1}{\ep^{3}}+\Big(-\frac{103}{3}-\\
 & -\frac{26}{3}\zeta_{2}\Big)\,\frac{1}{\ep^{2}}+\Big(\frac{772}{9}+\frac{113}{9}\zeta_{2}-\frac{599}{18}\zeta_{3}\Big)\,\frac{1}{\ep}+\Big(\frac{881}{2}+320\zeta_{2}-\frac{907}{36}\zeta_{3}+\nonumber \\
 & +\frac{707}{18}\zeta_{2}^{2}\Big)+\Big(-\frac{12587}{3}-\frac{3751}{3}\zeta_{2}+1836\zeta_{3}-\frac{85421}{180}\zeta_{2}^{2}+446\zeta_{2}\zeta_{3}+\nonumber \\
 & +\frac{9619}{18}\zeta_{5}\Big)\,\ep+\Big(\frac{39819}{2}-\frac{15985}{9}\zeta_{2}-\frac{52933}{9}\zeta_{3}+\frac{26666}{15}\zeta_{2}^{2}-\frac{10945}{9}\zeta_{2}\zeta_{3}-\nonumber \\
 & -\frac{248249}{36}\zeta_{5}+\frac{1307018}{945}\zeta_{2}^{3}+\frac{22225}{18}\zeta_{3}^{2}\Big)\,\ep^{2}+\mathcal{O}\!\left(\ep^{3}\right)\Big]\nonumber \\
\mathrm{VVRR}_{18} & =\smallfig{vvrr/18}=B^{2}\mathrm{PS}_{3}\left(q^{2}\right)^{-2-4\ep}\Big[\frac{1}{6}\,\frac{1}{\ep^{3}}-\frac{23}{12}\,\frac{1}{\ep^{2}}+\Big(\frac{61}{6}+2\zeta_{2}\Big)\,\frac{1}{\ep}+\\
 & +\Big(-\frac{73}{2}-23\zeta_{2}+\frac{29}{3}\zeta_{3}\Big)+\Big(\frac{225}{2}+122\zeta_{2}-\frac{667}{6}\zeta_{3}+\frac{79}{15}\zeta_{2}^{2}\Big)\,\ep+\Big(-\nonumber \\
 & -\frac{675}{2}-438\zeta_{2}+\frac{1769}{3}\zeta_{3}-\frac{1817}{30}\zeta_{2}^{2}+\frac{484}{3}\zeta_{2}\zeta_{3}-\frac{889}{3}\zeta_{5}\Big)\,\ep^{2}+\Big(\frac{2025}{2}+\nonumber \\
 & +1350\zeta_{2}-2117\zeta_{3}+\frac{4819}{15}\zeta_{2}^{2}-\frac{5566}{3}\zeta_{2}\zeta_{3}+\frac{20447}{6}\zeta_{5}-\frac{18932}{315}\zeta_{2}^{3}+\nonumber \\
 & +\frac{1057}{3}\zeta_{3}^{2}\Big)\,\ep^{3}+\mathcal{O}\!\left(\ep^{4}\right)\Big]\nonumber \\
\mathrm{VVRR}_{19} & =\smallfig{vvrr/19}=B^{2}\mathrm{PS}_{3}\left(q^{2}\right)^{-3-4\ep}\Big[\frac{43}{18}\,\frac{1}{\ep^{4}}-\frac{449}{36}\,\frac{1}{\ep^{3}}+\Big(-\frac{229}{9}-\\
 & -\frac{188}{9}\zeta_{2}\Big)\,\frac{1}{\ep^{2}}+\Big(\frac{2227}{6}+\frac{1106}{9}\zeta_{2}-\frac{1021}{9}\zeta_{3}\Big)\,\frac{1}{\ep}+\Big(-\frac{3017}{2}+\frac{496}{9}\zeta_{2}+\nonumber \\
 & +\frac{12719}{18}\zeta_{3}-\frac{608}{5}\zeta_{2}^{2}\Big)+\Big(\frac{8679}{2}-\frac{6818}{3}\zeta_{2}-\frac{604}{9}\zeta_{3}+\frac{12182}{15}\zeta_{2}^{2}+\nonumber \\
 & +\frac{7024}{9}\zeta_{2}\zeta_{3}-\frac{6185}{3}\zeta_{5}\Big)\,\ep+\Big(-\frac{22341}{2}+9136\zeta_{2}-\frac{33785}{3}\zeta_{3}-\frac{9301}{15}\zeta_{2}^{2}-\nonumber \\
 & -\frac{40312}{9}\zeta_{2}\zeta_{3}+\frac{76147}{6}\zeta_{5}-\frac{86417}{105}\zeta_{2}^{3}+\frac{16543}{9}\zeta_{3}^{2}\Big)\,\ep^{2}+\mathcal{O}\!\left(\ep^{3}\right)\Big]\nonumber \\
\mathrm{VVRR}_{20} & =\smallfig{vvrr/20}=B^{2}\mathrm{PS}_{3}\left(q^{2}\right)^{-3-4\ep}\Big[\frac{14}{9}\,\frac{1}{\ep^{4}}-\frac{17}{3}\,\frac{1}{\ep^{3}}+\Big(-\frac{133}{3}-\\
 & -\frac{143}{9}\zeta_{2}\Big)\,\frac{1}{\ep^{2}}+\Big(\frac{3458}{9}+\frac{437}{6}\zeta_{2}-\frac{1099}{9}\zeta_{3}\Big)\,\frac{1}{\ep}+\Big(-\frac{4388}{3}+\frac{533}{2}\zeta_{2}+\nonumber \\
 & +\frac{1239}{2}\zeta_{3}-\frac{3439}{15}\zeta_{2}^{2}\Big)+\Big(4152-\frac{25106}{9}\zeta_{2}+\frac{8641}{6}\zeta_{3}+\frac{37493}{30}\zeta_{2}^{2}+\nonumber \\
 & +\frac{6616}{9}\zeta_{2}\zeta_{3}-\frac{10861}{3}\zeta_{5}\Big)\,\ep+\Big(-10608+\frac{29690}{3}\zeta_{2}-\frac{171190}{9}\zeta_{3}+\nonumber \\
 & +\frac{8908}{5}\zeta_{2}^{2}-3968\zeta_{2}\zeta_{3}+\frac{40027}{2}\zeta_{5}-\frac{796774}{315}\zeta_{2}^{3}+\frac{21877}{9}\zeta_{3}^{2}\Big)\,\ep^{2}+\mathcal{O}\!\left(\ep^{3}\right)\Big]\nonumber \\
\mathrm{VVRR}_{21} & =\smallfig{vvrr/21}=B^{2}\mathrm{PS}_{3}\left(q^{2}\right)^{-3-4\ep}\Big[\frac{4}{3}\,\frac{1}{\ep^{3}}-\frac{20}{3}\,\frac{1}{\ep^{2}}+\Big(-\frac{137}{3}-\\
 & -\frac{179}{9}\zeta_{2}\Big)\,\frac{1}{\ep}+\Big(637+\frac{3185}{18}\zeta_{2}-\frac{1267}{9}\zeta_{3}\Big)+\Big(-\frac{11773}{3}-\frac{4030}{9}\zeta_{2}+\nonumber \\
 & +\frac{22597}{18}\zeta_{3}-\frac{12007}{45}\zeta_{2}^{2}\Big)\,\ep+\Big(18333-\frac{11806}{9}\zeta_{2}-\frac{29687}{9}\zeta_{3}+\frac{213787}{90}\zeta_{2}^{2}+\nonumber \\
 & +736\zeta_{2}\zeta_{3}-\frac{37105}{9}\zeta_{5}\Big)\,\ep^{2}+\Big(-75579+\frac{49348}{3}\zeta_{2}-\frac{67181}{9}\zeta_{3}-\nonumber \\
 & -\frac{113723}{18}\zeta_{2}^{2}-6768\zeta_{2}\zeta_{3}+\frac{669355}{18}\zeta_{5}-\frac{2858368}{945}\zeta_{2}^{3}+\frac{6749}{3}\zeta_{3}^{2}\Big)\,\ep^{3}+\nonumber \\
 & +\mathcal{O}\!\left(\ep^{4}\right)\Big]\nonumber \\
\mathrm{VVRR}_{22} & =\smallfig{vvrr/22}=\left(B^{*}\right)^{2}\mathrm{PS}_{3}\left(q^{2}\right)^{-2-4\ep}\Big[5\,\frac{1}{\ep^{3}}-\frac{65}{2}\,\frac{1}{\ep^{2}}+\Big(\frac{135}{2}-24\zeta_{2}\Big)\,\frac{1}{\ep}+\\
 & +\Big(-45+156\zeta_{2}-128\zeta_{3}\Big)+\Big(-324\zeta_{2}+832\zeta_{3}-\frac{896}{5}\zeta_{2}^{2}\Big)\,\ep+\Big(216\zeta_{2}-\nonumber \\
 & -1728\zeta_{3}+\frac{5824}{5}\zeta_{2}^{2}+320\zeta_{2}\zeta_{3}-2000\zeta_{5}\Big)\,\ep^{2}+\Big(1152\zeta_{3}-\frac{12096}{5}\zeta_{2}^{2}-\nonumber \\
 & -2080\zeta_{2}\zeta_{3}+13000\zeta_{5}-\frac{18208}{15}\zeta_{2}^{3}+832\zeta_{3}^{2}\Big)\,\ep^{3}+\mathcal{O}\!\left(\ep^{4}\right)\Big]\nonumber
\end{align}

\subsection{VRRV}
\label{sec:results-vrrv}

\begin{align}
\mathrm{VRRV}_{1} & =\smallfig{vrrv/1}=\frac{B}{B^{*}}\smallfig{vvrr/4}=\frac{B}{B^{*}}\mathrm{VVRR}_{4}=B\,B^{*}\mathrm{PS}_{3}\left(q^{2}\right)^{1-4\ep}\\
\mathrm{VRRV}_{2} & =\smallfig{vrrv/2}=\frac{B^{*}}{B}\smallfig{vvrr/6}=\frac{B^{*}}{B}\mathrm{VVRR}_{6}\\
\mathrm{VRRV}_{3} & =\smallfig{vrrv/3}=\frac{B}{B^{*}}\smallfig{vvrr/3}=\frac{B}{B^{*}}\mathrm{VVRR}_{3}\\
\mathrm{VRRV}_{4} & =\smallfig{vrrv/4}=B\,B^{*}\mathrm{PS}_{3}\left(q^{2}\right)^{1-4\ep}\Big[1+3\,\ep+\Big(\frac{29}{2}-5\zeta_{2}\Big)\,\ep^{2}+\Big(\frac{285}{4}-\\
 & -15\zeta_{2}-28\zeta_{3}\Big)\,\ep^{3}+\Big(\frac{2825}{8}-\frac{145}{2}\zeta_{2}-84\zeta_{3}-\frac{389}{10}\zeta_{2}^{2}\Big)\,\ep^{4}+\Big(\frac{28125}{16}-\nonumber \\
 & -\frac{1425}{4}\zeta_{2}-406\zeta_{3}-\frac{1167}{10}\zeta_{2}^{2}+140\zeta_{2}\zeta_{3}-564\zeta_{5}\Big)\,\ep^{5}+\Big(\frac{280625}{32}-\nonumber \\
 & -\frac{14125}{8}\zeta_{2}-1995\zeta_{3}-\frac{11281}{20}\zeta_{2}^{2}+420\zeta_{2}\zeta_{3}-1692\zeta_{5}-\frac{653}{2}\zeta_{2}^{3}+\nonumber \\
 & +392\zeta_{3}^{2}\Big)\,\ep^{6}+\mathcal{O}\!\left(\ep^{7}\right)\Big]\nonumber \\
\mathrm{VRRV}_{5} & =\smallfig{vrrv/5}=B\,B^{*}\mathrm{PS}_{3}\left(q^{2}\right)^{-1-4\ep}\Big[\frac{1}{\ep^{3}}-\frac{13}{2}\,\frac{1}{\ep^{2}}+\Big(\frac{27}{2}-8\zeta_{2}\Big)\,\frac{1}{\ep}+\\
 & +\Big(-9+52\zeta_{2}-48\zeta_{3}\Big)+\Big(-108\zeta_{2}+312\zeta_{3}-\frac{334}{5}\zeta_{2}^{2}\Big)\,\ep+\Big(72\zeta_{2}-\nonumber \\
 & -648\zeta_{3}+\frac{2171}{5}\zeta_{2}^{2}+300\zeta_{2}\zeta_{3}-1058\zeta_{5}\Big)\,\ep^{2}+\Big(432\zeta_{3}-\frac{4509}{5}\zeta_{2}^{2}-\nonumber \\
 & -1950\zeta_{2}\zeta_{3}+6877\zeta_{5}-\frac{20522}{35}\zeta_{2}^{3}+880\zeta_{3}^{2}\Big)\,\ep^{3}+\mathcal{O}\!\left(\ep^{4}\right)\Big]\nonumber \\
\mathrm{VRRV}_{6} & =\smallfig{vrrv/6}=B\,B^{*}\mathrm{PS}_{3}\left(q^{2}\right)^{-1-4\ep}\Big[\frac{7}{5}\zeta_{2}^{2}+\Big(-\frac{119}{10}\zeta_{2}^{2}-38\zeta_{2}\zeta_{3}+\\
 & +97\zeta_{5}\Big)\,\ep+\Big(\frac{371}{10}\zeta_{2}^{2}+323\zeta_{2}\zeta_{3}-\frac{1649}{2}\zeta_{5}+\frac{2361}{35}\zeta_{2}^{3}-128\zeta_{3}^{2}\Big)\,\ep^{2}+\nonumber \\
 & +\mathcal{O}\!\left(\ep^{3}\right)\Big]\nonumber \\
\mathrm{VRRV}_{7} & =\smallfig{vrrv/7}=B\,B^{*}\mathrm{PS}_{3}\left(q^{2}\right)^{-3-4\ep}\Big[\frac{52}{9}\,\frac{1}{\ep^{3}}-\frac{314}{9}\,\frac{1}{\ep^{2}}+\Big(-\frac{946}{9}-\\
 & -\frac{736}{9}\zeta_{2}\Big)\,\frac{1}{\ep}+\Big(\frac{18290}{9}+\frac{6092}{9}\zeta_{2}-\frac{5008}{9}\zeta_{3}\Big)+\Big(-\frac{38710}{3}-\frac{10502}{9}\zeta_{2}+\nonumber \\
 & +\frac{42824}{9}\zeta_{3}-\frac{4424}{5}\zeta_{2}^{2}\Big)\,\ep+\Big(60010-\frac{86690}{9}\zeta_{2}-\frac{91352}{9}\zeta_{3}+\frac{38402}{5}\zeta_{2}^{2}+\nonumber \\
 & +\frac{35536}{9}\zeta_{2}\zeta_{3}-15192\zeta_{5}\Big)\,\ep^{2}+\Big(-243930+\frac{255430}{3}\zeta_{2}-\frac{446504}{9}\zeta_{3}-\nonumber \\
 & -\frac{88991}{5}\zeta_{2}^{2}-\frac{323768}{9}\zeta_{2}\zeta_{3}+135836\zeta_{5}-\frac{3043568}{315}\zeta_{2}^{3}+\frac{106448}{9}\zeta_{3}^{2}\Big)\,\ep^{3}+\nonumber \\
 & +\mathcal{O}\!\left(\ep^{4}\right)\Big]\nonumber \\
\mathrm{VRRV}_{8} & =\smallfig{vrrv/8}=B\,B^{*}\mathrm{PS}_{3}\left(q^{2}\right)^{-3-4\ep}\Big[\frac{14}{3}\,\frac{1}{\ep^{4}}-\frac{55}{3}\,\frac{1}{\ep^{3}}+\Big(-\frac{353}{3}-\\
 & -58\zeta_{2}\Big)\,\frac{1}{\ep^{2}}+\Big(\frac{3214}{3}+273\zeta_{2}-400\zeta_{3}\Big)\,\frac{1}{\ep}+\Big(-4096+881\zeta_{2}+\nonumber \\
 & +1960\zeta_{3}-\frac{1799}{3}\zeta_{2}^{2}\Big)+\Big(11556-9596\zeta_{2}+5224\zeta_{3}+\frac{18035}{6}\zeta_{2}^{2}+\nonumber \\
 & +3320\zeta_{2}\zeta_{3}-10760\zeta_{5}\Big)\,\ep+\Big(-29124+33360\zeta_{2}-62032\zeta_{3}+\nonumber \\
 & +\frac{214823}{30}\zeta_{2}^{2}-17340\zeta_{2}\zeta_{3}+55500\zeta_{5}-\frac{211783}{35}\zeta_{2}^{3}+10600\zeta_{3}^{2}\Big)\,\ep^{2}+\nonumber \\
 & +\mathcal{O}\!\left(\ep^{3}\right)\Big]\nonumber \\
\mathrm{VRRV}_{9} & =\smallfig{vrrv/9}=\frac{B}{B^{*}}\smallfig{vvrr/22}=\frac{B}{B^{*}}\mathrm{VVRR}_{22}
\end{align}

\section{Table of loop integrals\label{sec:Table-of-loop-integrals}}

Here we collect integrals used during the 1$\to$3 amplitude calculation.
Most of them are taken from~\cite{GR99}, eq.~\eqref{eq:known-tricross}
is from~\cite{GHM05}.

To obtain a series in $\ep$ from these integrals, the hypergeometric
function $\!_{p}F_{q}$ needs to be expanded about its parameters.
This can be conveniently done using the Mathematica package \noun{HypExp}~\cite{HM07}.

\begin{align}
\smallfig{known/bubble} & =\frac{i\pi^{\frac{d}{2}}}{\left(2\pi\right)^{d}}\frac{\Gamma^{2}\!\left(\frac{d}{2}-1\right)\Gamma\!\left(2-\frac{d}{2}\right)}{\Gamma\!\left(d-2\right)}\left(-q^{2}-i0\right)^{\frac{d}{2}-2}\label{eq:known-bubble}\\
\smallfig{known/box} & =\frac{i\pi^{\frac{d}{2}-1}}{\left(2\pi\right)^{d-1}}\frac{\Gamma^{2}\!\left(\frac{d}{2}-2\right)\Gamma\!\left(3-\frac{d}{2}\right)}{\Gamma\!\left(d-3\right)}\left(-q^{2}-i0\right)^{\frac{d}{2}-4}\frac{1}{s_{12}s_{23}}\Big[\label{eq:known-box}\\
 & +\left(\frac{s_{12}s_{23}}{1-s_{12}}\right)^{\frac{d-4}{2}}\!\!_{2}F_{1}\!\left(\frac{d}{2}-2,\frac{d}{2}-2;\frac{d}{2}-1;\frac{1-s_{12}-s_{23}}{1-s_{12}}\right)+\nonumber \\
 & +\left(\frac{s_{12}s_{23}}{1-s_{23}}\right)^{\frac{d-4}{2}}\!\!_{2}F_{1}\!\left(\frac{d}{2}-2,\frac{d}{2}-2;\frac{d}{2}-1;\frac{1-s_{12}-s_{23}}{1-s_{23}}\right)+\nonumber \\
 & -\left(\frac{s_{12}s_{23}}{\left(1-s_{12}\right)\left(1-s_{23}\right)}\right)^{\frac{d-4}{2}}\!\!_{2}F_{1}\!\left(\frac{d}{2}-2,\frac{d}{2}-2;\frac{d}{2}-1;\frac{1-s_{12}-s_{23}}{\left(1-s_{12}\right)\left(1-s_{23}\right)}\right)\Big]\nonumber \\
\smallfig{known/sunset} & =\frac{1}{\left(4\pi\right)^{d}}\frac{\Gamma^{3}\!\left(\frac{d}{2}-1\right)\Gamma\!\left(3-d\right)}{\Gamma\!\left(3\frac{d}{2}-3\right)}\left(-q^{2}-i0\right)^{d-3}\\
\smallfig{known/triangle2l-1} & =\frac{-2}{\left(4\pi\right)^{d}}\frac{\Gamma^{2}\!\left(\frac{d}{2}-1\right)\Gamma\!\left(3-\frac{d}{2}\right)\Gamma\!\left(3-d\right)\Gamma\!\left(d-4\right)}{\Gamma\!\left(\frac{3d}{2}-4\right)}\left(-q^{2}-i0\right)^{d-4}\\
\smallfig{known/triangle2l-2} & =\frac{1}{\left(4\pi\right)^{d}}\left(-q^{2}-i0\right)^{d-4}\Big[\left(1-s_{12}\right)^{\frac{d-4}{2}}\frac{\Gamma^{2}\!\left(\frac{d}{2}-1\right)\Gamma\!\left(3-d\right)\Gamma\!\left(2-\frac{d}{2}\right)\Gamma\!\left(d-3\right)}{\Gamma\!\left(\frac{3d}{2}-4\right)}-\nonumber \\
 & -\frac{1}{2}\frac{3d-8}{d-3}\frac{\Gamma^{3}\!\left(\frac{d}{2}-1\right)\Gamma\!\left(3-d\right)}{\Gamma\!\left(\frac{3d}{2}-3\right)}\left(s_{12}\right)^{d-3}\,_{2}F_{1}\!\left(1,\frac{d}{2}-1;d-2;\frac{p_{12}^{2}}{q^{2}}\right)\Big]\\
\smallfig{known/triangle2l-3} & =\frac{-1}{\left(4\pi\right)^{d}}\frac{\Gamma\!\left(\frac{d}{2}-1\right)^{2}\Gamma\!\left(\frac{d}{2}-2\right)\Gamma\!\left(3-d\right)}{\Gamma\left(3\frac{d}{2}-4\right)}\times\\
 & \times\!_{2}F_{1}\!\left(\frac{d-2}{2},2-\frac{d}{2};3-\frac{d}{2};1-\frac{p_{12}}{q^{2}}\right)\left(s_{12}\right)^{\frac{d-4}{2}}\left(-q^{2}-i0\right)^{d-4}\nonumber \\
\smallfig{known/tricross} & =\frac{1}{\left(4\pi\right)^{d}}\left(-q^{2}-i0\right)^{d-6}\Big[-16\frac{\Gamma^{4}\!\left(d-4\right)\Gamma^{3}\!\left(5-d\right)\Gamma\!\left(3-\frac{d}{2}\right)\Gamma\!\left(\frac{d}{2}-1\right)}{\Gamma^{2}\!\left(2d-7\right)\Gamma\!\left(9-2d\right)}+\label{eq:known-tricross}\\
 & +32\frac{\Gamma^{2}\!\left(\frac{d}{2}-1\right)\Gamma\!\left(4-d\right)\Gamma\!\left(d-6\right)}{\Gamma\!\left(2d-7\right)}\,_{3}F_{2}\!\left(1,1,5-d;6-d,4-\frac{d}{2};1\right)+\nonumber \\
 & +\frac{\Gamma^{2}\!\left(\frac{d}{2}-2\right)\Gamma\!\left(4-d\right)\Gamma\!\left(2-\frac{d}{2}\right)\Gamma\!\left(d-3\right)}{\Gamma\!\left(3\frac{d}{2}-5\right)}\,_{3}F_{2}\!\left(1,d-4,2d-8;d-3,3\frac{d}{2}-5;1\right)-\nonumber \\
 & -\frac{\Gamma^{3}\!\left(\frac{d}{2}-2\right)\Gamma\!\left(4-d\right)}{\Gamma\!\left(3\frac{d}{2}-5\right)}\,_{4}F_{3}\!\left(1,\frac{d}{2}-1,d-4,2d-8;d-3,d-3,3\frac{d}{2}-5;1\right)\Big]\nonumber
\end{align}

\section{Multiple zeta values basis up to weight 12}
\label{sec:mzv}

For the purposes of reconstructing the analytical expressions
from high precision numerical results we use the following linear
basis of irrational MZV combinations:

\begin{tabular}{ll}
    Weight 0: & $1$ \\
    Weight 2: & $\zeta_2$ \\
    Weight 3: & $\zeta_3$ \\
    Weight 4: & $\zeta_2^2$ \\
    Weight 5: & $\zeta_2\,\zeta_3$, $\zeta_5$ \\
    Weight 6: & $\zeta_2^3$, $\zeta_3^2$ \\
    Weight 7: & $\zeta_2^2\,\zeta_3$, $\zeta_2\,\zeta_5$, $\zeta_7$ \\
    Weight 8: & $\zeta_2^4$, $\zeta_2\,\zeta_3^2$, $\zeta_3\,\zeta_5$, $\zeta_{5,3}$ \\
    Weight 9: & $\zeta_2^3\,\zeta_3$, $\zeta_3^3$, $\zeta_2^2\,\zeta_5$, $\zeta_2\,\zeta_7$, $\zeta_9$ \\
    Weight 10: & $\zeta_2^5$, $\zeta_2^2\,\zeta_3^2$, $\zeta_2\,\zeta_3\,\zeta_5$, $\zeta_5^2$, $\zeta_3\,\zeta_7$, $\zeta_2\,\zeta_{5,3}$, $\zeta_{7,3}$ \\
    Weight 11: & $\zeta_2^4\,\zeta_3$, $\zeta_2\,\zeta_3^3$, $\zeta_2^3\,\zeta_5$, $\zeta_3^2\,\zeta_5$, $\zeta_2^2\,\zeta_7$, $\zeta_2\,\zeta_9$, $\zeta_{11}$, $\zeta_3\,\zeta_{5,3}$, $\zeta_{5,3,3}$ \\
    Weight 12: & $\zeta_2^6$, $\zeta_2^3\,\zeta_3^2$, $\zeta_3^4$, $\zeta_2^2\,\zeta_3\,\zeta_5$, $\zeta_2\,\zeta_5^2$, $\zeta_2\,\zeta_3\,\zeta_7$, $\zeta_5\,\zeta_7$, $\zeta_3\,\zeta_9$, $\zeta_2^2\,\zeta_{5,3}$, $\zeta_2\,\zeta_{7,3}$, $\zeta_{9,3}$, $\zeta_{6,4,1,1}$ \\
\end{tabular}

The MZVs here are defined as in eq.~\eqref{eq:mzv}; the basis
itself is extracted from the files provided in~\cite{BBV09}.

\section{Ancillary files\label{sec:Ancillary-files}}

Along with this article we provide ancillary files with all the results
in machine-readable (Mathematica) form. To quickly summarize their
content, we provide:
\begin{description}
\item [{{*}.d4}] ~\\
The values of master integrals for VVVV, VVVR, VVRV, VVRR, VRRV,
VRRR, and RRRR cut structures as series in $\ep$, expanded around
$d=4-2\ep$ up to MZVs of weight~12.
The notation ``\code{Mzv{[}n,...{]}}''
in these files stands for $\zeta_{n,\dots}$, as defined by eq.~\eqref{eq:mzv}.
These values correspond
to those from \autoref{sec:Results}, with the prefactors of
$B$, $B^{*}$, $\mathrm{PS}_{n}$, and $\left(q^{2}\right)^{k}$
omitted.
\item [{topologies}] ~\\
A mapping from a topology name (``\code{L}'', ``\code{J}'', ``\code{H}'',
``\code{M}'', and ``\code{N}'', as listed in \autoref{tab:topologies}), into a list of propagators.
Here ``\code{p1}''\ldots ''\code{p4}'' denote
loop momenta, and ``\code{q}'' denotes the incoming momenta.
\item [{masters}] ~\\
A mapping from master names, for example ``\code{VVVR{[}2{]}}''
into integral definition via the topologies, for example ``\code{H{[}1,0,x,x,0,1,1,1,1,1,0{]}}'',
where integers denote powers of the corresponding propagators, and
``\code{x}'' denotes which propagators have been cut.
\item [{{*}.ldrr}] ~\\
Lowering dimensional recurrence relation matrices as in
eq.~\eqref{eq:ldrr} for each of the master sets.
In these files ``\code{nu}'' stands for $\frac{d}{2}$.
\item [{{*}.st}] ~\\
\noun{SummerTime}~\cite{LM15} files for each of the master sets.
One can use these to calculate series expansion of any master set
around arbitrary $d$, with arbitrary precision. For example, to calculate
the values of the VRRV master integrals as series around $d=4-2\ep$
up to order $\ep^{2}$ with 30 digits of precision, use this
command:\\
\code{Get["VRRV.st"] /. nu->2-ep // Map[TriangleSumsSeries[\#,\{ep,2\},30]\&]}
\item [{1to3/{*}.d4}] ~\\
The values of the 2-loop 1$\to$3 master integrals in topologies PA, NA,
and NB (defined in \autoref{sec:1to3-2l}) as series in~$\ep$, expanded around $d=4-2\ep$ up to MZVs of weight~8.
In these files the notation \code{Hlog{[}x,\{w,...\}{]}}
stands for $G\!\left(w,\dots;x\right)$ as defined in eq.~\eqref{eq:hlog-definition}.
Variables $y$ and $z$ are as defined in eq.~\eqref{eq:1to3-yz}.
\item [{1to3/{*}.ldrr}] ~\\
Lowering dimensional recurrence relation matrices for PA, NA, and
NB topologies of the 1$\to$3 master integrals.
\item [{1to3/{*}.my,~1to3/{*}.mz}] ~\\
Differential equation matrices $M_{ij}^{\left(I,y\right)}$ and $M_{ij}^{\left(I,z\right)}$,
as defined in eq.~\eqref{eq:de-system}.
\item [{1to3/{*}.t}] ~\\
Transformation matrices $T_{ij}^{\left(I\right)}$ as defined in eq.~\eqref{eq:epsilon-form-transformation}.
\end{description}

\bibliographystyle{JHEPmod}
\phantomsection\addcontentsline{toc}{section}{\refname}\bibliography{main}

\end{document}